\documentclass[sigconf]{acmart}

\AtBeginDocument{%
  }

\copyrightyear{2025}
\acmYear{2025}
\setcopyright{cc}
\setcctype{by}
\acmConference[CHI '25]{CHI Conference on Human Factors in Computing Systems}{April 26-May 1, 2025}{Yokohama, Japan}
\acmBooktitle{CHI Conference on Human Factors in Computing Systems (CHI '25), April 26-May 1, 2025, Yokohama, Japan}
\acmDOI{10.1145/3706598.3713865}
\acmISBN{979-8-4007-1394-1/25/04}

\begin{document}

\title{PolicyCraft: Supporting Collaborative and Participatory Policy Design through Case-Grounded Deliberation}

\author{Tzu-Sheng Kuo}
\email{tzushenk@cs.cmu.edu}
\orcid{0000-0002-1504-7640}
\affiliation{
  \institution{Carnegie Mellon University}
  \city{Pittsburgh}
  \state{PA}
  \country{USA}
}

\author{Quan Ze Chen}
\email{cqz@cs.washington.edu}
\orcid{0000-0002-6500-8922}
\affiliation{
  \institution{University of Washington}
  \city{Seattle}
  \state{WA}
  \country{USA}
}

\author{Amy X. Zhang}
\email{axz@cs.uw.edu}
\orcid{0000-0001-9462-9835}
\affiliation{
  \institution{University of Washington}
  \city{Seattle}
  \state{WA}
  \country{USA}
}

\author{Jane Hsieh}
\email{jhsieh2@cs.cmu.edu}
\orcid{0000-0002-4933-1156}
\affiliation{
  \institution{Carnegie Mellon University}
  \city{Pittsburgh}
  \state{PA}
  \country{USA}
}

\author{Haiyi Zhu}
\authornote{Co-senior authors contributed equally.}
\email{haiyiz@cs.cmu.edu}
\orcid{0000-0001-7271-9100}
\affiliation{
  \institution{Carnegie Mellon University}
  \city{Pittsburgh}
  \state{PA}
  \country{USA}
}

\author{Kenneth Holstein}
\authornotemark[1]
\email{kjholste@cs.cmu.edu}
\orcid{0000-0001-6730-922X}
\affiliation{
  \institution{Carnegie Mellon University}
  \city{Pittsburgh}
  \state{PA}
  \country{USA}
}

\renewcommand{\shortauthors}{Kuo et al.}

\begin{abstract}
Community and organizational policies are typically designed in a top-down, centralized fashion, with limited input from impacted stakeholders. This can result in policies that are misaligned with community needs or perceived as illegitimate. How can we support more collaborative, participatory approaches to policy design? In this paper, we present PolicyCraft, a system that structures collaborative policy design through \textit{case-grounded deliberation}. Building on past research that highlights the value of concrete cases in establishing common ground, PolicyCraft supports users in collaboratively proposing, critiquing, and revising policies through discussion and voting on cases. A field study across two university courses showed that students using PolicyCraft reached greater consensus and developed better-supported course policies, compared with those using a baseline system that did not scaffold their use of concrete cases. Reflecting on our findings, we discuss opportunities for future HCI systems to help groups more effectively bridge between abstract policies and concrete cases.
\end{abstract}


\begin{teaserfigure}
  \centering
  \includegraphics[width=0.92\textwidth]{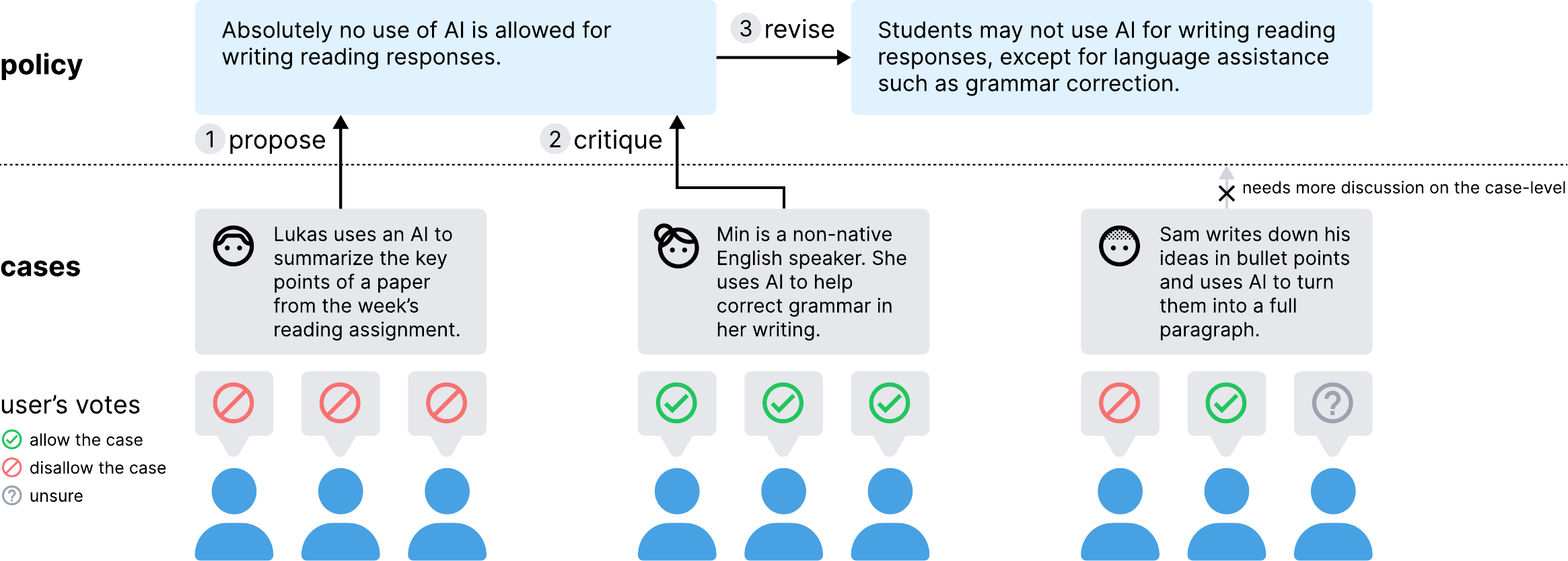}
  \caption{PolicyCraft is a system that supports users in collaboratively proposing, critiquing, and revising policies through discussion and voting on concrete cases. Throughout the iterative policy design process, PolicyCraft helps users understand whether they disagree about the wording of a policy or whether they have an underlying disagreement about how specific cases should be handled. The system then supports users in discussing and addressing disagreements accordingly.}
  \Description{The figure illustrates the conceptual workflow of PolicyCraft, a system designed to help users propose, critique, and revise policies based on cases they vote on and discuss. For example, imagine a group of students using PolicyCraft to develop course policies on generative AI usage in their class. Suppose a case is presented: "Lukas uses an AI to summarize the key points of a paper from the week’s reading assignment." Students might vote to disallow this case and propose a policy such as: "Absolutely no use of AI is allowed for writing reading responses." However, recognizing that this policy might be too restrictive, students could create another case to critique it: "Min is a non-native English speaker. She uses AI to help correct grammar in her writing." While most students might vote to allow this case, the existing policy would prohibit it. This misalignment between the policy and students’ opinions highlights the need for policy revision. In response, students might revise the policy to: "Students may not use AI for writing reading responses, except for language assistance such as grammar correction." This process can be repeated iteratively, using concrete cases to systematically refine policies. In some cases, people may not reach a consensus on their votes. For example: "Sam writes down his ideas in bullet points and uses AI to turn them into a full paragraph." While some may believe this case should be allowed, others may disagree or remain unsure. This lack of consensus suggests that further discussion at the case level is necessary before using it to inform policy revisions.}
  \label{fig:teaser}
\end{teaserfigure}


\maketitle

\section{Introduction}

Communities and organizations of all types, whether small or large, online or offline, often rely on \textit{regulatory policies}\footnote{Regulatory policies focus on regulating the behavior and practices of individuals within communities and organizations, while other policies like constituent, distributive, and redistributive focus on how resources, services, or responsibilities are shared or organized \cite{lowi1972four}.} to support community governance \cite{lowi1972four}. For example, online communities like Reddit have content moderation rules that govern what content is allowed in each subreddit \cite{fiesler2018reddit, chandrasekharan2018internet}. Small groups like university classes have course policies that guide how students may or may not use generative AI in their coursework \cite{tan2024more}. Local governments have zoning regulations that determine what kinds of buildings can or cannot be developed in specific geographic areas \cite{kim2024integrating}, such as restricting industrial development in residential neighborhoods. Such regulatory policies are essential for guiding the behavior of community members, upholding their shared values, and supporting their collective goals \cite{ostrom1990governing}.

Today, policy development processes are typically top-down and centralized, lacking input from the communities and stakeholders who they will impact \cite{zuckerman2023community, jhaver2023decentralizing, siddarth2021ai}. This can lead to the development of policies that are poorly fit to community needs \cite{zhang2024data}. Furthermore, community members may perceive these policies as illegitimate \cite{zuckerman2023community}, leading them to reject the policies \cite{boehner2016data}, organize strikes \cite{mithical2023moderation}, or even leave the community \cite{fiesler2020moving}. For example, volunteer moderators on Stack Overflow organized a strike to protest a new policy about moderation of AI-generated content, which the platform enforced without first communicating with the community \cite{mithical2023moderation}.

While more participatory, bottom-up approaches to policy design hold potential to support greater alignment with community needs and values, it can be challenging to realize this goal in practice. Even when communities share overall norms and values \cite{fiesler2018reddit, chandrasekharan2018internet}, individual community members may hold differing perspectives \cite{kuo2024wikibench, weld2024making, de2010cohere}. Given that policy proposals are often high-level and abstract, it can be challenging for community members to identify the root sources of their disagreements and to collaboratively refine policies to resolve these disagreements \cite{chen2023judgment}.

Past research has shown that the use of concrete cases or scenarios is critical for establishing common ground in discussions about policy design~\cite{feng2023case, chen2023case, kuo2024wikibench, chen2023judgment, cheong2024not, kirkham2023legal}. However, connections between concrete cases and abstract policies are often made in an ad-hoc and inconsistent manner during policy design conversations~\cite{koshy2023measuring, centivany2016values,epstein2014value,park2015toward,r2013regulation}. For example, while people are often inspired to propose policies based on specific, concrete scenarios (whether real or imagined), they do not always discuss how proposed policies might have unintended consequences in \textit{other} scenarios~\cite{epstein2014value,koshy2023measuring,lehoux2020anticipatory,wright2020policy}. Similarly, while people often iterate on the wording of policies to better address imagined scenarios or edge cases, the cases themselves are not always explicitly communicated as objects for discussion~\cite{epstein2014value,hwang2022rules,iandoli2014socially,park2015toward,wright2020policy}.

In this paper, we present PolicyCraft, a system designed to structure collaborative policy design through \textit{case-grounded deliberation}. As illustrated in Figure~\ref{fig:teaser}, PolicyCraft supports users in collaboratively proposing, critiquing, and revising policies through discussion and voting on concrete cases. In doing so, PolicyCraft is designed to help users pinpoint where their disagreements come from---whether they are about the wording of a policy or a more fundamental disagreement about how specific cases should be handled---and then discuss and address these disagreements accordingly. Using PolicyCraft, users share and discuss their perspectives on whether specific concrete cases should be allowed or disallowed, regardless of what current policies say. Building upon consensus at the case-level, users \textit{propose} new policies or \textit{critique} and \textit{refine} current policies in order to better reflect their collective viewpoints.

We evaluated PolicyCraft through a field study across two university classes, where students used PolicyCraft to collaboratively design course policies regarding which generative AI use cases should be allowed in their classes. We find that policies developed using PolicyCraft received stronger support and achieved greater consensus among the students who developed them, compared with a baseline system that did not scaffold students in using concrete cases to support policy design. Students using PolicyCraft also found it easier to understand each other’s perspectives, and were more motivated to iterate on policies based on cases shared and discussed by others. The policies developed by students in our study were refined by course instructors and students and adopted as official course policies.

Overall, this work contributes PolicyCraft, a system that supports collaborative and participatory policy design through the systematic use of cases. We also show that policies created with PolicyCraft receive stronger support and consensus from those involved in the policy development process. Reflecting on our findings, we discuss opportunities for future HCI research and design to support groups in more effectively bridging between abstract policies and concrete cases during collaborative policy design.

\section{Related Work}\label{sec:relatedwork}
Policy design has been a key area of focus in HCI research \cite{yang2024future, spaa2019understanding, jackson2014policy, lazar2012hci}. Following prior HCI scholarship, in this paper we understand policies as principles, guidelines, and written rules to guide behavior in communities and organizations \cite{yang2024future, jackson2014policy}. From investigating how policies are designed and implemented in practice \cite{butler2008don, centivany2016policy, fiesler2018reddit, viegas2007hidden} to creating tools that help develop policies \cite{zhang2020policykit, kriplean2012supporting}, HCI researchers view policy design as a critical topic of inquiry because policies directly impact how technology shapes communities and society \cite{yang2024future, jackson2014policy}. In this section, we first discuss the importance of collaborative and participatory approaches to policy design. We then review existing tools that aim to support such approaches. Finally, we introduce the concept of case-based reasoning \cite{aamodt1994case}, which inspires the design of PolicyCraft.

\subsection{Collaborative and Participatory Approaches to Policy Design}\label{sec:relatedwork_need}
Policies often face a crisis of legitimacy because they don't meet the needs and values of impacted communities. For example, online social platforms are often criticized as arbitrary and censoring free speech because their centralized trust and safety team enforces content moderation policies without community input \cite{zuckerman2023community, jhaver2023decentralizing}. Elected governments that impose top-down policies solely based on their own agendas also risk losing public support and future elections \cite{reynante2021framework, asad2015illegitimate}. To address the legitimacy crisis, researchers suggest adopting more collaborative and participatory approaches to policy design \cite{zuckerman2023community, jhaver2023decentralizing, corbett2023power, whitney2021hci, tandon2022hostile, reynante2021framework, ostrom1990governing, viegas2007hidden, krafft2021action}. As Ostrom stated in her influential principles on governing the commons \cite{ostrom1990governing}: \textit{"individuals affected by the operational rules should be able to participate in modifying those rules."} By incorporating the local knowledge of impacted individuals and communities, collaborative and participatory approaches ensure that rules and policies are well-suited to local circumstances \cite{viegas2007hidden}. These approaches also help stakeholders view the policies as legitimate, even if they disagree, because of the inclusive and deliberative way in which they were settled \cite{zittrain2019three}. Policies developed collaboratively by impacted communities are essential for maintaining stability, building trust and legitimacy, and reflecting collective goals and values \cite{ostrom1990governing}.

However, engaging the community in collaborative and participatory policy design poses several open challenges. First, even within a community with shared norms and values, individual members may still have differing and often conflicting perspectives \cite{reynante2021framework}. For example, on Wikipedia, even with clear policies on vandalism, individual Wikipedians can still interpret and apply the policy differently when moderating article edits \cite{kuo2024wikibench, halfaker2025collective}. These different interpretations exist because policies are often high-level and abstract, making it difficult for people to pinpoint the source of their disagreements without referring to concrete examples \cite{chen2023judgment}. Besides accounting for different community perspectives, another challenge arises when community members collaborate on policy iterations. Without a structured, coordinated process, it can be challenging for community members to achieve meaningful collective action, even if they are eager to contribute \cite{shaw2014computer}. Finally, individual community members often have different preferences, availability, and capacity, and thus require varying levels of support to participate effectively \cite{erete2017empowered, boehner2016data, kuo2023understanding, tang2024ai}. These challenges emphasize the need for research into tools and processes to support collaborative and participatory policy design, which we discuss in the next subsection \cite{jackson2014policy, yang2024future}.

\subsection{Tools for Collaborative and Participatory Policy Design}
Existing tools for collaborative and participatory policy design range from those that mainly gather community perspectives to \textit{inform} policy design to those that directly engage participants in drafting policies~\cite{arnstein1969ladder}.
In the first category, researchers have developed various tools to gather community and public perspectives. For example, Polis is an online platform where participants submit short statements expressing their views on a topic and vote to agree or disagree with statements from others \cite{small2021polis}. It is widely used in policy design worldwide, such as by the Taiwanese government to craft UberX regulations based on input from citizens, taxi drivers, Uber Inc., and other stakeholders \cite{hsiao2018vtaiwan}. ConsiderIt is another prominent platform where people submit the pros and cons of a policy and compare their views with those of others who agree or disagree with the policy \cite{kriplean2012supporting}. The City of Seattle has used ConsiderIt to gather public input on policy proposals, such as the plan to increase affordable housing in certain residential areas. Several other tools use visualizations \cite{liu2018consensus, kim2021starrythoughts,marnette2024talk,zhang2024data} or chatbots \cite{marnette2024talk, kim2021moderator, shin2022chatbots} to identify disagreements for consensus building.

In the second category are tools that allow individuals and communities to contribute to the drafting of actual policies. For example, drawing inspiration from microtasks in crowdsourcing \cite{little2010exploring}, researchers developed CommunityCrit, a system where users can propose policies and comment on others' policy proposals through micro-activities \cite{mahyar2018communitycrit}. Using CommunityCrit, the researchers worked with a local planning group in San Diego to redesign an intersection based on crowdsourced proposals \cite{jasim2021communitypulse}. Other works have crowdsourced policy demands against entities responsible for privacy or labor rights violations \cite{wu2022reasonable, salehi2015we}. For example, building on the find-fix-verify crowd programming pattern \cite{bernstein2010soylent}, Wu et al. designed questionnaires that ask independent crowd workers to find privacy concerns, propose potential fixes, and verify and rank the proposed fixes \cite{wu2022reasonable}. With the rise of generative AI, some tools like Remesh use GPT-4 to automatically synthesize crowdsourced public views into initial policies and then iteratively refine them through expert and public feedback \cite{konya2023democratic}. Other tools like PolicyKit and Pika provide infrastructures that allow online community members to set up their own process for policy design \cite{zhang2020policykit, wang2024pika}. Overall, these tools enable individuals to share and gather feedback on policy ideas, or to contribute inputs for others to synthesize into policies. However, they are not focused on supporting communities in directly collaborating to iteratively and deliberatively craft policy proposals.

PolicyCraft bridges between these two categories of tools by supporting communities in both \textit{sharing perspectives} and collaboratively \textit{authoring policies}, in an integrated, iterative workflow. 
Unlike Polis, ConsiderIt, and other tools in the first category that mainly focus on gathering participants' perspectives to inform policymakers, PolicyCraft further empowers participants to draft actual policies based on the cases they vote on and discuss. This elevates participants from a consultative role to a more collaborative role in policy design \cite{arnstein1969ladder}. Meanwhile, PolicyCraft differs from related work in the second category by supporting groups in directly collaborating and deliberating on policy drafts. To scaffold participants in doing so more systematically, PolicyCraft promotes \textit{case-grounded deliberation}, structuring users' discussions and contributions around the dual abstractions of \textit{cases} and \textit{policies}. This structuring aims to support both individuals and groups in effectively transitioning between case-level and policy-level discussions during policy design---including by helping them tease apart whether they disagree at the case-level versus the policy-level, and by helping ground policy iterations in consensus on concrete cases.

\subsection{Cases as a Medium for Policy Design and Deliberation}\label{sec:relatedwork_case}
Cases have been used as a medium for reasoning and deliberation across many fields \cite{aamodt1994case}. Following prior HCI scholarship and literature in case-based reasoning, in this paper we understand cases as concrete scenarios within specific problem situations that make abstract concepts tangible for reasoning, communication, and negotiation \cite{aamodt1994case, iandoli2014socially, chen2023case, feng2023case, de2010cohere}. For example, cases involving ethical dilemmas, such as the trolley problem \cite{thomson1984trolley}, are frequently used to guide moral deliberation and explore ethical principles \cite{molewijk2008teaching, awad2018moral, nanayakkara2020anticipatory}. Legal theories also often rely upon case-by-case judgments and deliberations to guide decision-making, rather than directly applying top-down rules \cite{cardozo2010nature, grey1983langdell, cheong2024not}. Inspired by jury trials in the legal domain, HCI researchers have proposed case-based discussion tools such as digital juries to adjudicate content moderation cases \cite{fan2020digital, koshy2024venire}. While these tools do not surface cases to the policy level for policymaking, the concrete cases offer a common ground that enables shared understanding and facilitates meaningful deliberation around what decisions and behaviors are desirable \cite{chen2023case, feng2023case}.

Research has shown that the use of concrete cases plays a major role in supporting effective policy design \cite{epstein2014value, lehoux2020anticipatory, wright2020policy}. For example, Wikipedians deliberate on and iteratively refine their definition of vandalism based on specific article edits they come across while moderating Wikipedia articles \cite{kuo2024wikibench, halfaker2025collective}. Other work also uses concrete vignettes and scenarios to support stakeholder reflection and deliberation around the ethical, legal, and policy implications of emerging technologies (e.g., \cite{das2024we, kieslich2024anticipating}). These concrete cases help people identify the source of their disagreements, whether it's due to ambiguous policies or genuine differences in their perspectives about how specific cases should be handled \cite{chen2023judgment}. When making collective decisions, deliberation grounded on concrete cases further provides procedural legitimacy and helps build consensus, even in the face of disagreements \cite{fan2020digital}.

However, connections between concrete cases and abstract policies are currently made in an ad-hoc and inconsistent manner during policy design ~\cite{koshy2023measuring, centivany2016values,epstein2014value,park2015toward,r2013regulation}. For example, people often propose policies based on specific cases and scenarios, but they do not always discuss how these policies might create unintended effects in \textit{other} scenarios \cite{epstein2014value,lehoux2020anticipatory,wright2020policy}. This can cause issues, as in online content moderation, where people usually agree with the general rules proposed based on past incidents but often feel frustrated when those rules are applied to situations they had not previously considered \cite{koshy2023measuring}. Meanwhile, although people with situated knowledge of the policy context often provide policy suggestions based on cases grounded in their lived experiences, the cases themselves---which provide important rationale for the policy suggestions---are not often explicitly shared in policy discussions \cite{epstein2014value,hwang2022rules,iandoli2014socially,park2015toward,wright2020policy}. Researchers have called for a more systematic approach to constructing and using concrete cases to support policy deliberation \cite{nanayakkara2020anticipatory}. For example, researchers have introduced the concept of \textit{research through litigation}, which involves carefully selecting cases to surface serious concerns and drive policy change \cite{kirkham2023legal}. PolicyCraft aims to better support this process in the context of collaborative and participatory policy design, by enabling users to create and iterate on policies through the systematic use of cases.

\section{Design Goals}\label{sec:designgoal}
We established the design goals for PolicyCraft based on our review of prior work in Section \ref{sec:relatedwork}. To ensure that these design goals aligned with real-world needs for community participation in policy design, we then validated them through one-hour, semi-structured interviews with six community organizers who had previously engaged in community policy design. These included two senior moderators responsible for shaping content moderation policies on Wikipedia and Stack Overflow, and four course- and university-level policy designers at a major US university. The discussion topics used to guide these semi-structured interviews can be found in Appendix \ref{sec:informal_conversation}. Overall, we distilled the following four design goals for systems that aim to support collaborative policy design.

\begin{itemize}

\item [\textbf{D1.}] \textbf{The system should encourage users to develop policies based on concrete cases.} Given evidence that discussing concrete cases can help people establish common ground during policy design conversations \cite{epstein2014value, lehoux2020anticipatory, wright2020policy, kuo2024wikibench}, systems should support users in making systematic use of cases. Systems could draw inspiration from prior work, such as the idea of research through litigation \cite{kirkham2023legal,yang2024future}, to support users in using cases to surface potential flaws in current policies. These cases can provide users with shared context for policy development.
\vfill

\item [\textbf{D2.}] \textbf{The system should help users identify and address underlying sources of disagreements.} People may disagree with each other during policy design for various reasons \cite{reynante2021framework}. For example, they may substantively disagree about how different cases should be handled, or they may simply have differing interpretations of a policy's wording \cite{kuo2024wikibench, chen2023judgment}. Systems should help users distinguish whether they disagree at the policy-level, case-level, or both. Systems should further assist users in discussing and addressing disagreements at the appropriate level.
\vfill

\item [\textbf{D3.}] \textbf{The system should support users in easily building upon and referencing each other's work.} While community-driven approaches to policy development ensure the policies are better aligned with local community perspectives \cite{ostrom1990governing}, it can be challenging for community members to achieve meaningful collective action without a coordinated process \cite{shaw2014computer}. Systems should make it easy for users to build upon each other's contributions, leveraging their complementary knowledge, experiences, and backgrounds \cite{salehi2015we, reynante2021framework}. Systems should also enable users to track how others build on their contributions and to collaborate on planning future actions \cite{shaw2014computer}.
\vfill

\item [\textbf{D4.}] \textbf{The system should accommodate different levels of engagement in policy design.} People have different amounts of time, preferences, and capacities for participating in the policy development process \cite{erete2017empowered, boehner2016data}. Systems should offer different ways for users to participate, including light-weight options like voting or more involved ones like creating a new policy \cite{mahyar2018communitycrit}. For more involved tasks like editing or creating policies, systems should offer additional scaffolding and assistance to reduce barriers to participation.
\vfill

\end{itemize}

\section{PolicyCraft}\label{sec:policycraft}

Based on these design goals, we developed PolicyCraft, a web-based system that supports collaborative and participatory policy design through case-grounded deliberation. PolicyCraft is designed to support policy design across a broad range of community contexts. PolicyCraft focuses on supporting the design of \textit{regulatory} policies~\cite{lowi1972four}: policies that guide what is allowed or disallowed within a community context. For example, in online communities like subreddits \cite{jhaver2023decentralizing}, we envision that PolicyCraft may assist a community in designing its content moderation rules (policies) based on community members' posts (cases). In local communities such as vTaiwan \cite{hsiao2018vtaiwan}, PolicyCraft may help refine regulations for emerging technologies (policies), such as the use of UberX in specific local contexts (cases).

\subsection{System Overview}

\begin{figure*}[t!]
  \centering
  \includegraphics[width=0.95\linewidth]{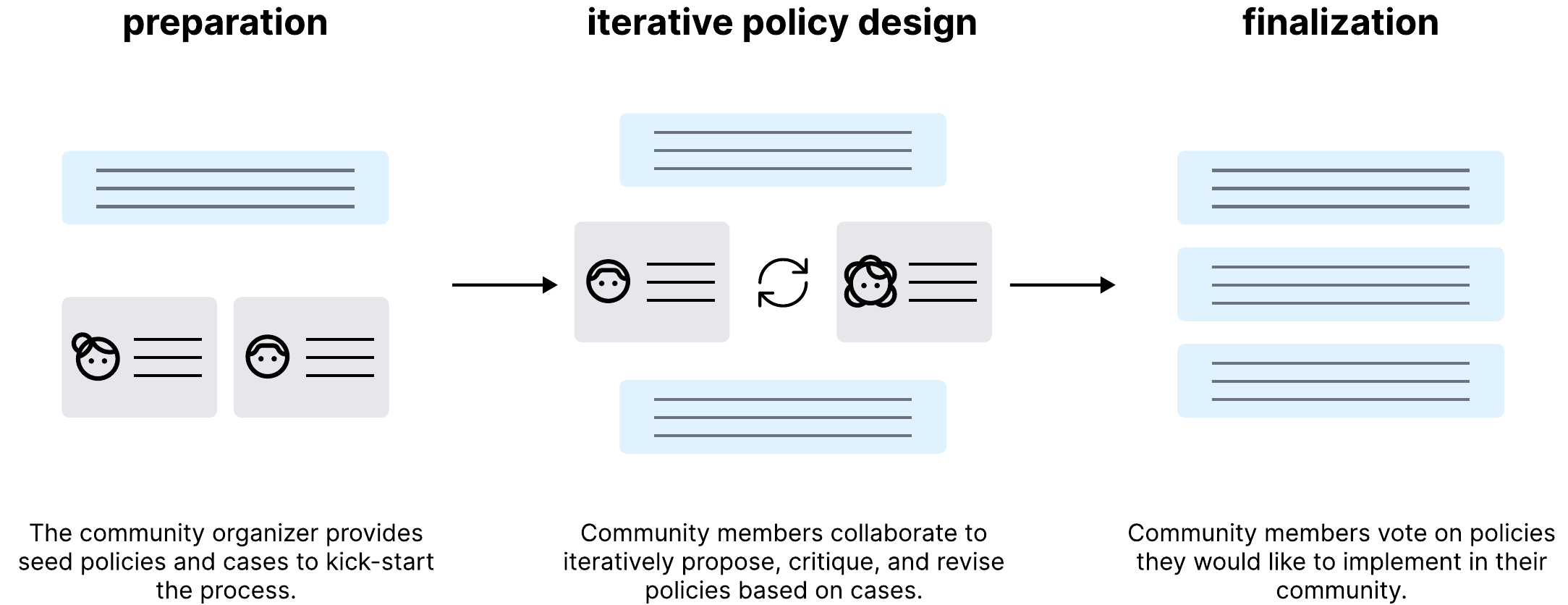}
  \caption{PolicyCraft's overall process.}
  \Description{The overall process of PolicyCraft consists of three phases. In the preparation phase, the community organizer provides seed policies and cases to kick-start the process. During the iterative policy design phase, community members collaborate to iteratively propose, critique, and revise policies based on cases. Finally, in the finalization phase, community members vote on policies they would like to implement in their community.}
  \label{fig:process}
\end{figure*}

PolicyCraft supports policy design through the overall process illustrated in Figure~\ref{fig:process}. Community organizers start by entering a few initial seed policies and cases into the system. These seed policies/cases can serve to minimize the cold-start problem and establish norms around aspects such as length, formatting, and level of abstraction~\cite{kiene2016surviving, chen2021cold}. Community members then collaborate to iteratively critique and revise the initial policies or propose new policies based on concrete cases. During this stage, community members can share their perspectives on whether specific concrete cases \textit{should} be allowed or disallowed, regardless of what current policies say, through case-level voting and discussion. In the final stage, community members consider which of the policies they believe should be implemented in their community. They express their perspectives by anonymously upvoting or downvoting policies and then optionally providing publicly visible reasons for their votes. During the voting process, policies cannot be edited. This overall process is flexible and allows for customization based on community-specific needs and goals.

In the following subsections, we describe PolicyCraft's core functionality: supporting users in collaboratively \textit{critiquing}, \textit{revising}, and \textit{proposing} policies through cases. We illustrate this functionality through a running example, describing how PolicyCraft would be used to support collaborative policy design in a university classroom setting. In this setting, students and instructors collaboratively develop course policies regarding which ways of using generative AI should be allowed in the course. The instructor serves as the community organizer and the students are community members. We also share additional features that help lower participation barriers and facilitate collaboration. Throughout the section, we connect specific features of PolicyCraft's design to the design goals outlined in the previous section, denoted as D1 to D4. Finally, we conclude with implementation details.

\begin{figure*}[t!]
  \centering
  \includegraphics[width=\linewidth]{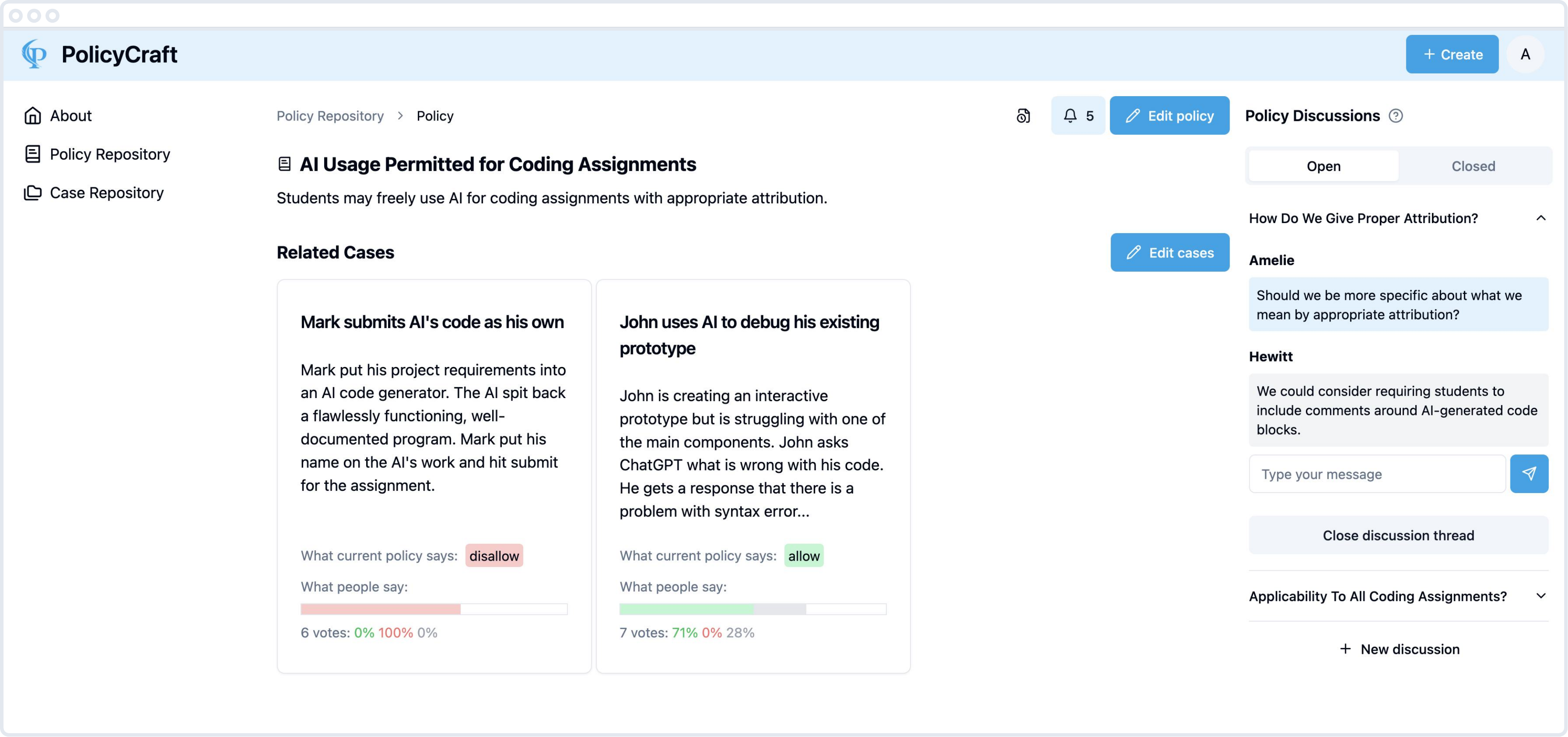}
  \caption{PolicyCraft's ``policy page'' for a given policy. On the left side is the navigation bar, where users can click to visit the Policy Repository to see all current policies, and the Case Repository to see all cases. Users can also visit the About page to view information about the current policy development campaign and to discuss general topics, not specific to a particular policy or case. In the center of the screen, the title and description of the selected policy is shown, along with its related cases. The visualization of users' votes on the cards is inspired by the design of Polis \cite{small2021polis}. The bar represents how many people have voted out of the total number of users, while the percentages reflect the distribution of votes—whether to allow, disallow, or indicate unsure—among those who have already voted. Users may add new related cases by clicking the ``Edit cases'' button, which brings them to the page shown in Figure~\ref{fig:edit_case_page}. They can also revise the policy by clicking the ``Edit policy'' button, which brings them to the page shown in Figure~\ref{fig:edit_policy_page}. Users can also start a discussion thread about this policy and close it once the topic has been resolved. Finally, users can propose new policies by clicking the ``Create'' button in the top-right corner.}
  \Description{A screenshot of PolicyCraft's policy page for a given policy.}
  \label{fig:policy_page}
\end{figure*}

\subsection{Critiquing Policies Through Cases}
\subsubsection{\textbf{Users can critique a policy by creating cases that highlight ambiguities or potential flaws.}} 
\begin{figure*}[t!]
  \centering
  \includegraphics[width=\linewidth]{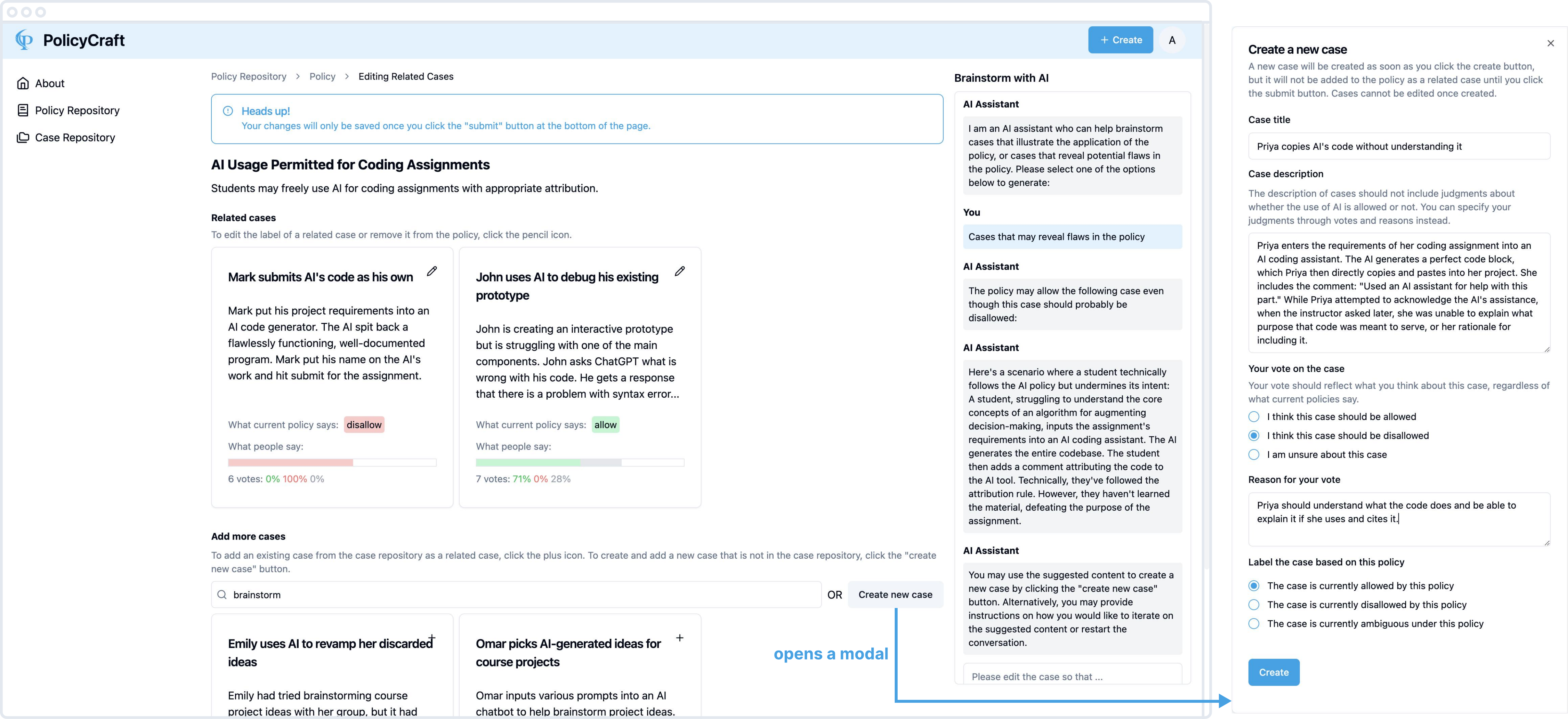}
  \caption{The workflow for editing a policy's ``related cases'' in PolicyCraft. The upper half of the screen shows the cases currently associated with a policy. The bottom section allows users to add additional cases either by searching the case repository with keywords (e.g., "brainstorm," as shown in this figure) or by authoring a new case. When adding a case to a policy, users label it to indicate whether they believe the current version of the policy would allow it, disallow it, or whether it is ambiguous how the policy would treat this case. Users can optionally brainstorm with the built-in AI assistant to generate cases that illustrate the policy or identify its potential flaws.}
  \Description{A screenshot of PolicyCraft's editing related case page for a given policy.}
  \vspace{10mm}
  \label{fig:edit_case_page}
\end{figure*}

After logging into PolicyCraft through their browser, users can start by going to the \textbf{policy repository}, where they can see an overview of all current policies. If users find a particular policy they think can be improved, they can click on it to see an expanded view on its \textbf{policy page}, as shown in Figure~\ref{fig:policy_page}. As a concrete example, consider a user reading a course policy about whether AI use is permitted for coding assignments. The current policy says: \textit{``Students may freely use AI for coding assignments with appropriate attribution.''} Users can critique a policy by creating a \textit{case}: a description of a concrete usage scenario that highlights ambiguities or potential flaws of the current policy (D1). For example, as illustrated in Figure~\ref{fig:edit_case_page} to highlight a potential flaw, a user may create a case that they believe is \textit{currently permitted} under the policy, but which they believe \textit{an ideal policy should disallow}.

\begin{figure*}[t!]
  \centering
  \includegraphics[width=\linewidth]{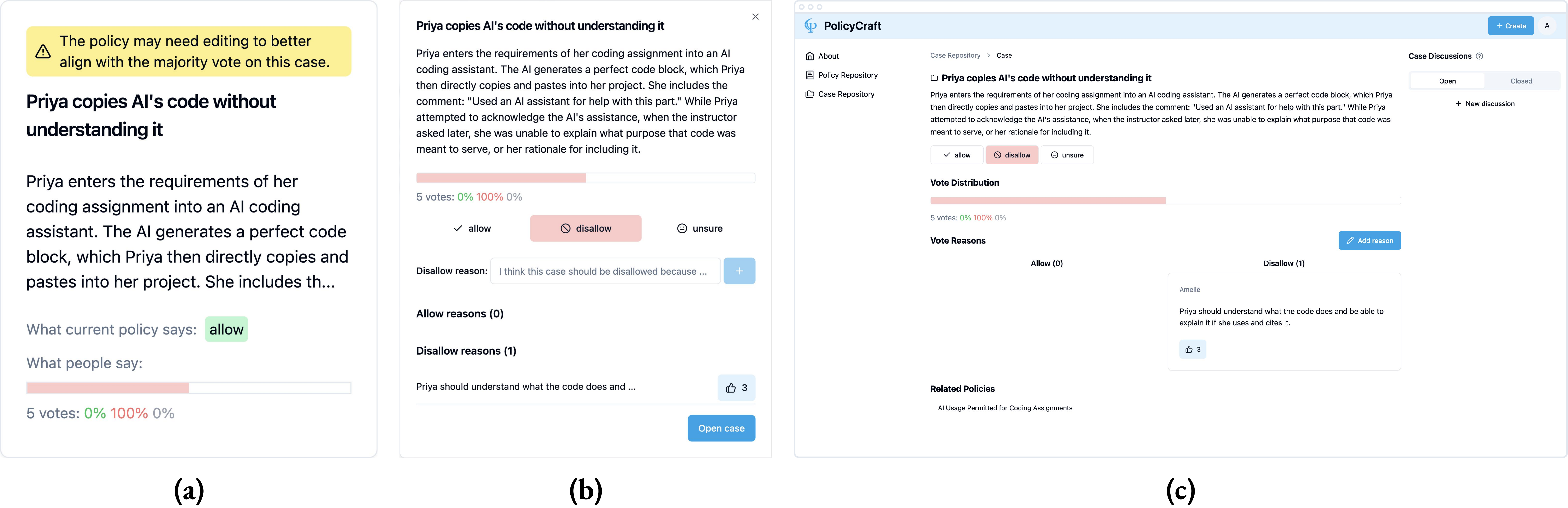}
  \caption{(a) Once a user adds the case shown in Figure~\ref{fig:edit_case_page} to the policy, it will appear as a card in the related cases section of the policy. The yellow message highlights the misalignment between the label linking the case to the policy (allow) and the majority vote on the case (disallow). (b) Users can click on the card to open a modal with additional details about the case, including other users' reasons for wanting to allow or disallow it. (c) For more in-depth discussions, users can click the "open case" button to visit the case page.}
  \Description{Three screenshots of PolicyCraft's cases are shown: (a) as a card, (b) as a pop-up modal when the card is clicked, and (c) as a detailed case page when the case is opened.}
  \vspace{10mm}
  \label{fig:case}
\end{figure*}

When the user creates and adds a case to the policy, they \textbf{label} whether they think the current draft of the policy would \textit{allow} or \textit{disallow} the given case. Alternatively, if they think the policy is ambiguous regarding how the case should be treated, they can label it as \textit{ambiguous}. Meanwhile, they also cast their personal \textbf{vote} on the case to indicate whether they think it ideally should be \textit{allowed} or \textit{disallowed}, or whether they are currently \textit{unsure}. Finally, users provide a brief \textbf{reason} for their vote.\footnote{We require users to provide a reason for their vote whenever they create a new case and vote to allow or disallow it. This design builds upon prior research showing that disagreement contributes to group ideation most when people elaborate and justify their stances \cite{aitamurto2023disagreement}. Inspired by the design of ConsiderIt~\cite{kriplean2012supporting}, if a user votes that they are ``unsure'', they can still explain their reasoning by providing separate allow and disallow reasons representing pros and cons they see.} Note that votes and reasons are meant to reflect a user's personal opinion on whether a case should be allowed, regardless of what current policies say. 

Once a user has added a case, it becomes visible in the \textbf{related cases} section of the policy page. Other users can then (1) edit the case's label if they have a different interpretation; (2) remove the case if they believe it is not relevant to the given policy; or (3) add additional related cases to the policy. The added case also becomes available in the \textbf{case repository}, where users can browse all created cases.\footnote{The same case can be added to multiple policies as a related case. The label of a case represents a link between that case and a specific policy, so labels may vary across different policies. By contrast, users' votes are attached only to the case, independent of current policies.} Anyone can read these cases and vote on whether they believe each case should ideally be allowed or not.\footnote{Users only see other people's votes after they cast their own. This design helps encourage more participation and prevent lurkers \cite{nonnecke2000lurker}. Cases cannot be edited, so people's votes stay relevant.} These votes and reasons help users understand what others think about specific cases (D2, D3).

\subsubsection{\textbf{PolicyCraft highlights misalignments between case labels and users' votes.}} Consider another user who visits the same policy page later and reads the case shown in Figure~\ref{fig:edit_case_page}. Like the previous user, many others have also voted to disallow the case, although the current policy would allow it. As shown in Figure~\ref{fig:case}, whenever the label linking a case and a policy is misaligned with the majority vote on a case, a yellow alert message automatically appears to highlight it as a potential issue. In such cases of misalignment, the yellow alert message will suggest that \textit{``The policy may need editing to better align with the majority vote on this case.''} If most people vote to either allow or disallow the case, but the label indicates that the current policy is ``ambiguous'' with regard to whether that case should be allowed/disallowed, the yellow alert message will suggest that \textit{``The policy may need editing to clarify whether this case is allowed or not.''} These alert messages are intended to help people focus their policy iteration and discussion around cases where there is a misalignment between what the policy says about a case and what people believe (D1). If most people vote that they are ``unsure'' about whether a given case should be allowed, an alert message will not appear, regardless of the label, because this indicates that more discussion is needed at the case-level before using it to iterate on policy (D2).

\subsection{Revising Policies Through Cases}\label{sec:revise_policy}
\subsubsection{\textbf{Users can revise a policy to better align it with people's votes on cases.}} 
\begin{figure*}[t!]
  \centering
  \includegraphics[width=\linewidth]{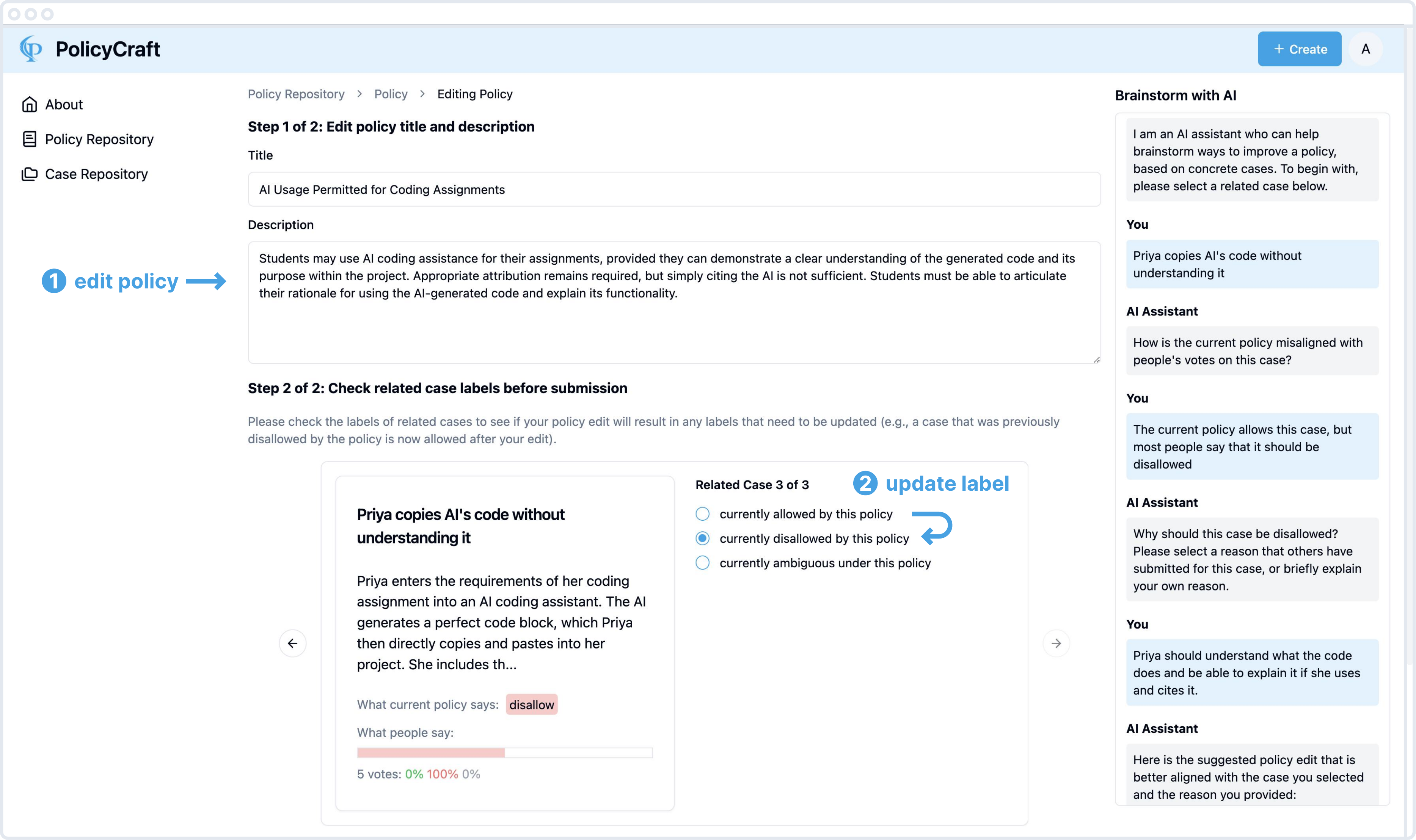}
  \caption{The workflow for editing a policy in PolicyCraft. Users first update the policy's wording and then check if the edit requires updating the labels for any of the related cases. Users may also brainstorm with the built-in AI assistant to improve the policy. For example, the policy description shown in this figure is suggested by the AI assistant based on the user's interactions shown in the right-side panel.}
  \Description{A screenshot of PolicyCraft's editing policy page for a given policy.}
  \label{fig:edit_policy_page}
\end{figure*}

Upon noticing a misalignment between what a policy says and what people believe at the case-level (yellow alert messages), a user may decide to revise the policy by clicking the \textbf{edit policy} button on the policy page. This takes them to the policy editing page, as shown in Figure~\ref{fig:edit_policy_page}, where they can edit the policy in two steps. First, users can edit the policy's title and description as needed to better align it with people's votes on cases (D1). Then, before submitting their changes, users are asked to review the labels of the policy's related cases to see if their policy edit results in any updates to case labels. For example, a case that was previously allowed by the policy might now be disallowed following their edit. After a user updates the labels as needed and submits their edits, the revised policy is made immediately visible to others. If a user's edits resolve a misalignment between the label and people's votes on a case, the alert on that case will disappear.

\subsubsection{\textbf{Users are encouraged to be bold in editing.}}
The current version of PolicyCraft enables any user to revise policies and labels, and includes a feature to detect editing conflicts in Wikipedia style \cite{kittur2007he}. Specifically, if a user submits a change to a policy while another user is still editing it, the second user will be asked to review the new changes and decide whether to include them with their own edits before submitting. This design aims to encourage participation (D3), rather than placing too many restrictions on users. To support coordination, the system makes the evolution of a policy transparent \cite{dabbish2014transparency, stuart2012social}: all users can see the full edit history and revert changes as needed.

\subsection{Proposing Policies Through Cases}
If users notice a gap in current policies, they can create a new policy by clicking the \textbf{create policy} button, which takes them to the \textbf{creation page}. To encourage users to design and deliberate around policies based on concrete cases (D1), users are reminded on the creation page that policies must include at least one related case in order to eventually be included in the policy finalization stage. On the creation page, users can also choose to create a new case that is not yet related to any policies. Users can first create a case, wait for it to receive votes from others, and then create a policy based on people's perspectives on that case.

\subsection{Additional Features to Support Participation and Collaboration}
Below, we briefly describe two additional features of PolicyCraft, aimed at facilitating participation and collaboration.

\subsubsection{\textbf{Scaffolding policy design with built-in AI assistants.}}\label{sec:system_ai} 
\begin{figure*}[p]
  \centering
  \includegraphics[width=\linewidth]{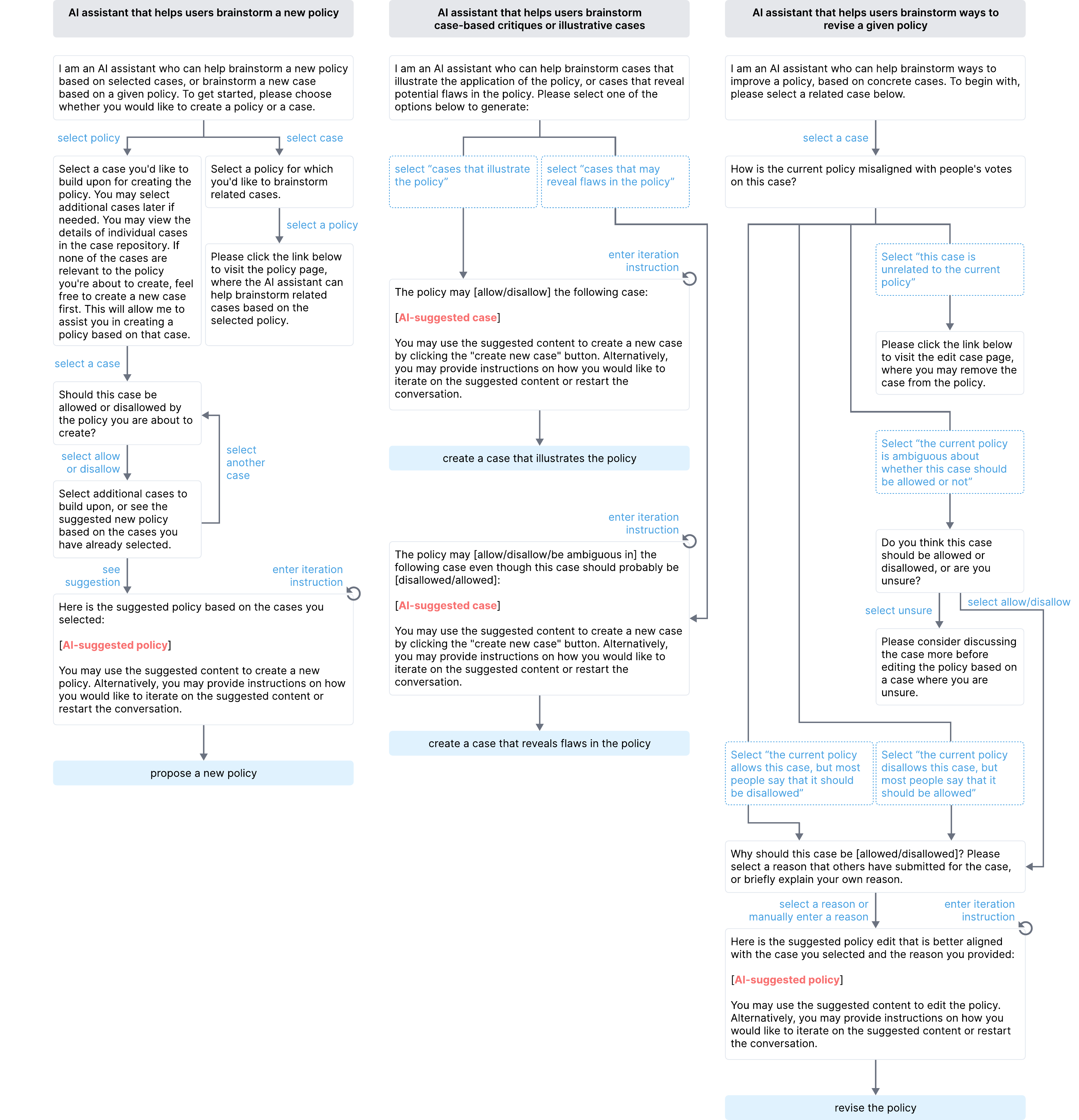}
  \caption{The conversational flows of PolicyCraft's three AI assistants. The blue text represents the user's selection or input, while the black text in boxes shows the AI's response. The red text represents placeholders for AI-suggested policies or cases, based on the information provided by users during the conversation. Users can optionally provide additional instructions to iteratively refine initial AI generations. Users can choose to restart the conversation at any point, but the chat history remains available for their reference. See Appendix \ref{sec:AI_assistants} for a more detailed walkthrough of the conversation workflow, including prompts.}
  \Description{The conversational flows of PolicyCraft’s three AI assistants are displayed across three columns. The left column illustrates how an AI assistant helps users brainstorm a new policy. The middle column demonstrates how an AI assistant helps users brainstorm case-based critiques or illustrative cases. The right column shows how an AI assistant helps users brainstorm ways to revise a given policy. For a detailed walkthrough of the conversation workflow, including prompts, refer to Appendix B.}
  \label{fig:ai_workflow}
\end{figure*}

To help reduce barriers to participation in policy design (D4), PolicyCraft has three built-in, LLM-based AI assistants that users can optionally use to support the critique, revision, or creation of new policies based on cases. The first AI assistant is available on the page where users edit a policy's related cases. Given a policy as input, this AI assistant can help users brainstorm \textit{case-based critiques} (cases that reveal potential flaws or ambiguities in the current policy) or \textit{illustrative cases} (cases that can help illustrate the application of the policy). The second AI assistant is available on the page where users edit a policy. Given one or more user-selected cases, along with the user's reasons for wanting to allow or disallow those cases, this AI assistant helps users brainstorm ways to \textit{revise} a given policy. The third AI assistant is available on the creation page where users can propose a new policy. Given one or more user-selected cases, along with the user's reasons for wanting to allow or disallow those cases, this AI assistant helps users brainstorm a \textit{new policy}. The design of the AI assistant's conversational flow can be found in Figure~\ref{fig:ai_workflow}.

\subsubsection{\textbf{Users can participate in discussions, receive notifications, and track their activities within the system.}} To facilitate user collaboration, PolicyCraft includes built-in features for discussion, notifications, and activity tracking (D3). Each policy and case has its own discussion panel where users can start and reply to discussion threads on various topics. We separate discussions about policies and cases to help users clarify if disagreements are about a policy or specific cases (D2). PolicyCraft also has a panel for meta-discussions on the main page (the \textbf{about page}) where users can talk and coordinate about general topics that go beyond individual policies and cases. This design is similar to Meta Stack Overflow, where users can have higher-level discussions beyond individual posts \cite{fang2023people}. Users receive notifications whenever new discussions are added to threads they have participated in, or when there are changes to policies they are following. Finally, users can keep track of their own activities on a dedicated activity page and quickly revisit the policies, cases, and discussions where they have previously contributed.

\subsection{Implementation}
PolicyCraft is a full-stack web application that people can set up and invite others to join via a URL. The current implementation of PolicyCraft is built with SvelteKit and uses Firestore databases. On the front end, we use shadcn-svelte components and customize them with Tailwind CSS. On the back end, in addition to the database, we also use Firebase for user authentication, hosting the server, and tracking database changes to send user notifications. Last but not least, we power the AI assistants with Gemini 1.5 Pro. PolicyCraft is open-source and available on GitHub.\footnote{\url{https://github.com/tskuo/PolicyCraft}}

\section{Field Study}

To understand how people use PolicyCraft and how its case-grounded approach shapes collaborative policy development, we conducted an evaluation study in university classes. In this context, we aim to support students in collaboratively designing course policies regarding which uses of generative AI should be allowed in class. Since students are directly impacted by course policies, engaging them in the design process, rather than leaving it solely to instructors, has potential to produce policies that students find more reasonable and legitimate~\cite{hess2007collaborative,moreno2005sharing,shor1996students}. This process also provides students to learn more about each other's perspectives and build consensus around course practices and policies.

Our evaluation primarily consists of a three-day field study. We ran the study in two classes taught by two of the co-authors who were open to incorporating students' input into their course policies on generative AI. Both courses were electives open to undergraduate and graduate students, focusing on the study and design of human interactions with technologies. In each class, we randomly divided the students into two groups. One group of students collaboratively designed course policies using the \textbf{full version of PolicyCraft}. The other group of students used a \textbf{baseline version}, which only included the policy-level features (policy creation, editing, and discussion), without any case-level features such as case creation, the case repository, or the AI assistants that scaffold case-grounded policy design. Students in the baseline condition were still able to discuss cases, if they so chose, using the system's discussion features. However, they were not explicitly supported by the system in discussing and using cases to drive iterative policy design. Students in each class designed their course policies independently from the other class, as each class had its unique learning goals. The number of students in each class and condition is shown in Table \ref{table:policy_stats}. In addition to the main field study, we conducted two brief surveys to better understand students' perceptions of the process and external raters' opinions on the final policies. All studies were approved by the university's institutional review board (IRB) where studies were conducted.

\subsection{Study Procedures}

On the first day of the study, students spent 45 minutes onboarding and beginning to use the system during class time. In each class the instructor introduced the study and randomly divided the students into groups, seated in separate areas of the classroom. Students in both groups then individually reviewed onboarding materials illustrating how to use the version of the system they were assigned to use (either the full version of PolicyCraft or the baseline version). 
Across classes, both versions of the system were initialized with the same set of three initial policies provided by the instructors to kick-start the discussion. In the full version of PolicyCraft each policy included two initial, illustrative cases provided and labeled by the instructors. The initial policies and cases are available in Appendix~\ref{sec:initial_policy_and_case}. Students were encouraged to use the system for discussions instead of discussing verbally, to leverage the system's collaboration features and avoid cross-group influence. By the end of the first class, all students had begun using the system.

Between the first and second class sessions, which were separated by one day for both participating classes (e.g., Monday and Wednesday), students were asked to use the system during a minimum of three periods: first, as homework after the first class, due by the end of the day; second, anytime during the day between classes; and finally, a third time anytime before the start of the second class. During each period, students were asked to complete at least seven actions, including creating or editing at least one policy. For the full version of PolicyCraft, all actions related to policies, cases, reasons, and discussions counted, except for voting on cases or `liking' reasons added by others, since these actions required only a single click. For the baseline version, all actions related to policies and discussions counted. We expected students to meet these minimum participation requirements easily. For example, a student using the full version of PolicyCraft could quickly complete three actions by creating a case, explaining the reason behind their vote on that case, and adding it to a policy. We set minimal participation requirements to ensure that students would have ample opportunities for interaction across the field study period, while also providing students with flexibility to decide when and how much they want to contribute (cf. \cite{zhang2018making, kuo2024wikibench}). Students were graded on whether they met these minimum participation requirements, but not on the content of their contributions. All students met or exceeded the requirements, knowing that the policies they collaboratively developed would inform the official course policies.

Each time students created or edited a policy, they answered a multiple-choice question about what inspired them to create or edit the policy. By analyzing how often each option was chosen, we could compare inspirations for policy creation and edits in each of our study conditions. Students could choose one or more of the following five options:
\begin{itemize}
\item [1.] To address a specific case/scenario that I thought of
\item [2.] To address a specific case/scenario that someone else shared
\item [3.] To address a general issue that I thought of
\item [4.] To address a general issue that someone else shared
\item [5.] Other
\end{itemize}

In the second class session, students within each group voted on which policies should be incorporated into the official course policy using upvotes and downvotes (part of the ``finalization'' stage of PolicyCraft as shown in Figure~\ref{fig:process}). Students then individually completed a post-study survey (Appendix~\ref{sec:poststudysurvey}) to understand their experiences with the process. Finally, students participated in a full-class discussion about the resulting policies, facilitated by their instructors. Instructors then used students' feedback to refine the best-supported policies into official course policies on generative AI use for the rest of the semester.

Finally, to understand how people who did not participate in policy development perceived the quality of the resulting policies, we conducted a short online survey after the field study. We recruited university students not involved in the field study as external raters to rate the policies that received majority upvotes from participants in each group. Each rater evaluated two sets of policies---one from the baseline condition and one from the PolicyCraft condition---selected at random from one of the
two classes. They were not told how the policy sets were developed. They rated individual policies based on \textit{quality}, \textit{clarity}, and \textit{feasibility} (for implementation in class). They also rated each policy set holistically for  \textit{comprehensiveness} and \textit{coherence}. Lastly, they selected which policy set (if any) they felt was higher quality overall, and briefly explained their response. As an incentive for participation, survey participants had the option to enter a raffle for a \$100 gift card.

\begin{table*}[t]
  \caption{Descriptive statistics of the resulting policies and the votes they received from participants.}
  \label{table:policy_stats}
  \begin{tabular}{lcccc}
    \toprule
     & \multicolumn{2}{c}{Class 1} & \multicolumn{2}{c}{Class 2} \\
    & Baseline & PolicyCraft & Baseline & PolicyCraft \\
    \midrule
    Number of participants & 22 & 20 & 14 & 12 \\
    Number of policies & 31 & 19 & 19 & 11 \\
    Number of policies with majority upvotes & 7 & 14 & 7 & 8 \\
    Percentage of policies with majority upvotes & 23\% & 74\% & 37\%  & 73\% \\
    Entropy based on votes (lower indicates higher consensus) & 0.64 & 0.47 & 0.50 & 0.27 \\
    \bottomrule
  \end{tabular}
\end{table*}

\section{Study Results}
We present findings from our evaluation study in the following subsections. Section~\ref{sec:result_1} presents the resulting policies and analyzes participants' votes on these policies. Our results show that policies developed using PolicyCraft received stronger support and achieved greater consensus among the students who developed them. Section~\ref{sec:result_2} explores possible mechanisms to explain this finding. From our post-study survey, we find that students using PolicyCraft found it easier to understand each other's perspectives. We also find evidence that PolicyCraft succeeded in promoting case-grounded, collaborative policy design: Students using PolicyCraft were more likely to iterate on policies based on cases raised and discussed by others. Section~\ref{sec:result_3} presents perceptions of the resulting policies from both course instructors and external raters. Overall, instructors and raters found the policies developed with PolicyCraft to be clearer and more nuanced regarding when generative AI use is appropriate, but also less concise and potentially \textit{too} comprehensive for direct use as course policies. 

\subsection{Resulting Policies and Vote Distributions}\label{sec:result_1}

\begin{table*}[t]
  \caption{Illustrative examples of policies that received majority upvotes and policies that did not, from Class 1. The number next to each arrow shows how many upvotes or downvotes each policy received.}
  \label{table:policy_class_1_examples}
  \begin{tabular}
  {p{0.14\textwidth}p{0.40\textwidth}p{0.40\textwidth}}
    \toprule
     & Baseline & PolicyCraft \\
    \midrule
    \small{Examples of Policies \linebreak with \linebreak Majority Upvotes}
    &
    \small{\textbf{AI as a Tool, Not a Substitute:} Students could be taught to use AI as a resource to enhance their learning, rather than relying on it to do their work for them. AI can be used for tasks such as research, data analysis, and language translation, but it should not replace critical thinking, problem-solving, or creativity. In addition, it would be useful to know which AI tools and prompts were used that helped with the research to give credit to the tool.}
    &
    \small{\textbf{AI for Course Understanding:} Students are permitted to utilize AI to enhance their understanding of course material, such as clarifying complex topics or visualizing key concepts. However, all submitted work must reflect the student's own analysis and understanding. While AI tools can provide guidance and support, direct copying or paraphrasing of AI-generated content is strictly prohibited (this includes drawing your comments on readings from any summary/analysis content that the AI provides.)}
    \\
    & \small{Votes: $16 \uparrow \; 0 \downarrow$} & \small{Votes: $18 \uparrow \; 1 \downarrow$}
    \\
    \midrule
    \small{Examples of Policies \linebreak without \linebreak Majority Upvotes}
    &
    \small{\textbf{Open AI Conversation History Policy:} Any generative / conversational AI used in the completion of an assignment should have detailed and chronologically marked logs of any exchange with said AI tool. This log (chat history, prompts, individual instructions for the tool, supplemental information...) should be made available to the instructor as part of the hand-in of the assignment if AI-generated content was directly used in an assignment.}
    &
    \small{\textbf{Including ChatGPT Chat Log of the Related Usage:} When students used GenAI for this course's learning purposes, they should cite the usage with the chatlog by pasting the chatlog url into the assignment submission to reinforce more self-regulated usage of LLM tools, even if GenAI was just used for brainstorming. When you work is directly a result of LLM tools either in direct idea (including summary or analysis of a reading), you must submit the chatlog url as a citation.)}
    \\
    & \small{Votes: $4 \uparrow \; 10 \downarrow$} & \small{Votes: $4 \uparrow \; 13 \downarrow$}
    \\
    \bottomrule
  \end{tabular}
\end{table*}

Overall, participants using the baseline system proposed a larger number of policies than participants using PolicyCraft. However, in the baseline condition, most policies did not receive majority support in the final voting stage. Table~\ref{table:policy_stats} shows the total number of policies created by each group and the number of those policies that received majority votes. To provide a glimpse of the resulting policies, in Table~\ref{table:policy_class_1_examples} we sample a few policies developed with each system that received a majority of upvotes and policies that did not from their groups. In Appendix~\ref{sec:finalpolicies}, we include the full set of policies that received majority support from each group within each class, for reference.

\subsubsection{\textbf{Policies developed through PolicyCraft received stronger support.}}
To understand whether PolicyCraft helped groups collaboratively develop better-supported policies, compared with the baseline system, we analyzed the number of policies that received a majority vote from participants. Specifically, we counted the number of policies where the difference between upvotes and downvotes exceeded half the number of participants involved in their development. As shown in Table~\ref{table:policy_stats}, 74\% of the policies (14 out of 19) developed with PolicyCraft in Class 1 received majority support from participants, compared to just 23\% in the baseline condition (7 out of 31). Similarly, 73\% of the policies developed with PolicyCraft in Class 2 received majority support from participants, compared to 37\% in the baseline condition. These results suggest that participants using PolicyCraft produced better-supported policies overall. In contrast, participants using the baseline system spent more effort creating policies that eventually failed to gain substantial support from the group.

\subsubsection{\textbf{Policies developed through PolicyCraft achieved greater consensus during voting.}}
In addition to comparing the percentage of policies that received a majority vote, we also wanted to understand whether PolicyCraft helped participants reach greater consensus in their voting across \textit{all} policies. A higher level of consensus on a policy occurs when most participants either upvote or downvote it. In contrast, a lower level of consensus is reflected by roughly equal numbers of upvotes and downvotes. To quantitatively assess the level of consensus on policies, we used Shannon Entropy, a metric from information theory \cite{shannon1948mathematical} commonly used to measure the degree of consensus in groups \cite{akiyama2016method, aranzales2021scientists, tapia2022entropy}. A lower entropy indicates a higher level of consensus. We first computed the entropy for each policy based on its vote distribution:
\begin{displaymath}
- (p_u \log_2 p_u + p_d \log_2 p_d)
\end{displaymath}
where $p_u$ and $p_d$ denote the proportion of upvotes and downvotes a policy received. For example, if a policy receives an equal number of upvotes and downvotes, both $p_u$ and $p_d$ would be $0.5$. In this case, the entropy is $1$, indicating the \textit{lowest} level of consensus. In contrast, when all votes for a policy are either upvotes or downvotes, the entropy is $0$, indicating the \textit{highest} level of consensus.\footnote{The value at 0 are given by the limit $0 \log 0 := \lim_{x \to 0^+} x \log x = 0$} Table~\ref{table:policy_stats} shows the mean entropy across policies for each condition.\footnote{The distribution of entropies for each condition is provided in Appendix~\ref{sec:entropy_analysis}.} In both classes policies developed using \textit{PolicyCraft} had a lower entropy compared to those developed using the baseline system, indicating a higher level of consensus in their voting on policies.

\subsection{Reasons for Greater Policy Support and Consensus}\label{sec:result_2}
To understand potential mechanisms behind these trends, we examined the results from the post-study survey that students completed at the end of the field study, as well as the multiple-choice question that participants answered each time they edited or created a policy.

\begin{figure}[t]
  \centering
  \includegraphics[width=0.95\linewidth]{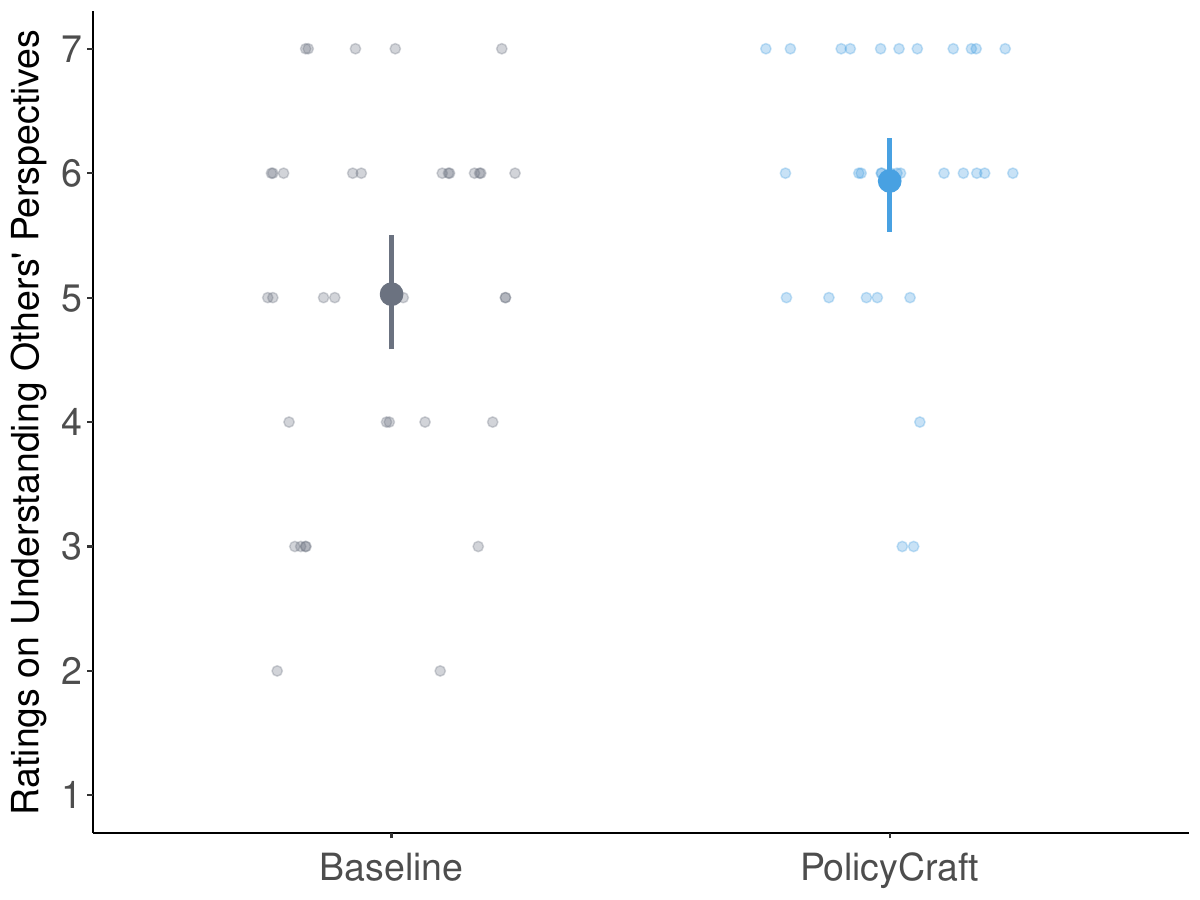}
  \caption{Participants' ratings on whether they could easily understand why people agree or disagree with each other while using the system. There is a significant relationship between the system a participant used and their ratings.}
  \Description{A scatter plot showing the system participants used during the study on the x-axis (Baseline or PolicyCraft) and their ratings on the y-axis for understanding others' perspectives. PolicyCraft's rating is significantly higher than the baseline.}
  \label{fig:post_study_rating_understanding}
\end{figure}

\subsubsection{\textbf{Participants who used PolicyCraft found it easier to understand why people agreed or disagreed with each other.}} 
In the post-study survey, participants rated their agreement, on a 7-point scale, with each of five statements about their experiences in collaborative policy design (see Appendix \ref{sec:poststudysurvey}). As shown in Figure~\ref{fig:post_study_rating_understanding}, we observe that participants who used PolicyCraft provided significantly higher ratings regarding whether they could \textit{``easily understand why people agree or disagree with each other''}, controlling for between-class differences ($p<0.01$, see Appendix \ref{sec:likert_data_analysis} for further details on our regression model\footnote{In Appendix \ref{sec:likert_data_analysis}, we present results from both a linear and an ordinal regression. The appropriateness of using ordinal versus linear regression for the analysis of Likert scale data has been a subject of wide debate, with recent scholarship showing that each approach has complementary benefits and drawbacks~\cite{robitzsch2020ordinal}. Our reported findings are robust to the choice of ordinal or linear regression.}). This result suggests that participants who used PolicyCraft found it easier to understand each other's perspectives during the process of iteratively developing policies, which may have contributed to greater consensus and support for policies observed during voting. As a participant reported in the post-study survey: \textit{``We can argue over cases and everyone's comment is clearly visible. The voting system provides a clear way for people to share opinion towards a specific case/policy. The voting system along with the comments makes the agreements/disagreements very clear.''}

\subsubsection{\textbf{Participants using PolicyCraft were more frequently inspired to edit or create policies based on cases shared by others.}}
In the multiple-choice question that participants answered each time they edited or created a policy, participants were asked whether they were inspired to take this action based on a \textit{specific case} or a \textit{general issue}, and whether the case or general issue was something they had \textit{thought of themselves} or that \textit{someone else had shared}. Both PolicyCraft and the baseline system had a similar proportion of policy editing and creation driven by specific cases (58\% and 53\%, respectively). This result is unsurprising, given that people are known to reason about concrete cases in order to support iterative policy development (see Section \ref{sec:relatedwork_case}). However, a one-sided hypothesis test reveals that in PolicyCraft, a larger proportion ($p<0.05$) of policy editing and creation is aimed at addressing specific cases shared by others (32\%), compared to the baseline system (19\%). This result suggests that PolicyCraft helped participants more often edit and create policies based on cases they collaboratively developed and discussed, with visibility into other participants' perspectives.

\begin{figure}[t]
  \centering
  \includegraphics[width=0.95\linewidth]{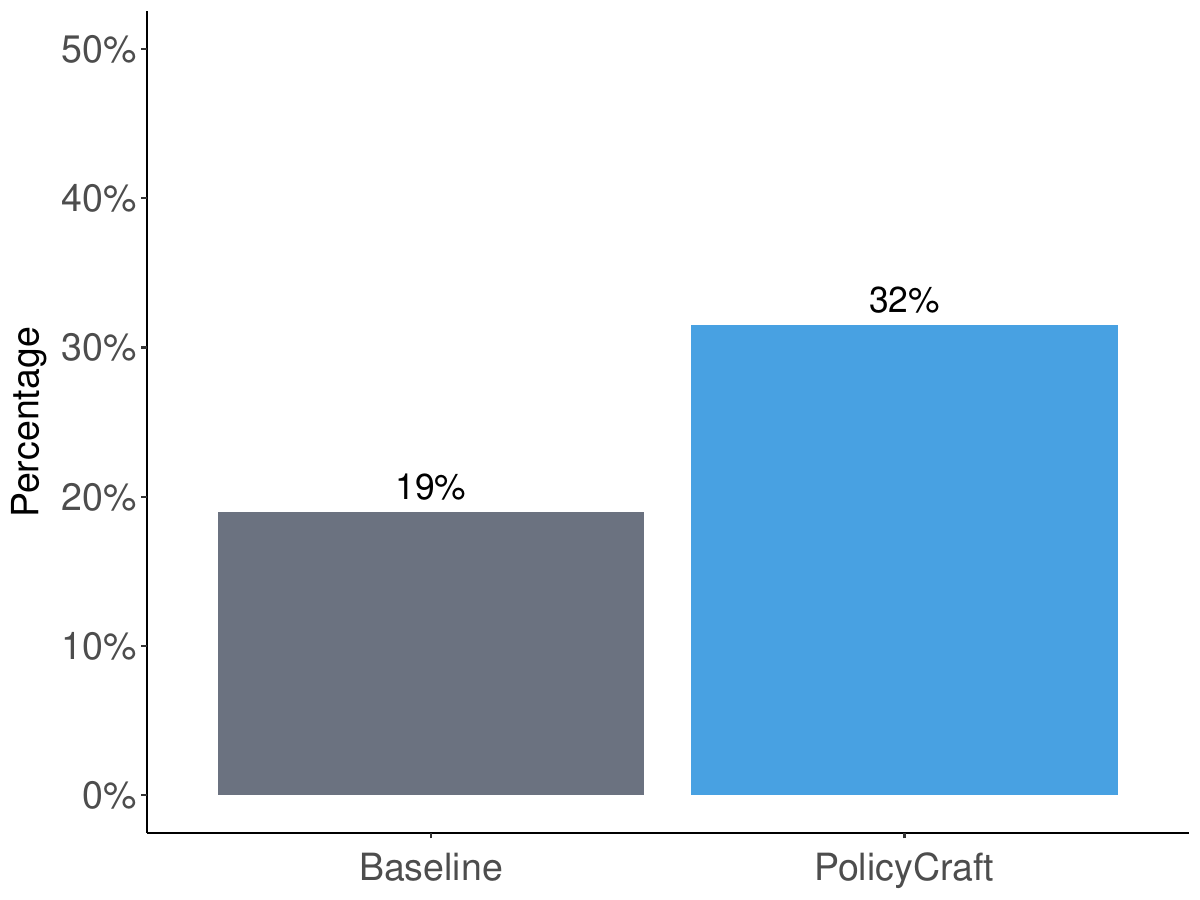}
  \caption{The percentage of policy editing and creation aimed at addressing \textit{specific cases shared by others} during the policy development process. PolicyCraft has a significantly larger proportion than the baseline system.}
  \Description{A bar chart displaying the system participants used during the study on the x-axis (Baseline or PolicyCraft) and their percentages on the y-axis for editing and creating policies to address specific cases shared by others during the policy development process. The percentage for PolicyCraft is significantly higher than that of the baseline.}
  \label{fig:percent_case_discuss}
\end{figure}

\begin{figure*}[t!]
  \centering
  \includegraphics[width=\linewidth]{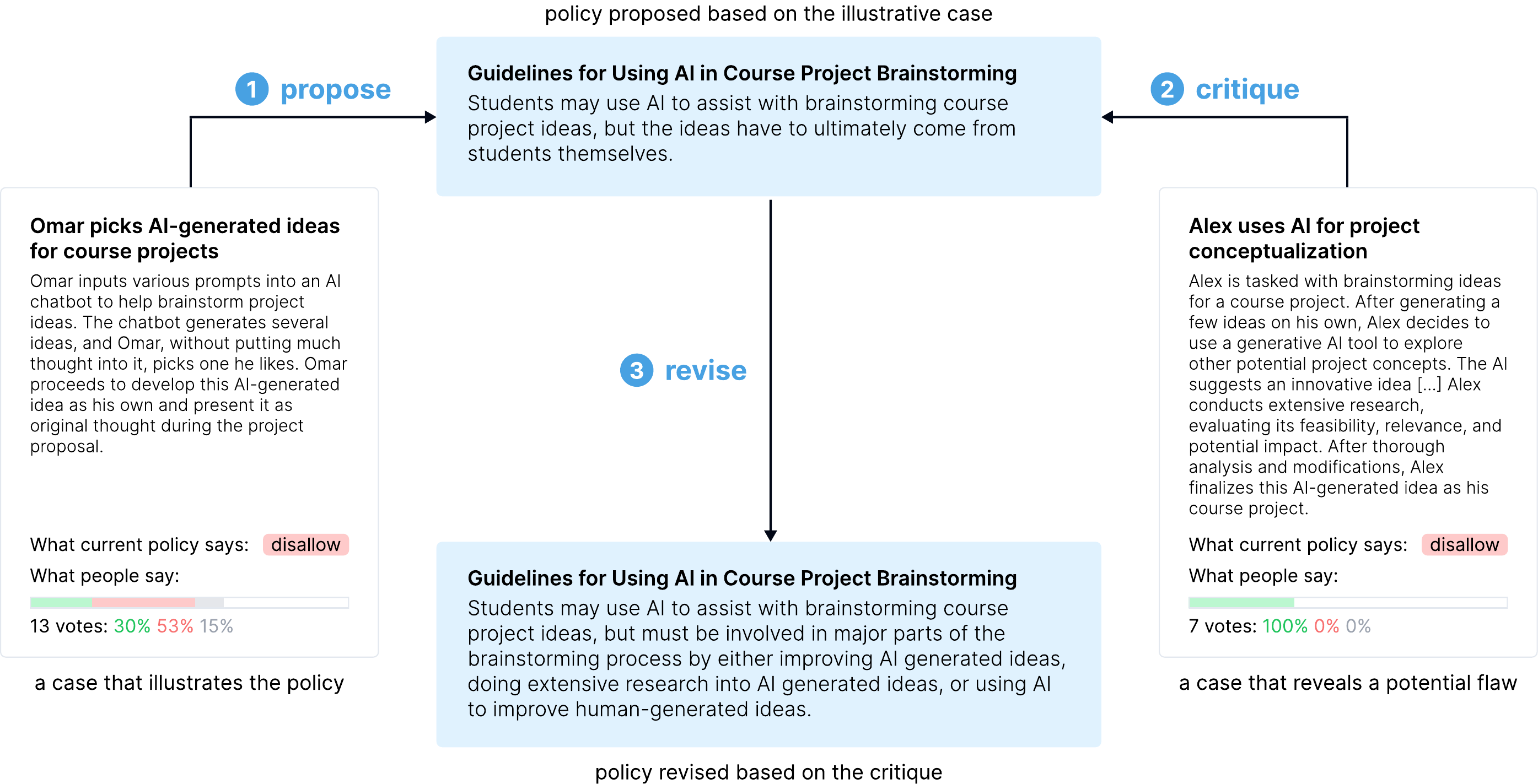}
  \caption{An illustrative example of how participants used cases to iterate on a policy during our field study, showing real policies and cases that our participants developed, voted on, and discussed.}
  \Description{An illustrative example of how participants used cases to iterate on a policy during our field study. In response to a case where Omar picks AI-generated ideas for course projects, participants proposed a policy outlining guidelines for using AI in course project brainstorming. This policy was then critiqued by a case in which Alex uses AI for project conceptualization. To address the critique, participants revised the policy to make it more specific.}
  \label{fig:three_actions}
\end{figure*}

Figure~\ref{fig:three_actions} illustrates how participants used cases to drive collaborative policy iteration, using a real example from our study. As shown, the initial version of the policy was drafted by a participant based on a case that had received some votes and discussion, indicating mixed views among participants. The initial policy allowed students to use AI for course project brainstorming, so long as the ideas ``ultimately come from students themselves''. Soon after the policy was created, another participant added a case that they believed should be allowed but which they believed would be disallowed under the current policy (shown in the right side of Figure~\ref{fig:three_actions}). Other participants voted in agreement. Because the majority vote on this case was misaligned with the policy's label, PolicyCraft highlighted the case with a yellow message noting: \textit{``The policy may need editing to better align with the majority vote on this case.''} Finally, one participant revised the initial policy description to address the misalignment. After additional iterations, this policy eventually received majority support. This concrete example shows how PolicyCraft helped participants iterate on policies by discussing cases and building consensus, even when they shared different perspectives.

\subsection{External Perceptions of Policies}\label{sec:result_3}

\subsubsection{\textbf{The instructors found the policies comprehensive and creative, but wished students had additional support in synthesizing and generalizing policies.}}
To understand the instructors' views on the resulting policies, compared with those they created without students' inputs in previous semesters, we asked them to review the policies and write down their observations and reflections prior to reviewing our study results (cf. \cite{smith2020keeping}).

The instructor of Class 1 found that the policies students collaboratively created \textit{``comprehensively address all aspects relevant to their class experiences, from AI use for conceptual understanding, original work, group work, reading reflections, presentations, coding, and the use of sensitive information.''} This instructor also found some policies particularly creative and unexpected, such as one that regulated instructors' behaviors by \textit{``prohibiting AI-generated grading and feedback''} and another meta-policy called the \textit{``three-strike system''} that governed the enforcement of all the policies. This instructor found that \textit{``the PolicyCraft version created a policy set that has a much higher quality than the baseline condition.''}

The instructor of Class 2 agreed but felt the final set of policies \textit{``did not quite feel like a finished set of policies yet.''} This instructor found that policies developed using PolicyCraft \textit{``reflected more use-case-specific considerations.''} While the instructor found this valuable \textit{``because the policies captured really important nuances''} it also led to \textit{``similar policies that should be combined''.} As a result, the instructor found it \textit{``difficult to say which policy set is better overall because one set of policies was more concise and worded more generally, while the other set was more nuanced but also more redundant.''}

Both instructors wished PolicyCraft had more support for students to synthesize and generalize the policies they developed. For example, the instructor of class two wanted a \textit{``dedicated clustering phase for participants to do some synthesis of their policies''} before voting. In lieu of having this support within the current version of PolicyCraft, each instructor took an initial pass at synthesizing the student-developed policies into final policy sets themselves after the study, and then shared the result with students for any additional feedback.

\subsubsection{\textbf{External raters found the policies developed with PolicyCraft clearer on appropriate uses of generative AI, but also less concise and potentially too comprehensive.}}
Finally, to understand how students outside of these classes, who did not participate in policy development, perceived the quality of the resulting policies, we analyzed the survey responses from external raters, who evaluated the policies that received a majority vote within each class and condition (without knowledge of how these policy sets were generated). While the ratings did not show a statistically significant difference in external raters' assessments across conditions, feedback from external raters provides insight into the qualitative differences they perceived between the policy set, revealing broad alignment with the course instructors' assessments. For example, a rater who preferred the set of policies developed using PolicyCraft  in Class 1 commented that this set of policies \textit{``is clearer about what are appropriate uses of generative AI. [The other set] seems to have a broader interpretation of which uses are okay, leaving it mostly up to interpretation, which creates more of a gray area.''} Similarly, another rater noted that the set of policies developed using PolicyCraft: \textit{``breaks down GenAI policies by particular use making it easy to answer question in the form, `If I am doing X, am I allowed to do Y?' whereas [the other set] comes off more as a list of sentiments about GenAI lightly adapted into an assortment of policies.''} In contrast, a rater who preferred the policies developed with the baseline system mentioned: \textit{``I feel [this set] is clearer and easier to implement for the students and instructors.''} Another rater also mentioned that the set developed using PolicyCraft seemed: \textit{``a little too comprehensive, to the point where students and instructors may struggle to keep track of what is and isn't allowed.''} The raters who evaluated policies in Class 2 provided similar feedback. For example, one rater preferred the PolicyCraft set because \textit{``[the other set] is vague in the exact moments it purports to be specific,''} while another rater preferred the baseline set because they felt the PolicyCraft set was \textit{``harder to implement and less concise.''} This overall perception is consistent with the instructors' feedback. In the Discussion section, we discuss how future systems might better support participants in striking a balance between specificity versus generality and conciseness.

\begin{table*}[t]
  \caption{Illustrative examples of policies on similar topics developed in each class. Both instructors agreed that the policies developed with PolicyCraft were more grounded in specific use cases (highlighted in bold).}
  \label{table:policy_specificity_compare}
  \begin{tabular}
  {p{0.10\textwidth}p{0.42\textwidth}p{0.42\textwidth}}
    \toprule
     & Baseline & PolicyCraft \\
    \midrule
    \small{Using AI for Programming \linebreak (Class 1)}
    &
    \small{\textit{AI for Resolving Coding Errors:} Students should be allowed to use specific AI tools to \textbf{fix the coding errors they come across}.}
    &
    \small{\textit{AI Usage Permitted for Coding Assignments:}
    Students may use AI to aid in coding assignments, but must use AI to augment their work, not create the solution for them. Students cannot use AI to \textbf{create large chunks of code without verifying it themselves}. AI generation of \textbf{very broad high-level pseudocode} is permitted, but \textbf{not step-by-step pseudocode or detailed lines of code}. AI can be used to \textbf{add comments/documentation to already written code} but students should review over them. AI usage must be appropriately attributed.}
    \\
    \midrule
    \small{Acknowledging Use of AI \linebreak (Class 2)}
    &
    \small{\textit{Universal AI Attribution Policy:} If AI was used for \textbf{a particular assignment}, written notice must be given to the professor using the appropriate technology of submission for that assignment (e.g. comments in one's code if programming, the comment box of a canvas submission, etc.) outlining how AI was used in a particular work. This policy takes precedence over all other policy and is necessary to prevent any legal copyright/cheating issues.}
    &
    \small{\textit{Usage of GenAI Tools should be Referenced and Cited:} In \textbf{assignments or presentations}, students should declare and cite the GenAI tools and prompts that they have used to create content or help that they have received openly. While some cases involve \textbf{grammar checks and simple paraphrasing for fluency}, the original idea comes from the student. However, when students use GenAI to \textbf{generate code or ideas for responses}, it is essential to add a reference.}
    \\
    \bottomrule
  \end{tabular}
\end{table*}

\section{Discussion}
It is crucial that community policies reflect the values and needs of the communities they impact, and that they are viewed as legitimate by community members. In this paper, we present PolicyCraft, a system that supports communities in collaboratively proposing, critiquing, and revising policies through discussion and voting on cases. We conducted a field study across two university courses to understand how people use PolicyCraft in practice. Overall, we found that students using PolicyCraft reached greater consensus and developed course policies with greater community support, compared with those using a baseline system (Section~\ref{sec:result_1}). Students using PolicyCraft found it easier to understand each other’s perspectives, and were more likely to iterate on policies based on concrete cases shared and discussed by others (Section~\ref{sec:result_2}). Finally, we present external views on the differences between policies developed with PolicyCraft and those created with the baseline system (Section~\ref{sec:result_3}).

Taken together, while community-external raters' quantitative evaluations did not show a statistically significant difference in the perceived quality of the resulting policies across conditions, their qualitative feedback revealed that policy quality is a complex concept with multiple dimensions shaped by individual preferences. For example, when considering the dimension of policy specificity versus generality, external raters expressed differing preferences that significantly influenced their perception of a policy's quality. Still, even if policies developed using PolicyCraft are not necessarily ``higher quality'' in a global sense, they received stronger support and consensus from community members who would be impacted by the resulting policies. This indicates that PolicyCraft's case-grounded approach to collaborative policy design can support the development of policies that are better aligned with local community perspectives, and that may be viewed as more legitimate by community members. It is possible that policies received greater community support, in part, simply because participants felt they had influence in shaping the policies. Indeed, this is a well-documented benefit of participation in design. However, given that participants in both the baseline and PolicyCraft conditions were directly engaged in collaboratively shaping policies, such participation effects cannot fully explain the observed advantages of PolicyCraft's case-grounded approach. In this section, we discuss future directions for HCI systems to support collaborative and participatory policy design.

\subsection{Balancing Specificity and Generality in Policy Design}
Our overall focus in designing PolicyCraft was to support groups in effectively bridging between abstract policies and concrete cases during collaborative policy design. Policy proposals are often high-level and abstract, leaving much open to interpretation, which can make it challenging for groups of people to understand where and why they disagree. By contrast, PolicyCraft assists users in developing policies based on collaborative discussion and consensus-building around specific cases. As mentioned by external raters and course instructors, and illustrated in Table~\ref{table:policy_specificity_compare}, policies developed with PolicyCraft in our study included more case-specific distinctions, making it easier for readers to answer questions like: \textit{``If I am doing X, am I allowed to do Y?’'} However, as a consequence, the policies also became less concise and harder for readers to keep track of. 

A key insight from our study is that scaffolding for users to \textit{ground} policy development in concrete cases---as provided in the current version of PolicyCraft---needs to be balanced with corresponding scaffolding for users to \textit{abstract} more general policies from these cases. In particular, further research is needed to understand how best to guide participants in striking the right balance between specificity versus generality and conciseness. We expect that the right balance will vary depending on the specific context in which a set of policies will be used. For example, in contexts where only broad guidelines are needed, too much detail can make policies needlessly difficult to implement. It is also worth highlighting the complementary roles of policies and cases, as used in the US legal system, where decisions are based on \textit{both} written laws (statutory law) and past court decisions (case law) \cite{chen2023case}. Utilizing both policies and cases for decision-making may help to balance specificity and generality. For example, in the context of PolicyCraft, this could be achieved by presenting more concise versions of the finalized policies \textit{together with} illustrative cases generated by participants, rather than presenting the policies alone. Future systems should explore mechanisms to scaffold users in developing policies that appropriately balance specificity, generality, and conciseness, tailored to their intended use contexts.

\subsection{Supporting Collaborative Review and Synthesis of Policies During Finalization}
The current version of PolicyCraft has a ``finalization'' stage, where participants can upvote or downvote policies and provide reasons for their votes, to support the selection of a final set of policies. However, as the course instructors suggested, it would be helpful for systems like PolicyCraft to explicitly support participants not only in voting on policies during the finalization phase, but also in reviewing the full set of policies and collaboratively synthesizing the policies as needed. For example, future systems could nudge participants to review the entire policy set and discuss whether they want to merge similar policies or split complex policies into multiple simpler ones. This could not only support the removal of redundancies across policies, but could also help participants refine policies to strike a better balance between specificity versus generality and conciseness. In our study, instructors refined and synthesized policies after the voting stage and then shared them with students via Google Docs for additional feedback before finalizing them as the official course policies. Future systems could integrate this process more effectively to better support end-to-end collaborative policy design.

\subsection{Making Sense of an Evolving Space of Policies and Cases}
One challenge participants in our study faced was navigating and making sense of the space of current policies and cases as their numbers grew and the content of each evolved. Adding more support for review and synthesis of policies during policy finalization could help to address this problem. However, it would also be useful to support users in more effectively making sense of the evolving policy--case space \textit{throughout} the collaborative policy development process. Future systems could draw inspiration from prior research on collective sensemaking by including visualizations that provide an overview of current policies and cases, along with the ability to visually track how the joint space of policies and cases have evolved over time, from an aerial view. In addition, sensemaking might be supported through LLM-based summaries that complement such visualizations by supporting rapid, targeted question-answering \cite{marnette2024talk}. In line with prior work, the usefulness of these visualizations or automated summaries could potentially be supported through a user-driven tagging system that enables users to shape the criteria along which similarity between policies and cases is determined \cite{zhang2018making, paul2009cosense, morris2010wesearch, paul2010understanding, liu2023selenite}.

\subsection{Advancing AI Assistants in Policy Design}
The current implementation of PolicyCraft includes three built-in, LLM-based AI assistants that help users with policy design, as mentioned in Section~\ref{sec:system_ai}. Some participants found the AI assistants helpful for brainstorming cases because \textit{``it helps me think on both side[s], what to allow or disallow, and what should be the case scenarios to accept or reject the policy.''} Other participants found revising and creating policies \textit{``straightforward and easy, especially with the help of the embedded AI assistant.''} However, some participants preferred to \textit{``discuss the real cases in our lives regarding policy crafting because AI-generated cases are quite similar and less reliable than real cases.''} Overall, the AI assistants were not used extensively during our study, with only 26 interactions from the 32 students who had access to the full version of PolicyCraft, across the three-day study period. Future research should further explore how AI assistants can most effectively scaffold collaborative policy design processes, and investigate how different forms of AI-based support may shape both the policy design process and the resulting policies.

\subsection{Navigating Power Dynamics in Policy Design and Implementation}
PolicyCraft is designed to facilitate a collaborative approach to policy design, potentially involving participation and deliberation among various stakeholder groups within a community. It is important to note that PolicyCraft cannot, on its own, overcome power dynamics that may be present among different stakeholder groups. In communities with distributed power dynamics (e.g., Wikipedia), community members may directly implement the policies they collaboratively develop using PolicyCraft. However, in communities with a hierarchical power structure, approval from community leaders may be necessary to actually implement policies. Nonetheless, we envision that even in such communities, PolicyCraft can be used by community members in a bottom-up fashion to generate policy proposals that have strong community support and consensus. We hope PolicyCraft will serve as a source of inspiration for empowering participation to enable a more democratic approach to community governance \cite{zhang2020policykit, salehi2015we, irani2013turkopticon, kuo2024wikibench}.

\subsection{Supporting Collaborative Policy Design Across a Broader Range of Contexts}
PolicyCraft is designed to support policy design across a broad range of community contexts. While the current study focuses on supporting students in collaboratively designing course policies on the use of generative AI, we envision its application in online communities like subreddits to develop content moderation policies \cite{jhaver2023decentralizing}, or in local communities such as vTaiwan to refine regulations for emerging technologies \cite{hsiao2018vtaiwan}. Still, some implementations of PolicyCraft will require careful adaptation to align with specific communities' norms. For example, to encourage policy iteration while holding participants accountable for their edits, the current implementation of PolicyCraft adopts Wikipedia's approach by allowing all participants to edit policies while maintaining a transparent edit history \cite{kittur2007he}. While this approach has shown to be effective in supporting coordination and preventing vandalism in peer production \cite{dabbish2014transparency, stuart2012social} and the current study, a more robust moderation strategy may be required when applying PolicyCraft in different contexts. Future research should investigate which aspects of PolicyCraft's current design need to be adapted for use by other communities and identify which findings from the current study are most transferable across diverse contexts.

\section{Conclusion}
In this work, we have demonstrated how a system that scaffolds users in grounding policy design in concrete cases can support more effective collaborative policy design. Our findings show that policies developed using PolicyCraft receive stronger support and more consensus from those impacted by the policies, compared to policies developed with a baseline system that did not scaffold their use of concrete cases. Building on this work, future HCI research should explore the design of tools and processes to better support collaborative and participatory policy design. This includes helping users balance policy design trade-offs, make collaborative decisions about how to abstract and generalize from concrete cases, better track and understand the evolving space of policies and cases during collaboration, and more effectively identify areas for policy improvements.

\begin{acks}
The funding for this research was provided by UL Research Institutes through the Center for Advancing Safety of Machine Intelligence, CMU's Block Center for Technology and Society, and Metagov's Grant for Interoperable Deliberative Tools. We thank Dominik Moritz, Jodi Forlizzi, Joseph Seering, Maarten Sap, Motahhare Eslami, Suguru Ishizaki, and Tiffany Chih for their insightful feedback on the system design. We are also grateful to CMU's Eberly Center for their guidance on the study design and to Taiwan's g0v community for their assistance in testing the system prototype. Finally, we thank Isadora Krsek for designing the PolicyCraft logo.
\end{acks}

\bibliographystyle{ACM-Reference-Format}
\bibliography{reference}


\begin{thebibliography}{101}


\ifx \showCODEN    \undefined \def \showCODEN     #1{\unskip}     \fi
\ifx \showDOI      \undefined \def \showDOI       #1{#1}\fi
\ifx \showISBNx    \undefined \def \showISBNx     #1{\unskip}     \fi
\ifx \showISBNxiii \undefined \def \showISBNxiii  #1{\unskip}     \fi
\ifx \showISSN     \undefined \def \showISSN      #1{\unskip}     \fi
\ifx \showLCCN     \undefined \def \showLCCN      #1{\unskip}     \fi
\ifx \shownote     \undefined \def \shownote      #1{#1}          \fi
\ifx \showarticletitle \undefined \def \showarticletitle #1{#1}   \fi
\ifx \showURL      \undefined \def \showURL       {\relax}        \fi
\providecommand\bibfield[2]{#2}
\providecommand\bibinfo[2]{#2}
\providecommand\natexlab[1]{#1}
\providecommand\showeprint[2][]{arXiv:#2}

\bibitem[Aamodt and Plaza(1994)]%
        {aamodt1994case}
\bibfield{author}{\bibinfo{person}{Agnar Aamodt} {and} \bibinfo{person}{Enric Plaza}.} \bibinfo{year}{1994}\natexlab{}.
\newblock \showarticletitle{Case-based reasoning: foundational issues, methodological variations, and system approaches}.
\newblock \bibinfo{journal}{\emph{AI Commun.}} \bibinfo{volume}{7}, \bibinfo{number}{1} (\bibinfo{date}{mar} \bibinfo{year}{1994}), \bibinfo{pages}{39–59}.
\newblock
\showISSN{0921-7126}


\bibitem[Aitamurto et~al\mbox{.}(2023)]%
        {aitamurto2023disagreement}
\bibfield{author}{\bibinfo{person}{Tanja Aitamurto}, \bibinfo{person}{Peter~G Royal}, {and} \bibinfo{person}{Jorge Saldivar}.} \bibinfo{year}{2023}\natexlab{}.
\newblock \showarticletitle{Disagreement, Agreement, and Elaboration in Crowdsourced Deliberation: Ideation Through Elaborated Perspectives}. In \bibinfo{booktitle}{\emph{Extended Abstracts of the 2023 CHI Conference on Human Factors in Computing Systems}} (Hamburg, Germany) \emph{(\bibinfo{series}{CHI EA '23})}. \bibinfo{publisher}{Association for Computing Machinery}, \bibinfo{address}{New York, NY, USA}, Article \bibinfo{articleno}{88}, \bibinfo{numpages}{10}~pages.
\newblock
\showISBNx{9781450394222}
\urldef\tempurl%
\url{https://doi.org/10.1145/3544549.3585708}
\showDOI{\tempurl}


\bibitem[Akiyama et~al\mbox{.}(2016)]%
        {akiyama2016method}
\bibfield{author}{\bibinfo{person}{Yoshio Akiyama}, \bibinfo{person}{James Nolan}, \bibinfo{person}{Marjorie Darrah}, \bibinfo{person}{Mushtaq~Abdal Rahem}, {and} \bibinfo{person}{Lei Wang}.} \bibinfo{year}{2016}\natexlab{}.
\newblock \showarticletitle{A method for measuring consensus within groups: An index of disagreement via conditional probability}.
\newblock \bibinfo{journal}{\emph{Information Sciences}}  \bibinfo{volume}{345} (\bibinfo{year}{2016}), \bibinfo{pages}{116--128}.
\newblock


\bibitem[Aranzales et~al\mbox{.}(2021)]%
        {aranzales2021scientists}
\bibfield{author}{\bibinfo{person}{Iv{\'a}n Aranzales}, \bibinfo{person}{Ho~Fai Chan}, \bibinfo{person}{Reiner Eichenberger}, \bibinfo{person}{Rainer Hegselmann}, \bibinfo{person}{David Stadelmann}, {and} \bibinfo{person}{Benno Torgler}.} \bibinfo{year}{2021}\natexlab{}.
\newblock \showarticletitle{Scientists have favorable opinions on immunity certificates but raise concerns regarding fairness and inequality}.
\newblock \bibinfo{journal}{\emph{Scientific reports}} \bibinfo{volume}{11}, \bibinfo{number}{1} (\bibinfo{year}{2021}), \bibinfo{pages}{14016}.
\newblock


\bibitem[Arnstein(1969)]%
        {arnstein1969ladder}
\bibfield{author}{\bibinfo{person}{Sherry~R Arnstein}.} \bibinfo{year}{1969}\natexlab{}.
\newblock \showarticletitle{A ladder of citizen participation}.
\newblock \bibinfo{journal}{\emph{Journal of the American Institute of planners}} \bibinfo{volume}{35}, \bibinfo{number}{4} (\bibinfo{year}{1969}), \bibinfo{pages}{216--224}.
\newblock


\bibitem[Asad and Le~Dantec(2015)]%
        {asad2015illegitimate}
\bibfield{author}{\bibinfo{person}{Mariam Asad} {and} \bibinfo{person}{Christopher~A. Le~Dantec}.} \bibinfo{year}{2015}\natexlab{}.
\newblock \showarticletitle{Illegitimate Civic Participation: Supporting Community Activists on the Ground}. In \bibinfo{booktitle}{\emph{Proceedings of the 18th ACM Conference on Computer Supported Cooperative Work \& Social Computing}} (Vancouver, BC, Canada) \emph{(\bibinfo{series}{CSCW '15})}. \bibinfo{publisher}{Association for Computing Machinery}, \bibinfo{address}{New York, NY, USA}, \bibinfo{pages}{1694–1703}.
\newblock
\showISBNx{9781450329224}
\urldef\tempurl%
\url{https://doi.org/10.1145/2675133.2675156}
\showDOI{\tempurl}


\bibitem[Awad et~al\mbox{.}(2018)]%
        {awad2018moral}
\bibfield{author}{\bibinfo{person}{Edmond Awad}, \bibinfo{person}{Sohan Dsouza}, \bibinfo{person}{Richard Kim}, \bibinfo{person}{Jonathan Schulz}, \bibinfo{person}{Joseph Henrich}, \bibinfo{person}{Azim Shariff}, \bibinfo{person}{Jean-Fran{\c{c}}ois Bonnefon}, {and} \bibinfo{person}{Iyad Rahwan}.} \bibinfo{year}{2018}\natexlab{}.
\newblock \showarticletitle{The moral machine experiment}.
\newblock \bibinfo{journal}{\emph{Nature}} \bibinfo{volume}{563}, \bibinfo{number}{7729} (\bibinfo{year}{2018}), \bibinfo{pages}{59--64}.
\newblock


\bibitem[Bernstein et~al\mbox{.}(2010)]%
        {bernstein2010soylent}
\bibfield{author}{\bibinfo{person}{Michael~S. Bernstein}, \bibinfo{person}{Greg Little}, \bibinfo{person}{Robert~C. Miller}, \bibinfo{person}{Bj\"{o}rn Hartmann}, \bibinfo{person}{Mark~S. Ackerman}, \bibinfo{person}{David~R. Karger}, \bibinfo{person}{David Crowell}, {and} \bibinfo{person}{Katrina Panovich}.} \bibinfo{year}{2010}\natexlab{}.
\newblock \showarticletitle{Soylent: a word processor with a crowd inside}. In \bibinfo{booktitle}{\emph{Proceedings of the 23nd Annual ACM Symposium on User Interface Software and Technology}} (New York, New York, USA) \emph{(\bibinfo{series}{UIST '10})}. \bibinfo{publisher}{Association for Computing Machinery}, \bibinfo{address}{New York, NY, USA}, \bibinfo{pages}{313–322}.
\newblock
\showISBNx{9781450302715}
\urldef\tempurl%
\url{https://doi.org/10.1145/1866029.1866078}
\showDOI{\tempurl}


\bibitem[Boehner and DiSalvo(2016)]%
        {boehner2016data}
\bibfield{author}{\bibinfo{person}{Kirsten Boehner} {and} \bibinfo{person}{Carl DiSalvo}.} \bibinfo{year}{2016}\natexlab{}.
\newblock \showarticletitle{Data, Design and Civics: An Exploratory Study of Civic Tech}. In \bibinfo{booktitle}{\emph{Proceedings of the 2016 CHI Conference on Human Factors in Computing Systems}} (San Jose, California, USA) \emph{(\bibinfo{series}{CHI '16})}. \bibinfo{publisher}{Association for Computing Machinery}, \bibinfo{address}{New York, NY, USA}, \bibinfo{pages}{2970–2981}.
\newblock
\showISBNx{9781450333627}
\urldef\tempurl%
\url{https://doi.org/10.1145/2858036.2858326}
\showDOI{\tempurl}


\bibitem[Butler et~al\mbox{.}(2008)]%
        {butler2008don}
\bibfield{author}{\bibinfo{person}{Brian Butler}, \bibinfo{person}{Elisabeth Joyce}, {and} \bibinfo{person}{Jacqueline Pike}.} \bibinfo{year}{2008}\natexlab{}.
\newblock \showarticletitle{Don't look now, but we've created a bureaucracy: the nature and roles of policies and rules in wikipedia}. In \bibinfo{booktitle}{\emph{Proceedings of the SIGCHI Conference on Human Factors in Computing Systems}} (Florence, Italy) \emph{(\bibinfo{series}{CHI '08})}. \bibinfo{publisher}{Association for Computing Machinery}, \bibinfo{address}{New York, NY, USA}, \bibinfo{pages}{1101–1110}.
\newblock
\showISBNx{9781605580111}
\urldef\tempurl%
\url{https://doi.org/10.1145/1357054.1357227}
\showDOI{\tempurl}


\bibitem[Cardozo and Kaufman(2010)]%
        {cardozo2010nature}
\bibfield{author}{\bibinfo{person}{Benjamin~N Cardozo} {and} \bibinfo{person}{Andrew~L Kaufman}.} \bibinfo{year}{2010}\natexlab{}.
\newblock \bibinfo{booktitle}{\emph{The Nature of the Judicial Process}}.
\newblock \bibinfo{publisher}{Quid Pro Books}.
\newblock


\bibitem[Centivany(2016a)]%
        {centivany2016policy}
\bibfield{author}{\bibinfo{person}{Alissa Centivany}.} \bibinfo{year}{2016}\natexlab{a}.
\newblock \showarticletitle{Policy as Embedded Generativity: A Case Study of the Emergence and Evolution of HathiTrust}. In \bibinfo{booktitle}{\emph{Proceedings of the 19th ACM Conference on Computer-Supported Cooperative Work \& Social Computing}} (San Francisco, California, USA) \emph{(\bibinfo{series}{CSCW '16})}. \bibinfo{publisher}{Association for Computing Machinery}, \bibinfo{address}{New York, NY, USA}, \bibinfo{pages}{926–940}.
\newblock
\showISBNx{9781450335928}
\urldef\tempurl%
\url{https://doi.org/10.1145/2818048.2820069}
\showDOI{\tempurl}


\bibitem[Centivany(2016b)]%
        {centivany2016values}
\bibfield{author}{\bibinfo{person}{Alissa Centivany}.} \bibinfo{year}{2016}\natexlab{b}.
\newblock \showarticletitle{Values, ethics and participatory policymaking in online communities}.
\newblock \bibinfo{journal}{\emph{Proceedings of the Association for Information Science and Technology}} \bibinfo{volume}{53}, \bibinfo{number}{1} (\bibinfo{year}{2016}), \bibinfo{pages}{1--10}.
\newblock


\bibitem[Chandrasekharan et~al\mbox{.}(2018)]%
        {chandrasekharan2018internet}
\bibfield{author}{\bibinfo{person}{Eshwar Chandrasekharan}, \bibinfo{person}{Mattia Samory}, \bibinfo{person}{Shagun Jhaver}, \bibinfo{person}{Hunter Charvat}, \bibinfo{person}{Amy Bruckman}, \bibinfo{person}{Cliff Lampe}, \bibinfo{person}{Jacob Eisenstein}, {and} \bibinfo{person}{Eric Gilbert}.} \bibinfo{year}{2018}\natexlab{}.
\newblock \showarticletitle{The Internet's Hidden Rules: An Empirical Study of Reddit Norm Violations at Micro, Meso, and Macro Scales}.
\newblock \bibinfo{journal}{\emph{Proc. ACM Hum.-Comput. Interact.}} \bibinfo{volume}{2}, \bibinfo{number}{CSCW}, Article \bibinfo{articleno}{32} (\bibinfo{date}{nov} \bibinfo{year}{2018}), \bibinfo{numpages}{25}~pages.
\newblock
\urldef\tempurl%
\url{https://doi.org/10.1145/3274301}
\showDOI{\tempurl}


\bibitem[Chen(2021)]%
        {chen2021cold}
\bibfield{author}{\bibinfo{person}{Andrew Chen}.} \bibinfo{year}{2021}\natexlab{}.
\newblock \bibinfo{booktitle}{\emph{The cold start problem}}.
\newblock \bibinfo{publisher}{Harper Business}.
\newblock


\bibitem[Chen and Zhang(2023a)]%
        {chen2023case}
\bibfield{author}{\bibinfo{person}{Quan~Ze Chen} {and} \bibinfo{person}{Amy~X. Zhang}.} \bibinfo{year}{2023}\natexlab{a}.
\newblock \bibinfo{title}{Case Law Grounding: Aligning Judgments of Humans and AI on Socially-Constructed Concepts}.
\newblock
\newblock
\showeprint[arxiv]{2310.07019}~[cs.HC]
\urldef\tempurl%
\url{https://arxiv.org/abs/2310.07019}
\showURL{%
\tempurl}


\bibitem[Chen and Zhang(2023b)]%
        {chen2023judgment}
\bibfield{author}{\bibinfo{person}{Quan~Ze Chen} {and} \bibinfo{person}{Amy~X. Zhang}.} \bibinfo{year}{2023}\natexlab{b}.
\newblock \showarticletitle{Judgment Sieve: Reducing Uncertainty in Group Judgments through Interventions Targeting Ambiguity versus Disagreement}.
\newblock \bibinfo{journal}{\emph{Proc. ACM Hum.-Comput. Interact.}} \bibinfo{volume}{7}, \bibinfo{number}{CSCW2}, Article \bibinfo{articleno}{283} (\bibinfo{date}{oct} \bibinfo{year}{2023}), \bibinfo{numpages}{26}~pages.
\newblock
\urldef\tempurl%
\url{https://doi.org/10.1145/3610074}
\showDOI{\tempurl}


\bibitem[Cheong et~al\mbox{.}(2024)]%
        {cheong2024not}
\bibfield{author}{\bibinfo{person}{Inyoung Cheong}, \bibinfo{person}{King Xia}, \bibinfo{person}{K.~J.~Kevin Feng}, \bibinfo{person}{Quan~Ze Chen}, {and} \bibinfo{person}{Amy~X. Zhang}.} \bibinfo{year}{2024}\natexlab{}.
\newblock \showarticletitle{(A)I Am Not a Lawyer, But...: Engaging Legal Experts towards Responsible LLM Policies for Legal Advice}. In \bibinfo{booktitle}{\emph{Proceedings of the 2024 ACM Conference on Fairness, Accountability, and Transparency}} (Rio de Janeiro, Brazil) \emph{(\bibinfo{series}{FAccT '24})}. \bibinfo{publisher}{Association for Computing Machinery}, \bibinfo{address}{New York, NY, USA}, \bibinfo{pages}{2454–2469}.
\newblock
\showISBNx{9798400704505}
\urldef\tempurl%
\url{https://doi.org/10.1145/3630106.3659048}
\showDOI{\tempurl}


\bibitem[Corbett et~al\mbox{.}(2023)]%
        {corbett2023power}
\bibfield{author}{\bibinfo{person}{Eric Corbett}, \bibinfo{person}{Emily Denton}, {and} \bibinfo{person}{Sheena Erete}.} \bibinfo{year}{2023}\natexlab{}.
\newblock \showarticletitle{Power and Public Participation in AI}. In \bibinfo{booktitle}{\emph{Proceedings of the 3rd ACM Conference on Equity and Access in Algorithms, Mechanisms, and Optimization}} (Boston, MA, USA) \emph{(\bibinfo{series}{EAAMO '23})}. \bibinfo{publisher}{Association for Computing Machinery}, \bibinfo{address}{New York, NY, USA}, Article \bibinfo{articleno}{8}, \bibinfo{numpages}{13}~pages.
\newblock
\showISBNx{9798400703812}
\urldef\tempurl%
\url{https://doi.org/10.1145/3617694.3623228}
\showDOI{\tempurl}


\bibitem[Dabbish et~al\mbox{.}(2014)]%
        {dabbish2014transparency}
\bibfield{author}{\bibinfo{person}{Laura Dabbish}, \bibinfo{person}{Colleen Stuart}, \bibinfo{person}{Jason Tsay}, {and} \bibinfo{person}{Jim Herbsleb}.} \bibinfo{year}{2014}\natexlab{}.
\newblock \showarticletitle{Transparency and coordination in peer production}.
\newblock \bibinfo{journal}{\emph{arXiv preprint arXiv:1407.0377}} (\bibinfo{year}{2014}).
\newblock


\bibitem[Das et~al\mbox{.}(0)]%
        {das2024we}
\bibfield{author}{\bibinfo{person}{Ranjana Das}, \bibinfo{person}{Yen~Nee Wong}, \bibinfo{person}{Rhianne Jones}, {and} \bibinfo{person}{Philip~JB Jackson}.} \bibinfo{year}{0}\natexlab{}.
\newblock \showarticletitle{How do we speak about algorithms and algorithmic media futures? Using vignettes and scenarios in a citizen council on data-driven media personalisation}.
\newblock \bibinfo{journal}{\emph{New Media \& Society}} \bibinfo{volume}{0}, \bibinfo{number}{0} (\bibinfo{year}{0}), \bibinfo{pages}{14614448241232589}.
\newblock
\urldef\tempurl%
\url{https://doi.org/10.1177/14614448241232589}
\showDOI{\tempurl}


\bibitem[De~Liddo and Buckingham~Shum(2010)]%
        {de2010cohere}
\bibfield{author}{\bibinfo{person}{Anna De~Liddo} {and} \bibinfo{person}{Simon Buckingham~Shum}.} \bibinfo{year}{2010}\natexlab{}.
\newblock \showarticletitle{Cohere: A prototype for contested collective intelligence}.
\newblock  (\bibinfo{year}{2010}).
\newblock


\bibitem[Epstein et~al\mbox{.}(2014)]%
        {epstein2014value}
\bibfield{author}{\bibinfo{person}{Dmitry Epstein}, \bibinfo{person}{Cynthia Farina}, {and} \bibinfo{person}{Josiah Heidt}.} \bibinfo{year}{2014}\natexlab{}.
\newblock \showarticletitle{The value of words: Narrative as evidence in policy making}.
\newblock \bibinfo{journal}{\emph{Evidence \& Policy}} \bibinfo{volume}{10}, \bibinfo{number}{2} (\bibinfo{year}{2014}), \bibinfo{pages}{243--258}.
\newblock


\bibitem[Erete and Burrell(2017)]%
        {erete2017empowered}
\bibfield{author}{\bibinfo{person}{Sheena Erete} {and} \bibinfo{person}{Jennifer~O. Burrell}.} \bibinfo{year}{2017}\natexlab{}.
\newblock \showarticletitle{Empowered Participation: How Citizens Use Technology in Local Governance}. In \bibinfo{booktitle}{\emph{Proceedings of the 2017 CHI Conference on Human Factors in Computing Systems}} (Denver, Colorado, USA) \emph{(\bibinfo{series}{CHI '17})}. \bibinfo{publisher}{Association for Computing Machinery}, \bibinfo{address}{New York, NY, USA}, \bibinfo{pages}{2307–2319}.
\newblock
\showISBNx{9781450346559}
\urldef\tempurl%
\url{https://doi.org/10.1145/3025453.3025996}
\showDOI{\tempurl}


\bibitem[Fan and Zhang(2020)]%
        {fan2020digital}
\bibfield{author}{\bibinfo{person}{Jenny Fan} {and} \bibinfo{person}{Amy~X. Zhang}.} \bibinfo{year}{2020}\natexlab{}.
\newblock \showarticletitle{Digital Juries: A Civics-Oriented Approach to Platform Governance}. In \bibinfo{booktitle}{\emph{Proceedings of the 2020 CHI Conference on Human Factors in Computing Systems}} (Honolulu, HI, USA) \emph{(\bibinfo{series}{CHI '20})}. \bibinfo{publisher}{Association for Computing Machinery}, \bibinfo{address}{New York, NY, USA}, \bibinfo{pages}{1–14}.
\newblock
\showISBNx{9781450367080}
\urldef\tempurl%
\url{https://doi.org/10.1145/3313831.3376293}
\showDOI{\tempurl}


\bibitem[Fang et~al\mbox{.}(2023)]%
        {fang2023people}
\bibfield{author}{\bibinfo{person}{Jingchao Fang}, \bibinfo{person}{Jia-Wei Liang}, {and} \bibinfo{person}{Hao-Chuan Wang}.} \bibinfo{year}{2023}\natexlab{}.
\newblock \showarticletitle{How People Initiate and Respond to Discussions Around Online Community Norms: A Preliminary Analysis on Meta Stack Overflow Discussions}. In \bibinfo{booktitle}{\emph{Companion Publication of the 2023 Conference on Computer Supported Cooperative Work and Social Computing}} (Minneapolis, MN, USA) \emph{(\bibinfo{series}{CSCW '23 Companion})}. \bibinfo{publisher}{Association for Computing Machinery}, \bibinfo{address}{New York, NY, USA}, \bibinfo{pages}{221–225}.
\newblock
\showISBNx{9798400701290}
\urldef\tempurl%
\url{https://doi.org/10.1145/3584931.3606966}
\showDOI{\tempurl}


\bibitem[Feng et~al\mbox{.}(2023)]%
        {feng2023case}
\bibfield{author}{\bibinfo{person}{K.~J.~Kevin Feng}, \bibinfo{person}{Quan~Ze Chen}, \bibinfo{person}{Inyoung Cheong}, \bibinfo{person}{King Xia}, {and} \bibinfo{person}{Amy~X. Zhang}.} \bibinfo{year}{2023}\natexlab{}.
\newblock \bibinfo{title}{Case Repositories: Towards Case-Based Reasoning for AI Alignment}.
\newblock
\newblock
\showeprint[arxiv]{2311.10934}~[cs.AI]
\urldef\tempurl%
\url{https://arxiv.org/abs/2311.10934}
\showURL{%
\tempurl}


\bibitem[Fiesler and Dym(2020)]%
        {fiesler2020moving}
\bibfield{author}{\bibinfo{person}{Casey Fiesler} {and} \bibinfo{person}{Brianna Dym}.} \bibinfo{year}{2020}\natexlab{}.
\newblock \showarticletitle{Moving Across Lands: Online Platform Migration in Fandom Communities}.
\newblock \bibinfo{journal}{\emph{Proc. ACM Hum.-Comput. Interact.}} \bibinfo{volume}{4}, \bibinfo{number}{CSCW1}, Article \bibinfo{articleno}{42} (\bibinfo{date}{may} \bibinfo{year}{2020}), \bibinfo{numpages}{25}~pages.
\newblock
\urldef\tempurl%
\url{https://doi.org/10.1145/3392847}
\showDOI{\tempurl}


\bibitem[Fiesler et~al\mbox{.}(2018)]%
        {fiesler2018reddit}
\bibfield{author}{\bibinfo{person}{Casey Fiesler}, \bibinfo{person}{Jialun Jiang}, \bibinfo{person}{Joshua McCann}, \bibinfo{person}{Kyle Frye}, {and} \bibinfo{person}{Jed Brubaker}.} \bibinfo{year}{2018}\natexlab{}.
\newblock \showarticletitle{Reddit Rules! Characterizing an Ecosystem of Governance}.
\newblock \bibinfo{journal}{\emph{Proceedings of the International AAAI Conference on Web and Social Media}} \bibinfo{volume}{12}, \bibinfo{number}{1} (\bibinfo{date}{Jun.} \bibinfo{year}{2018}).
\newblock
\urldef\tempurl%
\url{https://doi.org/10.1609/icwsm.v12i1.15033}
\showDOI{\tempurl}


\bibitem[Grey(1983)]%
        {grey1983langdell}
\bibfield{author}{\bibinfo{person}{Thomas~C Grey}.} \bibinfo{year}{1983}\natexlab{}.
\newblock \showarticletitle{Langdell's orthodoxy}.
\newblock \bibinfo{journal}{\emph{U. PiTT. l. rev.}}  \bibinfo{volume}{45} (\bibinfo{year}{1983}), \bibinfo{pages}{1}.
\newblock


\bibitem[Halfaker et~al\mbox{.}(2025)]%
        {halfaker2025collective}
\bibfield{author}{\bibinfo{person}{Aaron~L Halfaker}, \bibinfo{person}{Tzu-Sheng Kuo}, \bibinfo{person}{Ciell Brusse}, \bibinfo{person}{Kenneth Holstein}, {and} \bibinfo{person}{Haiyi Zhu}.} \bibinfo{year}{2025}\natexlab{}.
\newblock \showarticletitle{Collective Meaning Cascades but Strange Ducks Swim Upstream: Facilitating Collective Meaning-making through Co-development of AI Models}. In \bibinfo{booktitle}{\emph{Extended Abstracts of the 2025 CHI Conference on Human Factors in Computing Systems}} \emph{(\bibinfo{series}{CHI EA '25})}.
\newblock
\urldef\tempurl%
\url{https://doi.org/10.1145/3706599.3706683}
\showDOI{\tempurl}


\bibitem[Hess(2007)]%
        {hess2007collaborative}
\bibfield{author}{\bibinfo{person}{Gerald~F Hess}.} \bibinfo{year}{2007}\natexlab{}.
\newblock \showarticletitle{Collaborative course design: Not my course, not their course, but our course}.
\newblock \bibinfo{journal}{\emph{Washburn LJ}}  \bibinfo{volume}{47} (\bibinfo{year}{2007}), \bibinfo{pages}{367}.
\newblock


\bibitem[Hsiao et~al\mbox{.}(2018)]%
        {hsiao2018vtaiwan}
\bibfield{author}{\bibinfo{person}{Yu-Tang Hsiao}, \bibinfo{person}{Shu-Yang Lin}, \bibinfo{person}{Audrey Tang}, \bibinfo{person}{Darshana Narayanan}, {and} \bibinfo{person}{Claudina Sarahe}.} \bibinfo{year}{2018}\natexlab{}.
\newblock \showarticletitle{vTaiwan: An empirical study of open consultation process in Taiwan}.
\newblock \bibinfo{journal}{\emph{SocArXiv}}  \bibinfo{volume}{4} (\bibinfo{year}{2018}).
\newblock


\bibitem[Hwang and Shaw(2022)]%
        {hwang2022rules}
\bibfield{author}{\bibinfo{person}{Sohyeon Hwang} {and} \bibinfo{person}{Aaron Shaw}.} \bibinfo{year}{2022}\natexlab{}.
\newblock \showarticletitle{Rules and rule-making in the five largest Wikipedias}. In \bibinfo{booktitle}{\emph{Proceedings of the International AAAI Conference on Web and Social Media}}, Vol.~\bibinfo{volume}{16}. \bibinfo{pages}{347--357}.
\newblock


\bibitem[Iandoli et~al\mbox{.}(2014)]%
        {iandoli2014socially}
\bibfield{author}{\bibinfo{person}{Luca Iandoli}, \bibinfo{person}{Ivana Quinto}, \bibinfo{person}{Anna De~Liddo}, {and} \bibinfo{person}{Simon~Buckingham Shum}.} \bibinfo{year}{2014}\natexlab{}.
\newblock \showarticletitle{Socially augmented argumentation tools: Rationale, design and evaluation of a debate dashboard}.
\newblock \bibinfo{journal}{\emph{International Journal of Human-Computer Studies}} \bibinfo{volume}{72}, \bibinfo{number}{3} (\bibinfo{year}{2014}), \bibinfo{pages}{298--319}.
\newblock


\bibitem[Irani and Silberman(2013)]%
        {irani2013turkopticon}
\bibfield{author}{\bibinfo{person}{Lilly~C. Irani} {and} \bibinfo{person}{M.~Six Silberman}.} \bibinfo{year}{2013}\natexlab{}.
\newblock \showarticletitle{Turkopticon: interrupting worker invisibility in amazon mechanical turk}. In \bibinfo{booktitle}{\emph{Proceedings of the SIGCHI Conference on Human Factors in Computing Systems}} (Paris, France) \emph{(\bibinfo{series}{CHI '13})}. \bibinfo{publisher}{Association for Computing Machinery}, \bibinfo{address}{New York, NY, USA}, \bibinfo{pages}{611–620}.
\newblock
\showISBNx{9781450318990}
\urldef\tempurl%
\url{https://doi.org/10.1145/2470654.2470742}
\showDOI{\tempurl}


\bibitem[Jackson et~al\mbox{.}(2014)]%
        {jackson2014policy}
\bibfield{author}{\bibinfo{person}{Steven~J. Jackson}, \bibinfo{person}{Tarleton Gillespie}, {and} \bibinfo{person}{Sandy Payette}.} \bibinfo{year}{2014}\natexlab{}.
\newblock \showarticletitle{The policy knot: re-integrating policy, practice and design in cscw studies of social computing}. In \bibinfo{booktitle}{\emph{Proceedings of the 17th ACM Conference on Computer Supported Cooperative Work \& Social Computing}} (Baltimore, Maryland, USA) \emph{(\bibinfo{series}{CSCW '14})}. \bibinfo{publisher}{Association for Computing Machinery}, \bibinfo{address}{New York, NY, USA}, \bibinfo{pages}{588–602}.
\newblock
\showISBNx{9781450325400}
\urldef\tempurl%
\url{https://doi.org/10.1145/2531602.2531674}
\showDOI{\tempurl}


\bibitem[Jasim et~al\mbox{.}(2021)]%
        {jasim2021communitypulse}
\bibfield{author}{\bibinfo{person}{Mahmood Jasim}, \bibinfo{person}{Enamul Hoque}, \bibinfo{person}{Ali Sarvghad}, {and} \bibinfo{person}{Narges Mahyar}.} \bibinfo{year}{2021}\natexlab{}.
\newblock \showarticletitle{CommunityPulse: Facilitating Community Input Analysis by Surfacing Hidden Insights, Reflections, and Priorities}. In \bibinfo{booktitle}{\emph{Proceedings of the 2021 ACM Designing Interactive Systems Conference}} (Virtual Event, USA) \emph{(\bibinfo{series}{DIS '21})}. \bibinfo{publisher}{Association for Computing Machinery}, \bibinfo{address}{New York, NY, USA}, \bibinfo{pages}{846–863}.
\newblock
\showISBNx{9781450384766}
\urldef\tempurl%
\url{https://doi.org/10.1145/3461778.3462132}
\showDOI{\tempurl}


\bibitem[Jhaver et~al\mbox{.}(2023)]%
        {jhaver2023decentralizing}
\bibfield{author}{\bibinfo{person}{Shagun Jhaver}, \bibinfo{person}{Seth Frey}, {and} \bibinfo{person}{Amy~X. Zhang}.} \bibinfo{year}{2023}\natexlab{}.
\newblock \showarticletitle{Decentralizing Platform Power: A Design Space of Multi-Level Governance in Online Social Platforms}.
\newblock \bibinfo{journal}{\emph{Social Media + Society}} \bibinfo{volume}{9}, \bibinfo{number}{4} (\bibinfo{year}{2023}), \bibinfo{pages}{20563051231207857}.
\newblock
\urldef\tempurl%
\url{https://doi.org/10.1177/20563051231207857}
\showDOI{\tempurl}
\showeprint{https://doi.org/10.1177/20563051231207857}


\bibitem[Kiene et~al\mbox{.}(2016)]%
        {kiene2016surviving}
\bibfield{author}{\bibinfo{person}{Charles Kiene}, \bibinfo{person}{Andr\'{e}s Monroy-Hern\'{a}ndez}, {and} \bibinfo{person}{Benjamin~Mako Hill}.} \bibinfo{year}{2016}\natexlab{}.
\newblock \showarticletitle{Surviving an "Eternal September": How an Online Community Managed a Surge of Newcomers}. In \bibinfo{booktitle}{\emph{Proceedings of the 2016 CHI Conference on Human Factors in Computing Systems}} (San Jose, California, USA) \emph{(\bibinfo{series}{CHI '16})}. \bibinfo{publisher}{Association for Computing Machinery}, \bibinfo{address}{New York, NY, USA}, \bibinfo{pages}{1152–1156}.
\newblock
\showISBNx{9781450333627}
\urldef\tempurl%
\url{https://doi.org/10.1145/2858036.2858356}
\showDOI{\tempurl}


\bibitem[Kieslich et~al\mbox{.}(2024)]%
        {kieslich2024anticipating}
\bibfield{author}{\bibinfo{person}{Kimon Kieslich}, \bibinfo{person}{Nicholas Diakopoulos}, {and} \bibinfo{person}{Natali Helberger}.} \bibinfo{year}{2024}\natexlab{}.
\newblock \showarticletitle{Anticipating impacts: using large-scale scenario-writing to explore diverse implications of generative AI in the news environment}.
\newblock \bibinfo{journal}{\emph{AI and Ethics}} (\bibinfo{year}{2024}), \bibinfo{pages}{1--23}.
\newblock


\bibitem[Kim et~al\mbox{.}(2021b)]%
        {kim2021starrythoughts}
\bibfield{author}{\bibinfo{person}{Hyunwoo Kim}, \bibinfo{person}{Haesoo Kim}, \bibinfo{person}{Kyung~Je Jo}, {and} \bibinfo{person}{Juho Kim}.} \bibinfo{year}{2021}\natexlab{b}.
\newblock \showarticletitle{StarryThoughts: Facilitating Diverse Opinion Exploration on Social Issues}.
\newblock \bibinfo{journal}{\emph{Proc. ACM Hum.-Comput. Interact.}} \bibinfo{volume}{5}, \bibinfo{number}{CSCW1}, Article \bibinfo{articleno}{66} (\bibinfo{date}{apr} \bibinfo{year}{2021}), \bibinfo{numpages}{29}~pages.
\newblock
\urldef\tempurl%
\url{https://doi.org/10.1145/3449140}
\showDOI{\tempurl}


\bibitem[Kim et~al\mbox{.}(2021a)]%
        {kim2021moderator}
\bibfield{author}{\bibinfo{person}{Soomin Kim}, \bibinfo{person}{Jinsu Eun}, \bibinfo{person}{Joseph Seering}, {and} \bibinfo{person}{Joonhwan Lee}.} \bibinfo{year}{2021}\natexlab{a}.
\newblock \showarticletitle{Moderator Chatbot for Deliberative Discussion: Effects of Discussion Structure and Discussant Facilitation}.
\newblock \bibinfo{journal}{\emph{Proc. ACM Hum.-Comput. Interact.}} \bibinfo{volume}{5}, \bibinfo{number}{CSCW1}, Article \bibinfo{articleno}{87} (\bibinfo{date}{apr} \bibinfo{year}{2021}), \bibinfo{numpages}{26}~pages.
\newblock
\urldef\tempurl%
\url{https://doi.org/10.1145/3449161}
\showDOI{\tempurl}


\bibitem[Kim et~al\mbox{.}(2024)]%
        {kim2024integrating}
\bibfield{author}{\bibinfo{person}{Seyun Kim}, \bibinfo{person}{Jonathan Ho}, \bibinfo{person}{Yinan Li}, \bibinfo{person}{Bonnie Fan}, \bibinfo{person}{Willa~Yunqi Yang}, \bibinfo{person}{Jessie Ramey}, \bibinfo{person}{Sarah~E. Fox}, \bibinfo{person}{Haiyi Zhu}, \bibinfo{person}{John Zimmerman}, {and} \bibinfo{person}{Motahhare Eslami}.} \bibinfo{year}{2024}\natexlab{}.
\newblock \showarticletitle{Integrating Equity in Public Sector Data-Driven Decision Making: Exploring the Desired Futures of Underserved Stakeholders}.
\newblock \bibinfo{journal}{\emph{Proc. ACM Hum.-Comput. Interact.}} \bibinfo{volume}{8}, \bibinfo{number}{CSCW2}, Article \bibinfo{articleno}{366} (\bibinfo{date}{Nov.} \bibinfo{year}{2024}), \bibinfo{numpages}{39}~pages.
\newblock
\urldef\tempurl%
\url{https://doi.org/10.1145/3686905}
\showDOI{\tempurl}


\bibitem[Kirkham(2023)]%
        {kirkham2023legal}
\bibfield{author}{\bibinfo{person}{Reuben Kirkham}.} \bibinfo{year}{2023}\natexlab{}.
\newblock \bibinfo{title}{(Legal Design) Research through Litigation}.
\newblock
\newblock
\showeprint[arxiv]{2303.14336}~[cs.HC]
\urldef\tempurl%
\url{https://arxiv.org/abs/2303.14336}
\showURL{%
\tempurl}


\bibitem[Kittur et~al\mbox{.}(2007)]%
        {kittur2007he}
\bibfield{author}{\bibinfo{person}{Aniket Kittur}, \bibinfo{person}{Bongwon Suh}, \bibinfo{person}{Bryan~A. Pendleton}, {and} \bibinfo{person}{Ed~H. Chi}.} \bibinfo{year}{2007}\natexlab{}.
\newblock \showarticletitle{He says, she says: conflict and coordination in Wikipedia}. In \bibinfo{booktitle}{\emph{Proceedings of the SIGCHI Conference on Human Factors in Computing Systems}} (San Jose, California, USA) \emph{(\bibinfo{series}{CHI '07})}. \bibinfo{publisher}{Association for Computing Machinery}, \bibinfo{address}{New York, NY, USA}, \bibinfo{pages}{453–462}.
\newblock
\showISBNx{9781595935939}
\urldef\tempurl%
\url{https://doi.org/10.1145/1240624.1240698}
\showDOI{\tempurl}


\bibitem[Konya et~al\mbox{.}(2023)]%
        {konya2023democratic}
\bibfield{author}{\bibinfo{person}{Andrew Konya}, \bibinfo{person}{Lisa Schirch}, \bibinfo{person}{Colin Irwin}, {and} \bibinfo{person}{Aviv Ovadya}.} \bibinfo{year}{2023}\natexlab{}.
\newblock \bibinfo{title}{Democratic Policy Development using Collective Dialogues and AI}.
\newblock
\newblock
\showeprint[arxiv]{2311.02242}~[cs.CY]
\urldef\tempurl%
\url{https://arxiv.org/abs/2311.02242}
\showURL{%
\tempurl}


\bibitem[Koshy et~al\mbox{.}(2023)]%
        {koshy2023measuring}
\bibfield{author}{\bibinfo{person}{Vinay Koshy}, \bibinfo{person}{Tanvi Bajpai}, \bibinfo{person}{Eshwar Chandrasekharan}, \bibinfo{person}{Hari Sundaram}, {and} \bibinfo{person}{Karrie Karahalios}.} \bibinfo{year}{2023}\natexlab{}.
\newblock \showarticletitle{Measuring User-Moderator Alignment on r/ChangeMyView}.
\newblock \bibinfo{journal}{\emph{Proc. ACM Hum.-Comput. Interact.}} \bibinfo{volume}{7}, \bibinfo{number}{CSCW2}, Article \bibinfo{articleno}{286} (\bibinfo{date}{oct} \bibinfo{year}{2023}), \bibinfo{numpages}{36}~pages.
\newblock
\urldef\tempurl%
\url{https://doi.org/10.1145/3610077}
\showDOI{\tempurl}


\bibitem[Koshy et~al\mbox{.}(2024)]%
        {koshy2024venire}
\bibfield{author}{\bibinfo{person}{Vinay Koshy}, \bibinfo{person}{Frederick Choi}, \bibinfo{person}{Yi-Shyuan Chiang}, \bibinfo{person}{Hari Sundaram}, \bibinfo{person}{Eshwar Chandrasekharan}, {and} \bibinfo{person}{Karrie Karahalios}.} \bibinfo{year}{2024}\natexlab{}.
\newblock \showarticletitle{Venire: A Machine Learning-Guided Panel Review System for Community Content Moderation}.
\newblock \bibinfo{journal}{\emph{arXiv preprint arXiv:2410.23448}} (\bibinfo{year}{2024}).
\newblock


\bibitem[Krafft et~al\mbox{.}(2021)]%
        {krafft2021action}
\bibfield{author}{\bibinfo{person}{P.~M. Krafft}, \bibinfo{person}{Meg Young}, \bibinfo{person}{Michael Katell}, \bibinfo{person}{Jennifer~E. Lee}, \bibinfo{person}{Shankar Narayan}, \bibinfo{person}{Micah Epstein}, \bibinfo{person}{Dharma Dailey}, \bibinfo{person}{Bernease Herman}, \bibinfo{person}{Aaron Tam}, \bibinfo{person}{Vivian Guetler}, \bibinfo{person}{Corinne Bintz}, \bibinfo{person}{Daniella Raz}, \bibinfo{person}{Pa~Ousman Jobe}, \bibinfo{person}{Franziska Putz}, \bibinfo{person}{Brian Robick}, {and} \bibinfo{person}{Bissan Barghouti}.} \bibinfo{year}{2021}\natexlab{}.
\newblock \showarticletitle{An Action-Oriented AI Policy Toolkit for Technology Audits by Community Advocates and Activists}. In \bibinfo{booktitle}{\emph{Proceedings of the 2021 ACM Conference on Fairness, Accountability, and Transparency}} (Virtual Event, Canada) \emph{(\bibinfo{series}{FAccT '21})}. \bibinfo{publisher}{Association for Computing Machinery}, \bibinfo{address}{New York, NY, USA}, \bibinfo{pages}{772–781}.
\newblock
\showISBNx{9781450383097}
\urldef\tempurl%
\url{https://doi.org/10.1145/3442188.3445938}
\showDOI{\tempurl}


\bibitem[Kriplean et~al\mbox{.}(2012)]%
        {kriplean2012supporting}
\bibfield{author}{\bibinfo{person}{Travis Kriplean}, \bibinfo{person}{Jonathan Morgan}, \bibinfo{person}{Deen Freelon}, \bibinfo{person}{Alan Borning}, {and} \bibinfo{person}{Lance Bennett}.} \bibinfo{year}{2012}\natexlab{}.
\newblock \showarticletitle{Supporting reflective public thought with considerit}. In \bibinfo{booktitle}{\emph{Proceedings of the ACM 2012 Conference on Computer Supported Cooperative Work}} (Seattle, Washington, USA) \emph{(\bibinfo{series}{CSCW '12})}. \bibinfo{publisher}{Association for Computing Machinery}, \bibinfo{address}{New York, NY, USA}, \bibinfo{pages}{265–274}.
\newblock
\showISBNx{9781450310864}
\urldef\tempurl%
\url{https://doi.org/10.1145/2145204.2145249}
\showDOI{\tempurl}


\bibitem[Kuo et~al\mbox{.}(2024)]%
        {kuo2024wikibench}
\bibfield{author}{\bibinfo{person}{Tzu-Sheng Kuo}, \bibinfo{person}{Aaron~Lee Halfaker}, \bibinfo{person}{Zirui Cheng}, \bibinfo{person}{Jiwoo Kim}, \bibinfo{person}{Meng-Hsin Wu}, \bibinfo{person}{Tongshuang Wu}, \bibinfo{person}{Kenneth Holstein}, {and} \bibinfo{person}{Haiyi Zhu}.} \bibinfo{year}{2024}\natexlab{}.
\newblock \showarticletitle{Wikibench: Community-Driven Data Curation for AI Evaluation on Wikipedia}. In \bibinfo{booktitle}{\emph{Proceedings of the CHI Conference on Human Factors in Computing Systems}} (Honolulu, HI, USA) \emph{(\bibinfo{series}{CHI '24})}. \bibinfo{publisher}{Association for Computing Machinery}, \bibinfo{address}{New York, NY, USA}, Article \bibinfo{articleno}{193}, \bibinfo{numpages}{24}~pages.
\newblock
\showISBNx{9798400703300}
\urldef\tempurl%
\url{https://doi.org/10.1145/3613904.3642278}
\showDOI{\tempurl}


\bibitem[Kuo et~al\mbox{.}(2023)]%
        {kuo2023understanding}
\bibfield{author}{\bibinfo{person}{Tzu-Sheng Kuo}, \bibinfo{person}{Hong Shen}, \bibinfo{person}{Jisoo Geum}, \bibinfo{person}{Nev Jones}, \bibinfo{person}{Jason~I. Hong}, \bibinfo{person}{Haiyi Zhu}, {and} \bibinfo{person}{Kenneth Holstein}.} \bibinfo{year}{2023}\natexlab{}.
\newblock \showarticletitle{Understanding Frontline Workers’ and Unhoused Individuals’ Perspectives on AI Used in Homeless Services}. In \bibinfo{booktitle}{\emph{Proceedings of the 2023 CHI Conference on Human Factors in Computing Systems}} (Hamburg, Germany) \emph{(\bibinfo{series}{CHI '23})}. \bibinfo{publisher}{Association for Computing Machinery}, \bibinfo{address}{New York, NY, USA}, Article \bibinfo{articleno}{860}, \bibinfo{numpages}{17}~pages.
\newblock
\showISBNx{9781450394215}
\urldef\tempurl%
\url{https://doi.org/10.1145/3544548.3580882}
\showDOI{\tempurl}


\bibitem[Lazar et~al\mbox{.}(2012)]%
        {lazar2012hci}
\bibfield{author}{\bibinfo{person}{Jonathan Lazar}, \bibinfo{person}{Julio Abascal}, \bibinfo{person}{Janet Davis}, \bibinfo{person}{Vanessa Evers}, \bibinfo{person}{Jan Gulliksen}, \bibinfo{person}{Joaquim Jorge}, \bibinfo{person}{Tom McEwan}, \bibinfo{person}{Fabio Patern\`{o}}, \bibinfo{person}{Hans Persson}, \bibinfo{person}{Raquel Prates}, \bibinfo{person}{Hans von Axelson}, \bibinfo{person}{Marco Winckler}, {and} \bibinfo{person}{Volker Wulf}.} \bibinfo{year}{2012}\natexlab{}.
\newblock \showarticletitle{HCI public policy activities in 2012: a 10-country discussion}.
\newblock \bibinfo{journal}{\emph{Interactions}} \bibinfo{volume}{19}, \bibinfo{number}{3} (\bibinfo{date}{may} \bibinfo{year}{2012}), \bibinfo{pages}{78–81}.
\newblock
\showISSN{1072-5520}
\urldef\tempurl%
\url{https://doi.org/10.1145/2168931.2168947}
\showDOI{\tempurl}


\bibitem[Lehoux et~al\mbox{.}(2020)]%
        {lehoux2020anticipatory}
\bibfield{author}{\bibinfo{person}{Pascale Lehoux}, \bibinfo{person}{Fiona~Alice Miller}, {and} \bibinfo{person}{Bryn Williams-Jones}.} \bibinfo{year}{2020}\natexlab{}.
\newblock \showarticletitle{Anticipatory governance and moral imagination: Methodological insights from a scenario-based public deliberation study}.
\newblock \bibinfo{journal}{\emph{Technological Forecasting and Social Change}}  \bibinfo{volume}{151} (\bibinfo{year}{2020}), \bibinfo{pages}{119800}.
\newblock


\bibitem[Little et~al\mbox{.}(2010)]%
        {little2010exploring}
\bibfield{author}{\bibinfo{person}{Greg Little}, \bibinfo{person}{Lydia~B. Chilton}, \bibinfo{person}{Max Goldman}, {and} \bibinfo{person}{Robert~C. Miller}.} \bibinfo{year}{2010}\natexlab{}.
\newblock \showarticletitle{Exploring iterative and parallel human computation processes}. In \bibinfo{booktitle}{\emph{Proceedings of the ACM SIGKDD Workshop on Human Computation}} (Washington DC) \emph{(\bibinfo{series}{HCOMP '10})}. \bibinfo{publisher}{Association for Computing Machinery}, \bibinfo{address}{New York, NY, USA}, \bibinfo{pages}{68–76}.
\newblock
\showISBNx{9781450302227}
\urldef\tempurl%
\url{https://doi.org/10.1145/1837885.1837907}
\showDOI{\tempurl}


\bibitem[Liu et~al\mbox{.}(2023)]%
        {liu2023selenite}
\bibfield{author}{\bibinfo{person}{Michael~Xieyang Liu}, \bibinfo{person}{Tongshuang Wu}, \bibinfo{person}{Tianying Chen}, \bibinfo{person}{Franklin~Mingzhe Li}, \bibinfo{person}{Aniket Kittur}, {and} \bibinfo{person}{Brad~A Myers}.} \bibinfo{year}{2023}\natexlab{}.
\newblock \showarticletitle{Selenite: Scaffolding decision making with comprehensive overviews elicited from large language models}.
\newblock \bibinfo{journal}{\emph{arXiv preprint arXiv:2310.02161}} (\bibinfo{year}{2023}).
\newblock


\bibitem[Liu et~al\mbox{.}(2018)]%
        {liu2018consensus}
\bibfield{author}{\bibinfo{person}{Weichen Liu}, \bibinfo{person}{Sijia Xiao}, \bibinfo{person}{Jacob~T. Browne}, \bibinfo{person}{Ming Yang}, {and} \bibinfo{person}{Steven~P. Dow}.} \bibinfo{year}{2018}\natexlab{}.
\newblock \showarticletitle{ConsensUs: Supporting Multi-Criteria Group Decisions by Visualizing Points of Disagreement}.
\newblock \bibinfo{journal}{\emph{Trans. Soc. Comput.}} \bibinfo{volume}{1}, \bibinfo{number}{1}, Article \bibinfo{articleno}{4} (\bibinfo{date}{jan} \bibinfo{year}{2018}), \bibinfo{numpages}{26}~pages.
\newblock
\urldef\tempurl%
\url{https://doi.org/10.1145/3159649}
\showDOI{\tempurl}


\bibitem[Lowi(1972)]%
        {lowi1972four}
\bibfield{author}{\bibinfo{person}{Theodore~J. Lowi}.} \bibinfo{year}{1972}\natexlab{}.
\newblock \showarticletitle{Four Systems of Policy, Politics, and Choice}.
\newblock \bibinfo{journal}{\emph{Public Administration Review}} \bibinfo{volume}{32}, \bibinfo{number}{4} (\bibinfo{year}{1972}), \bibinfo{pages}{298--310}.
\newblock
\showISSN{00333352, 15406210}
\urldef\tempurl%
\url{http://www.jstor.org/stable/974990}
\showURL{%
\tempurl}


\bibitem[Mahyar et~al\mbox{.}(2018)]%
        {mahyar2018communitycrit}
\bibfield{author}{\bibinfo{person}{Narges Mahyar}, \bibinfo{person}{Michael~R. James}, \bibinfo{person}{Michelle~M. Ng}, \bibinfo{person}{Reginald~A. Wu}, {and} \bibinfo{person}{Steven~P. Dow}.} \bibinfo{year}{2018}\natexlab{}.
\newblock \showarticletitle{CommunityCrit: Inviting the Public to Improve and Evaluate Urban Design Ideas through Micro-Activities}. In \bibinfo{booktitle}{\emph{Proceedings of the 2018 CHI Conference on Human Factors in Computing Systems}} (Montreal QC, Canada) \emph{(\bibinfo{series}{CHI '18})}. \bibinfo{publisher}{Association for Computing Machinery}, \bibinfo{address}{New York, NY, USA}, \bibinfo{pages}{1–14}.
\newblock
\showISBNx{9781450356206}
\urldef\tempurl%
\url{https://doi.org/10.1145/3173574.3173769}
\showDOI{\tempurl}


\bibitem[Marnette and McKenzie({[n.\,d.]})]%
        {marnette2024talk}
\bibfield{author}{\bibinfo{person}{Bruno Marnette} {and} \bibinfo{person}{Colleen McKenzie}.} \bibinfo{year}{[n.\,d.]}\natexlab{}.
\newblock \bibinfo{booktitle}{\emph{Talk to the City: an open-source AI tool for scaling deliberation}}.
\newblock
\urldef\tempurl%
\url{https://ai.objectives.institute/blog/talk-to-the-city-an-open-source-ai-tool-to-scale-deliberation}
\showURL{%
Retrieved August 23, 2024 from \tempurl}


\bibitem[Mithical(2023)]%
        {mithical2023moderation}
\bibfield{author}{\bibinfo{person}{Mithical}.} \bibinfo{year}{2023}\natexlab{}.
\newblock \bibinfo{booktitle}{\emph{Moderation Strike: Stack Overflow, Inc. cannot consistently ignore, mistreat, and malign its volunteers}}.
\newblock
\urldef\tempurl%
\url{https://meta.stackexchange.com/questions/389811/moderation-strike-stack-overflow-inc-cannot-consistently-ignore-mistreat-an}
\showURL{%
Retrieved August 19, 2024 from \tempurl}


\bibitem[Molewijk et~al\mbox{.}(2008)]%
        {molewijk2008teaching}
\bibfield{author}{\bibinfo{person}{Albert~C Molewijk}, \bibinfo{person}{Tineke Abma}, \bibinfo{person}{Margreet Stolper}, {and} \bibinfo{person}{Guy Widdershoven}.} \bibinfo{year}{2008}\natexlab{}.
\newblock \showarticletitle{Teaching ethics in the clinic. The theory and practice of moral case deliberation}.
\newblock \bibinfo{journal}{\emph{Journal of Medical Ethics}} \bibinfo{volume}{34}, \bibinfo{number}{2} (\bibinfo{year}{2008}), \bibinfo{pages}{120--124}.
\newblock


\bibitem[Moreno-Lopez(2005)]%
        {moreno2005sharing}
\bibfield{author}{\bibinfo{person}{Isabel Moreno-Lopez}.} \bibinfo{year}{2005}\natexlab{}.
\newblock \showarticletitle{Sharing power with students: The critical language classroom}.
\newblock \bibinfo{journal}{\emph{Radical Pedagogy}} \bibinfo{volume}{7}, \bibinfo{number}{2} (\bibinfo{year}{2005}), \bibinfo{pages}{23--49}.
\newblock


\bibitem[Morris et~al\mbox{.}(2010)]%
        {morris2010wesearch}
\bibfield{author}{\bibinfo{person}{Meredith~Ringel Morris}, \bibinfo{person}{Jarrod Lombardo}, {and} \bibinfo{person}{Daniel Wigdor}.} \bibinfo{year}{2010}\natexlab{}.
\newblock \showarticletitle{WeSearch: supporting collaborative search and sensemaking on a tabletop display}. In \bibinfo{booktitle}{\emph{Proceedings of the 2010 ACM Conference on Computer Supported Cooperative Work}} (Savannah, Georgia, USA) \emph{(\bibinfo{series}{CSCW '10})}. \bibinfo{publisher}{Association for Computing Machinery}, \bibinfo{address}{New York, NY, USA}, \bibinfo{pages}{401–410}.
\newblock
\showISBNx{9781605587950}
\urldef\tempurl%
\url{https://doi.org/10.1145/1718918.1718987}
\showDOI{\tempurl}


\bibitem[Nanayakkara et~al\mbox{.}(2020)]%
        {nanayakkara2020anticipatory}
\bibfield{author}{\bibinfo{person}{Priyanka Nanayakkara}, \bibinfo{person}{Nicholas Diakopoulos}, {and} \bibinfo{person}{Jessica Hullman}.} \bibinfo{year}{2020}\natexlab{}.
\newblock \showarticletitle{Anticipatory ethics and the role of uncertainty}.
\newblock \bibinfo{journal}{\emph{arXiv preprint arXiv:2011.13170}} (\bibinfo{year}{2020}).
\newblock


\bibitem[Nonnecke and Preece(2000)]%
        {nonnecke2000lurker}
\bibfield{author}{\bibinfo{person}{Blair Nonnecke} {and} \bibinfo{person}{Jenny Preece}.} \bibinfo{year}{2000}\natexlab{}.
\newblock \showarticletitle{Lurker demographics: counting the silent}. In \bibinfo{booktitle}{\emph{Proceedings of the SIGCHI Conference on Human Factors in Computing Systems}} (The Hague, The Netherlands) \emph{(\bibinfo{series}{CHI '00})}. \bibinfo{publisher}{Association for Computing Machinery}, \bibinfo{address}{New York, NY, USA}, \bibinfo{pages}{73–80}.
\newblock
\showISBNx{1581132166}
\urldef\tempurl%
\url{https://doi.org/10.1145/332040.332409}
\showDOI{\tempurl}


\bibitem[Ostrom(1990)]%
        {ostrom1990governing}
\bibfield{author}{\bibinfo{person}{Elinor Ostrom}.} \bibinfo{year}{1990}\natexlab{}.
\newblock \bibinfo{booktitle}{\emph{Governing the commons: The evolution of institutions for collective action}}.
\newblock \bibinfo{publisher}{Cambridge University Press}.
\newblock


\bibitem[Park et~al\mbox{.}(2015)]%
        {park2015toward}
\bibfield{author}{\bibinfo{person}{Joonsuk Park}, \bibinfo{person}{Cheryl Blake}, {and} \bibinfo{person}{Claire Cardie}.} \bibinfo{year}{2015}\natexlab{}.
\newblock \showarticletitle{Toward machine-assisted participation in erulemaking: An argumentation model of evaluability}. In \bibinfo{booktitle}{\emph{Proceedings of the 15th International Conference on Artificial Intelligence and Law}}. \bibinfo{pages}{206--210}.
\newblock


\bibitem[Paul and Morris(2009)]%
        {paul2009cosense}
\bibfield{author}{\bibinfo{person}{Sharoda~A. Paul} {and} \bibinfo{person}{Meredith~Ringel Morris}.} \bibinfo{year}{2009}\natexlab{}.
\newblock \showarticletitle{CoSense: enhancing sensemaking for collaborative web search}. In \bibinfo{booktitle}{\emph{Proceedings of the SIGCHI Conference on Human Factors in Computing Systems}} (Boston, MA, USA) \emph{(\bibinfo{series}{CHI '09})}. \bibinfo{publisher}{Association for Computing Machinery}, \bibinfo{address}{New York, NY, USA}, \bibinfo{pages}{1771–1780}.
\newblock
\showISBNx{9781605582467}
\urldef\tempurl%
\url{https://doi.org/10.1145/1518701.1518974}
\showDOI{\tempurl}


\bibitem[Paul and Reddy(2010)]%
        {paul2010understanding}
\bibfield{author}{\bibinfo{person}{Sharoda~A. Paul} {and} \bibinfo{person}{Madhu~C. Reddy}.} \bibinfo{year}{2010}\natexlab{}.
\newblock \showarticletitle{Understanding together: sensemaking in collaborative information seeking}. In \bibinfo{booktitle}{\emph{Proceedings of the 2010 ACM Conference on Computer Supported Cooperative Work}} (Savannah, Georgia, USA) \emph{(\bibinfo{series}{CSCW '10})}. \bibinfo{publisher}{Association for Computing Machinery}, \bibinfo{address}{New York, NY, USA}, \bibinfo{pages}{321–330}.
\newblock
\showISBNx{9781605587950}
\urldef\tempurl%
\url{https://doi.org/10.1145/1718918.1718976}
\showDOI{\tempurl}


\bibitem[R.~Farina et~al\mbox{.}(2013)]%
        {r2013regulation}
\bibfield{author}{\bibinfo{person}{Cynthia R.~Farina}, \bibinfo{person}{Dmitry Epstein}, \bibinfo{person}{Josiah B.~Heidt}, {and} \bibinfo{person}{Mary J.~Newhart}.} \bibinfo{year}{2013}\natexlab{}.
\newblock \showarticletitle{Regulation Room: Getting “more, better” civic participation in complex government policymaking}.
\newblock \bibinfo{journal}{\emph{Transforming Government: People, Process and Policy}} \bibinfo{volume}{7}, \bibinfo{number}{4} (\bibinfo{year}{2013}), \bibinfo{pages}{501--516}.
\newblock


\bibitem[Reynante et~al\mbox{.}(2021)]%
        {reynante2021framework}
\bibfield{author}{\bibinfo{person}{Brandon Reynante}, \bibinfo{person}{Steven~P. Dow}, {and} \bibinfo{person}{Narges Mahyar}.} \bibinfo{year}{2021}\natexlab{}.
\newblock \showarticletitle{A Framework for Open Civic Design: Integrating Public Participation, Crowdsourcing, and Design Thinking}.
\newblock \bibinfo{journal}{\emph{Digit. Gov.: Res. Pract.}} \bibinfo{volume}{2}, \bibinfo{number}{4}, Article \bibinfo{articleno}{31} (\bibinfo{date}{dec} \bibinfo{year}{2021}), \bibinfo{numpages}{22}~pages.
\newblock
\urldef\tempurl%
\url{https://doi.org/10.1145/3487607}
\showDOI{\tempurl}


\bibitem[Robitzsch(2020)]%
        {robitzsch2020ordinal}
\bibfield{author}{\bibinfo{person}{Alexander Robitzsch}.} \bibinfo{year}{2020}\natexlab{}.
\newblock \showarticletitle{Why ordinal variables can (almost) always be treated as continuous variables: Clarifying assumptions of robust continuous and ordinal factor analysis estimation methods}. In \bibinfo{booktitle}{\emph{Frontiers in education}}, Vol.~\bibinfo{volume}{5}. Frontiers Media SA, \bibinfo{pages}{589965}.
\newblock


\bibitem[Salehi et~al\mbox{.}(2015)]%
        {salehi2015we}
\bibfield{author}{\bibinfo{person}{Niloufar Salehi}, \bibinfo{person}{Lilly~C. Irani}, \bibinfo{person}{Michael~S. Bernstein}, \bibinfo{person}{Ali Alkhatib}, \bibinfo{person}{Eva Ogbe}, \bibinfo{person}{Kristy Milland}, {and} \bibinfo{person}{Clickhappier}.} \bibinfo{year}{2015}\natexlab{}.
\newblock \showarticletitle{We Are Dynamo: Overcoming Stalling and Friction in Collective Action for Crowd Workers}. In \bibinfo{booktitle}{\emph{Proceedings of the 33rd Annual ACM Conference on Human Factors in Computing Systems}} (Seoul, Republic of Korea) \emph{(\bibinfo{series}{CHI '15})}. \bibinfo{publisher}{Association for Computing Machinery}, \bibinfo{address}{New York, NY, USA}, \bibinfo{pages}{1621–1630}.
\newblock
\showISBNx{9781450331456}
\urldef\tempurl%
\url{https://doi.org/10.1145/2702123.2702508}
\showDOI{\tempurl}


\bibitem[Shannon(1948)]%
        {shannon1948mathematical}
\bibfield{author}{\bibinfo{person}{C.~E. Shannon}.} \bibinfo{year}{1948}\natexlab{}.
\newblock \showarticletitle{A mathematical theory of communication}.
\newblock \bibinfo{journal}{\emph{The Bell System Technical Journal}} \bibinfo{volume}{27}, \bibinfo{number}{3} (\bibinfo{year}{1948}), \bibinfo{pages}{379--423}.
\newblock
\urldef\tempurl%
\url{https://doi.org/10.1002/j.1538-7305.1948.tb01338.x}
\showDOI{\tempurl}


\bibitem[Shaw et~al\mbox{.}(2014)]%
        {shaw2014computer}
\bibfield{author}{\bibinfo{person}{Aaron Shaw}, \bibinfo{person}{Haoqi Zhang}, \bibinfo{person}{Andr\'{e}s Monroy-Hern\'{a}ndez}, \bibinfo{person}{Sean Munson}, \bibinfo{person}{Benjamin~Mako Hill}, \bibinfo{person}{Elizabeth Gerber}, \bibinfo{person}{Peter Kinnaird}, {and} \bibinfo{person}{Patrick Minder}.} \bibinfo{year}{2014}\natexlab{}.
\newblock \showarticletitle{Computer supported collective action}.
\newblock \bibinfo{journal}{\emph{Interactions}} \bibinfo{volume}{21}, \bibinfo{number}{2} (\bibinfo{date}{mar} \bibinfo{year}{2014}), \bibinfo{pages}{74–77}.
\newblock
\showISSN{1072-5520}
\urldef\tempurl%
\url{https://doi.org/10.1145/2576875}
\showDOI{\tempurl}


\bibitem[Shin et~al\mbox{.}(2022)]%
        {shin2022chatbots}
\bibfield{author}{\bibinfo{person}{Joongi Shin}, \bibinfo{person}{Michael~A. Hedderich}, \bibinfo{person}{Andr\'{e}S Lucero}, {and} \bibinfo{person}{Antti Oulasvirta}.} \bibinfo{year}{2022}\natexlab{}.
\newblock \showarticletitle{Chatbots Facilitating Consensus-Building in Asynchronous Co-Design}. In \bibinfo{booktitle}{\emph{Proceedings of the 35th Annual ACM Symposium on User Interface Software and Technology}} (Bend, OR, USA) \emph{(\bibinfo{series}{UIST '22})}. \bibinfo{publisher}{Association for Computing Machinery}, \bibinfo{address}{New York, NY, USA}, Article \bibinfo{articleno}{78}, \bibinfo{numpages}{13}~pages.
\newblock
\showISBNx{9781450393201}
\urldef\tempurl%
\url{https://doi.org/10.1145/3526113.3545671}
\showDOI{\tempurl}


\bibitem[Shor(1996)]%
        {shor1996students}
\bibfield{author}{\bibinfo{person}{Ira Shor}.} \bibinfo{year}{1996}\natexlab{}.
\newblock \bibinfo{booktitle}{\emph{When students have power: Negotiating authority in a critical pedagogy}}.
\newblock \bibinfo{publisher}{University of Chicago Press}.
\newblock


\bibitem[Siddarth et~al\mbox{.}(2021)]%
        {siddarth2021ai}
\bibfield{author}{\bibinfo{person}{Divya Siddarth}, \bibinfo{person}{Daron Acemoglu}, \bibinfo{person}{Danielle Allen}, \bibinfo{person}{Kate Crawford}, \bibinfo{person}{James Evans}, \bibinfo{person}{Michael Jordan}, {and} \bibinfo{person}{E Weyl}.} \bibinfo{year}{2021}\natexlab{}.
\newblock \showarticletitle{How AI fails us}.
\newblock \bibinfo{journal}{\emph{arXiv preprint arXiv:2201.04200}} (\bibinfo{year}{2021}).
\newblock


\bibitem[Small et~al\mbox{.}(2021)]%
        {small2021polis}
\bibfield{author}{\bibinfo{person}{Christopher Small}, \bibinfo{person}{Michael Bjorkegren}, \bibinfo{person}{Timo Erkkil{\"a}}, \bibinfo{person}{Lynette Shaw}, {and} \bibinfo{person}{Colin Megill}.} \bibinfo{year}{2021}\natexlab{}.
\newblock \showarticletitle{Polis: Scaling deliberation by mapping high dimensional opinion spaces}.
\newblock \bibinfo{journal}{\emph{Recerca: revista de pensament i an{\`a}lisi}} \bibinfo{volume}{26}, \bibinfo{number}{2} (\bibinfo{year}{2021}).
\newblock


\bibitem[Smith et~al\mbox{.}(2020)]%
        {smith2020keeping}
\bibfield{author}{\bibinfo{person}{C.~Estelle Smith}, \bibinfo{person}{Bowen Yu}, \bibinfo{person}{Anjali Srivastava}, \bibinfo{person}{Aaron Halfaker}, \bibinfo{person}{Loren Terveen}, {and} \bibinfo{person}{Haiyi Zhu}.} \bibinfo{year}{2020}\natexlab{}.
\newblock \showarticletitle{Keeping Community in the Loop: Understanding Wikipedia Stakeholder Values for Machine Learning-Based Systems}. In \bibinfo{booktitle}{\emph{Proceedings of the 2020 CHI Conference on Human Factors in Computing Systems}} (Honolulu, HI, USA) \emph{(\bibinfo{series}{CHI '20})}. \bibinfo{publisher}{Association for Computing Machinery}, \bibinfo{address}{New York, NY, USA}, \bibinfo{pages}{1–14}.
\newblock
\showISBNx{9781450367080}
\urldef\tempurl%
\url{https://doi.org/10.1145/3313831.3376783}
\showDOI{\tempurl}


\bibitem[Spaa et~al\mbox{.}(2019)]%
        {spaa2019understanding}
\bibfield{author}{\bibinfo{person}{Anne Spaa}, \bibinfo{person}{Abigail Durrant}, \bibinfo{person}{Chris Elsden}, {and} \bibinfo{person}{John Vines}.} \bibinfo{year}{2019}\natexlab{}.
\newblock \showarticletitle{Understanding the Boundaries between Policymaking and HCI}. In \bibinfo{booktitle}{\emph{Proceedings of the 2019 CHI Conference on Human Factors in Computing Systems}} (Glasgow, Scotland Uk) \emph{(\bibinfo{series}{CHI '19})}. \bibinfo{publisher}{Association for Computing Machinery}, \bibinfo{address}{New York, NY, USA}, \bibinfo{pages}{1–15}.
\newblock
\showISBNx{9781450359702}
\urldef\tempurl%
\url{https://doi.org/10.1145/3290605.3300314}
\showDOI{\tempurl}


\bibitem[Stuart et~al\mbox{.}(2012)]%
        {stuart2012social}
\bibfield{author}{\bibinfo{person}{H.~Colleen Stuart}, \bibinfo{person}{Laura Dabbish}, \bibinfo{person}{Sara Kiesler}, \bibinfo{person}{Peter Kinnaird}, {and} \bibinfo{person}{Ruogu Kang}.} \bibinfo{year}{2012}\natexlab{}.
\newblock \showarticletitle{Social transparency in networked information exchange: a theoretical framework}. In \bibinfo{booktitle}{\emph{Proceedings of the ACM 2012 Conference on Computer Supported Cooperative Work}} (Seattle, Washington, USA) \emph{(\bibinfo{series}{CSCW '12})}. \bibinfo{publisher}{Association for Computing Machinery}, \bibinfo{address}{New York, NY, USA}, \bibinfo{pages}{451–460}.
\newblock
\showISBNx{9781450310864}
\urldef\tempurl%
\url{https://doi.org/10.1145/2145204.2145275}
\showDOI{\tempurl}


\bibitem[Tan and Subramonyam(2024)]%
        {tan2024more}
\bibfield{author}{\bibinfo{person}{Mei Tan} {and} \bibinfo{person}{Hari Subramonyam}.} \bibinfo{year}{2024}\natexlab{}.
\newblock \showarticletitle{More than Model Documentation: Uncovering Teachers' Bespoke Information Needs for Informed Classroom Integration of ChatGPT}. In \bibinfo{booktitle}{\emph{Proceedings of the CHI Conference on Human Factors in Computing Systems}} (Honolulu, HI, USA) \emph{(\bibinfo{series}{CHI '24})}. \bibinfo{publisher}{Association for Computing Machinery}, \bibinfo{address}{New York, NY, USA}, Article \bibinfo{articleno}{269}, \bibinfo{numpages}{19}~pages.
\newblock
\showISBNx{9798400703300}
\urldef\tempurl%
\url{https://doi.org/10.1145/3613904.3642592}
\showDOI{\tempurl}


\bibitem[Tandon et~al\mbox{.}(2022)]%
        {tandon2022hostile}
\bibfield{author}{\bibinfo{person}{Udayan Tandon}, \bibinfo{person}{Vera Khovanskaya}, \bibinfo{person}{Enrique Arcilla}, \bibinfo{person}{Mikaiil~Haji Hussein}, \bibinfo{person}{Peter Zschiesche}, {and} \bibinfo{person}{Lilly Irani}.} \bibinfo{year}{2022}\natexlab{}.
\newblock \showarticletitle{Hostile Ecologies: Navigating the Barriers to Community-Led Innovation}.
\newblock \bibinfo{journal}{\emph{Proc. ACM Hum.-Comput. Interact.}} \bibinfo{volume}{6}, \bibinfo{number}{CSCW2}, Article \bibinfo{articleno}{443} (\bibinfo{date}{nov} \bibinfo{year}{2022}), \bibinfo{numpages}{26}~pages.
\newblock
\urldef\tempurl%
\url{https://doi.org/10.1145/3555544}
\showDOI{\tempurl}


\bibitem[Tang et~al\mbox{.}(2024)]%
        {tang2024ai}
\bibfield{author}{\bibinfo{person}{Ningjing Tang}, \bibinfo{person}{Jiayin Zhi}, \bibinfo{person}{Tzu-Sheng Kuo}, \bibinfo{person}{Calla Kainaroi}, \bibinfo{person}{Jeremy~J. Northup}, \bibinfo{person}{Kenneth Holstein}, \bibinfo{person}{Haiyi Zhu}, \bibinfo{person}{Hoda Heidari}, {and} \bibinfo{person}{Hong Shen}.} \bibinfo{year}{2024}\natexlab{}.
\newblock \showarticletitle{AI Failure Cards: Understanding and Supporting Grassroots Efforts to Mitigate AI Failures in Homeless Services}. In \bibinfo{booktitle}{\emph{Proceedings of the 2024 ACM Conference on Fairness, Accountability, and Transparency}} (Rio de Janeiro, Brazil) \emph{(\bibinfo{series}{FAccT '24})}. \bibinfo{publisher}{Association for Computing Machinery}, \bibinfo{address}{New York, NY, USA}, \bibinfo{pages}{713–732}.
\newblock
\showISBNx{9798400704505}
\urldef\tempurl%
\url{https://doi.org/10.1145/3630106.3658935}
\showDOI{\tempurl}


\bibitem[Tapia et~al\mbox{.}(2022)]%
        {tapia2022entropy}
\bibfield{author}{\bibinfo{person}{JM Tapia}, \bibinfo{person}{Francisco Chiclana}, \bibinfo{person}{Maria~Jos{\'e} del Moral}, {and} \bibinfo{person}{Enrique Herrera-Viedma}.} \bibinfo{year}{2022}\natexlab{}.
\newblock \showarticletitle{Entropy Based Approach to Measuring Consensus in Group Decision-Making Problems}. In \bibinfo{booktitle}{\emph{International Conference on Industrial, Engineering and Other Applications of Applied Intelligent Systems}}. Springer, \bibinfo{pages}{409--415}.
\newblock


\bibitem[Thomson(1984)]%
        {thomson1984trolley}
\bibfield{author}{\bibinfo{person}{Judith~Jarvis Thomson}.} \bibinfo{year}{1984}\natexlab{}.
\newblock \showarticletitle{The Trolley Problem}.
\newblock \bibinfo{journal}{\emph{Yale Law Journal}}  \bibinfo{volume}{94} (\bibinfo{year}{1984}), \bibinfo{pages}{1395}.
\newblock


\bibitem[Vi\'{e}gas et~al\mbox{.}(2007)]%
        {viegas2007hidden}
\bibfield{author}{\bibinfo{person}{Fernanda~B. Vi\'{e}gas}, \bibinfo{person}{Martin Wattenberg}, {and} \bibinfo{person}{Matthew~M. McKeon}.} \bibinfo{year}{2007}\natexlab{}.
\newblock \showarticletitle{The hidden order of wikipedia}. In \bibinfo{booktitle}{\emph{Proceedings of the 2nd International Conference on Online Communities and Social Computing}} (Beijing, China) \emph{(\bibinfo{series}{OCSC'07})}. \bibinfo{publisher}{Springer-Verlag}, \bibinfo{address}{Berlin, Heidelberg}, \bibinfo{pages}{445–454}.
\newblock
\showISBNx{9783540732563}


\bibitem[Wang et~al\mbox{.}(2024)]%
        {wang2024pika}
\bibfield{author}{\bibinfo{person}{Leijie Wang}, \bibinfo{person}{Nicholas Vincent}, \bibinfo{person}{Julija Rukanskaitundefined}, {and} \bibinfo{person}{Amy~Xian Zhang}.} \bibinfo{year}{2024}\natexlab{}.
\newblock \showarticletitle{Pika: Empowering Non-Programmers to Author Executable Governance Policies in Online Communities}. In \bibinfo{booktitle}{\emph{Proceedings of the CHI Conference on Human Factors in Computing Systems}} (Honolulu, HI, USA) \emph{(\bibinfo{series}{CHI '24})}. \bibinfo{publisher}{Association for Computing Machinery}, \bibinfo{address}{New York, NY, USA}, Article \bibinfo{articleno}{925}, \bibinfo{numpages}{18}~pages.
\newblock
\showISBNx{9798400703300}
\urldef\tempurl%
\url{https://doi.org/10.1145/3613904.3642012}
\showDOI{\tempurl}


\bibitem[Weld et~al\mbox{.}(2024)]%
        {weld2024making}
\bibfield{author}{\bibinfo{person}{Galen Weld}, \bibinfo{person}{Amy~X. Zhang}, {and} \bibinfo{person}{Tim Althoff}.} \bibinfo{year}{2024}\natexlab{}.
\newblock \showarticletitle{Making Online Communities ‘Better’: A Taxonomy of Community Values on Reddit}.
\newblock \bibinfo{journal}{\emph{Proceedings of the International AAAI Conference on Web and Social Media}} \bibinfo{volume}{18}, \bibinfo{number}{1} (\bibinfo{date}{May} \bibinfo{year}{2024}), \bibinfo{pages}{1611--1633}.
\newblock
\urldef\tempurl%
\url{https://doi.org/10.1609/icwsm.v18i1.31413}
\showDOI{\tempurl}


\bibitem[Whitney et~al\mbox{.}(2021)]%
        {whitney2021hci}
\bibfield{author}{\bibinfo{person}{Cedric~Deslandes Whitney}, \bibinfo{person}{Teresa Naval}, \bibinfo{person}{Elizabeth Quepons}, \bibinfo{person}{Simrandeep Singh}, \bibinfo{person}{Steven~R Rick}, {and} \bibinfo{person}{Lilly Irani}.} \bibinfo{year}{2021}\natexlab{}.
\newblock \showarticletitle{HCI Tactics for Politics from Below: Meeting the Challenges of Smart Cities}. In \bibinfo{booktitle}{\emph{Proceedings of the 2021 CHI Conference on Human Factors in Computing Systems}} (Yokohama, Japan) \emph{(\bibinfo{series}{CHI '21})}. \bibinfo{publisher}{Association for Computing Machinery}, \bibinfo{address}{New York, NY, USA}, Article \bibinfo{articleno}{297}, \bibinfo{numpages}{15}~pages.
\newblock
\showISBNx{9781450380966}
\urldef\tempurl%
\url{https://doi.org/10.1145/3411764.3445314}
\showDOI{\tempurl}


\bibitem[Wright et~al\mbox{.}(2020)]%
        {wright2020policy}
\bibfield{author}{\bibinfo{person}{David Wright}, \bibinfo{person}{Bernd Stahl}, {and} \bibinfo{person}{Tally Hatzakis}.} \bibinfo{year}{2020}\natexlab{}.
\newblock \showarticletitle{Policy scenarios as an instrument for policymakers}.
\newblock \bibinfo{journal}{\emph{Technological Forecasting and Social Change}}  \bibinfo{volume}{154} (\bibinfo{year}{2020}), \bibinfo{pages}{119972}.
\newblock


\bibitem[Wu et~al\mbox{.}(2022)]%
        {wu2022reasonable}
\bibfield{author}{\bibinfo{person}{Yuxi Wu}, \bibinfo{person}{W.~Keith Edwards}, {and} \bibinfo{person}{Sauvik Das}.} \bibinfo{year}{2022}\natexlab{}.
\newblock \showarticletitle{“A Reasonable Thing to Ask For”: Towards a Unified Voice in Privacy Collective Action}. In \bibinfo{booktitle}{\emph{Proceedings of the 2022 CHI Conference on Human Factors in Computing Systems}} (New Orleans, LA, USA) \emph{(\bibinfo{series}{CHI '22})}. \bibinfo{publisher}{Association for Computing Machinery}, \bibinfo{address}{New York, NY, USA}, Article \bibinfo{articleno}{32}, \bibinfo{numpages}{17}~pages.
\newblock
\showISBNx{9781450391573}
\urldef\tempurl%
\url{https://doi.org/10.1145/3491102.3517467}
\showDOI{\tempurl}


\bibitem[Yang et~al\mbox{.}(2024)]%
        {yang2024future}
\bibfield{author}{\bibinfo{person}{Qian Yang}, \bibinfo{person}{Richmond~Y. Wong}, \bibinfo{person}{Steven Jackson}, \bibinfo{person}{Sabine Junginger}, \bibinfo{person}{Margaret~D. Hagan}, \bibinfo{person}{Thomas Gilbert}, {and} \bibinfo{person}{John Zimmerman}.} \bibinfo{year}{2024}\natexlab{}.
\newblock \showarticletitle{The Future of HCI-Policy Collaboration}. In \bibinfo{booktitle}{\emph{Proceedings of the CHI Conference on Human Factors in Computing Systems}} (Honolulu, HI, USA) \emph{(\bibinfo{series}{CHI '24})}. \bibinfo{publisher}{Association for Computing Machinery}, \bibinfo{address}{New York, NY, USA}, Article \bibinfo{articleno}{820}, \bibinfo{numpages}{15}~pages.
\newblock
\showISBNx{9798400703300}
\urldef\tempurl%
\url{https://doi.org/10.1145/3613904.3642771}
\showDOI{\tempurl}


\bibitem[Zhang et~al\mbox{.}(2024)]%
        {zhang2024data}
\bibfield{author}{\bibinfo{person}{Angie Zhang}, \bibinfo{person}{Rocita Rana}, \bibinfo{person}{Alexander Boltz}, \bibinfo{person}{Veena Dubal}, {and} \bibinfo{person}{Min~Kyung Lee}.} \bibinfo{year}{2024}\natexlab{}.
\newblock \showarticletitle{Data Probes as Boundary Objects for Technology Policy Design: Demystifying Technology for Policymakers and Aligning Stakeholder Objectives in Rideshare Gig Work}. In \bibinfo{booktitle}{\emph{Proceedings of the CHI Conference on Human Factors in Computing Systems}} (Honolulu, HI, USA) \emph{(\bibinfo{series}{CHI '24})}. \bibinfo{publisher}{Association for Computing Machinery}, \bibinfo{address}{New York, NY, USA}, Article \bibinfo{articleno}{388}, \bibinfo{numpages}{21}~pages.
\newblock
\showISBNx{9798400703300}
\urldef\tempurl%
\url{https://doi.org/10.1145/3613904.3642000}
\showDOI{\tempurl}


\bibitem[Zhang and Cranshaw(2018)]%
        {zhang2018making}
\bibfield{author}{\bibinfo{person}{Amy~X. Zhang} {and} \bibinfo{person}{Justin Cranshaw}.} \bibinfo{year}{2018}\natexlab{}.
\newblock \showarticletitle{Making Sense of Group Chat through Collaborative Tagging and Summarization}.
\newblock \bibinfo{journal}{\emph{Proc. ACM Hum.-Comput. Interact.}} \bibinfo{volume}{2}, \bibinfo{number}{CSCW}, Article \bibinfo{articleno}{196} (\bibinfo{date}{nov} \bibinfo{year}{2018}), \bibinfo{numpages}{27}~pages.
\newblock
\urldef\tempurl%
\url{https://doi.org/10.1145/3274465}
\showDOI{\tempurl}


\bibitem[Zhang et~al\mbox{.}(2020)]%
        {zhang2020policykit}
\bibfield{author}{\bibinfo{person}{Amy~X. Zhang}, \bibinfo{person}{Grant Hugh}, {and} \bibinfo{person}{Michael~S. Bernstein}.} \bibinfo{year}{2020}\natexlab{}.
\newblock \showarticletitle{PolicyKit: Building Governance in Online Communities}. In \bibinfo{booktitle}{\emph{Proceedings of the 33rd Annual ACM Symposium on User Interface Software and Technology}} (Virtual Event, USA) \emph{(\bibinfo{series}{UIST '20})}. \bibinfo{publisher}{Association for Computing Machinery}, \bibinfo{address}{New York, NY, USA}, \bibinfo{pages}{365–378}.
\newblock
\showISBNx{9781450375146}
\urldef\tempurl%
\url{https://doi.org/10.1145/3379337.3415858}
\showDOI{\tempurl}


\bibitem[Zittrain(2019)]%
        {zittrain2019three}
\bibfield{author}{\bibinfo{person}{Jonathan~L Zittrain}.} \bibinfo{year}{2019}\natexlab{}.
\newblock \showarticletitle{Three eras of digital governance}.
\newblock \bibinfo{journal}{\emph{Available at SSRN 3458435}} (\bibinfo{year}{2019}).
\newblock


\bibitem[Zuckerman and Rajendra-Nicolucci(2023)]%
        {zuckerman2023community}
\bibfield{author}{\bibinfo{person}{Ethan Zuckerman} {and} \bibinfo{person}{Chand Rajendra-Nicolucci}.} \bibinfo{year}{2023}\natexlab{}.
\newblock \showarticletitle{From Community Governance to Customer Service and Back Again: Re-Examining Pre-Web Models of Online Governance to Address Platforms’ Crisis of Legitimacy}.
\newblock \bibinfo{journal}{\emph{Social Media + Society}} \bibinfo{volume}{9}, \bibinfo{number}{3} (\bibinfo{year}{2023}), \bibinfo{pages}{20563051231196864}.
\newblock
\urldef\tempurl%
\url{https://doi.org/10.1177/20563051231196864}
\showDOI{\tempurl}
\showeprint{https://doi.org/10.1177/20563051231196864}


\end{thebibliography}

\appendix

\section{Interview Topics}\label{sec:informal_conversation}
As mentioned in Section \ref{sec:designgoal}, we conducted semi-structured interviews with community organizers from different contexts to understand whether and how the design goals we had derived from prior research literature aligned with real-world needs across a range of community contexts. In each interview, we spoke with community organizers, grounding our conversation in actual past experiences where the development of community policies was needed. Below are the discussion topics used to guide our semi-structure interviews:
\begin{itemize}
    \item \textbf{Context}: What is the community context? 
    \item \textbf{Roles}: What are the interviewees' roles and responsibilities as community organizers within their community?
    \item \textbf{Needs}: What is (or was) the reason a policy needed to be proposed or implemented in their community?
    \item \textbf{Processes}: What does the current process for policy development look like in their community? What do they envision as the \textit{ideal} process for the development of community policies?
    \item \textbf{Challenges}: What challenges do their communities currently encounter, (or what challenges do they foresee) in the policy development process?
\end{itemize}

\section{AI Assistant Details}\label{sec:AI_assistants}
As mentioned in Section \ref{sec:system_ai}, PolicyCraft has three built-in LLM-based AI assistants that users can optionally use to create, critique, or revise policies based on cases. This section provides the prompts and more details for these AI assistants. Note that the current implementation of the AI assistants, including the prompts and conversation flows, is not intended to be optimal but serves as a demonstration of how LLMs can be integrated into PolicyCraft's case-grounded deliberation approach to policy design. The current work aims to open up a research space for future studies to further develop LLM-based AI assistants that support policy design through the systematic use of cases.

\subsection{AI Assistant for Policy Creation}
The AI assistant on the creation page can help users brainstorm new policies based on selected cases. Its conversation flow is shown in the left column of Figure \ref{fig:ai_workflow}. Once a user chooses to create a policy, they can select one or more cases from the case repository and specify whether each case should be allowed or disallowed by the policy they are about to create. After the user finishes selecting cases, PolicyCraft will feed the following prompt to an LLM:
\begin{quote}
{\footnotesize\fontfamily{pcr}\selectfont You are a helpful assistant focusing on supporting users in creating a new policy. In a few sentences, propose a policy that meets the following criteria. 1. The policy should <allow|disallow> the following scenario: [selected case 1]. 2. The policy should <allow|disallow> the following scenario: [selected case 2]. [...] }
\end{quote}
The LLM-generated policy will replace the red text in the left column of Figure \ref{fig:ai_workflow} and be presented to the user, who can then provide additional instructions to refine the policy further. PolicyCraft will feed the current policy and user instructions to an LLM using the following prompt to generate a refined policy.
\begin{quote}
{\footnotesize\fontfamily{pcr}\selectfont You are a helpful assistant focusing on supporting users in editing the following policy: [current policy]. In a few sentences, slightly revise the policy without significant changes based on the following instructions: [user instruction].}
\end{quote}
When the user is satisfied with the policy, whether refined manually or with the help of the AI assistant, they can proceed to create a new policy.

\subsection{AI Assistant for Policy Critique}
The AI assistant on the page for editing cases related to a given policy (see Figure \ref{fig:edit_case_page}) can help users brainstorm cases that illustrate or reveal flaws in the policy. Its conversation flow is shown in the middle column of Figure \ref{fig:ai_workflow}.

When the user chooses to create a case that illustrates the policy, PolicyCraft feeds the following prompt to an LLM:
\begin{quote}
{\footnotesize\fontfamily{pcr}\selectfont You are a helpful assistant focusing on supporting users' reflections on \underline{context} policies. Here is an overview of the \underline{context}. [context details]. In a few sentences, provide an example scenario of a \underline{character} in this \underline{context} where the \underline{character} \underline{does something} and <abides by|violates> the following policy: [policy description].}
\end{quote}
The underlined content and context details are configured by community organizers during the initialization of PolicyCraft, based on their specific context. For example, in our study, the \underline{context} is ``course'', the \underline{character} is ``student'', and \underline{does something} is ``uses AI''. The system randomly generates cases that either abide by or violate the policy to ensure balanced coverage.

Similarly, when the user chooses to create a case that reveals flaws in the policy, PolicyCraft feeds the following prompt to an LLM:
\begin{quote}
{\footnotesize\fontfamily{pcr}\selectfont You are a helpful assistant focusing on supporting users' reflections on \underline{context} policies. Here is an overview of the \underline{context}. [context details]. In a few sentences, provide an example scenario of a \underline{character} in this \underline{context} where < the \underline{character} technically abides by the following policy but undermines the policy's intent | the \underline{character} technically violates the following policy despite genuinely trying to comply | it is unclear whether the \underline{character} violates the following policy or not >: [policy description].} 
\end{quote}
The system will randomly generate cases using one of the three prompt templates within the angle brackets to encourage exploration of potential flaws.

The LLM-generated cases will replace the red text in the middle column of Figure \ref{fig:ai_workflow} and be presented to the user, who can then provide further instructions to refine the case. PolicyCraft will feed the current case and user instructions to an LLM using the following prompt to generate a refined case:
\begin{quote}
{\footnotesize\fontfamily{pcr}\selectfont You are a helpful assistant focusing on supporting users in editing the following case: [current case]. In a few sentences, slightly revise the case without significant changes based on the following instructions: [user instruction].}
\end{quote}
When the user is satisfied with the case, whether refined manually or with the help of the AI assistant, they can proceed to create a new case.

\subsection{AI Assistant for Policy Revision}
The AI assistant on the page for editing a policy (see Figure \ref{fig:edit_policy_page}) can help users brainstorm ways to revise the policy based on a selected case. Its conversation flow is shown in the right column of Figure \ref{fig:ai_workflow}. The AI assistant will ask users to check whether the selected case is unrelated to the policy or inherently ambiguous. In such circumstances, the AI assistant will present the user with appropriate action options, rather than using the case to drive policy revision. When the policy allows or disallows the case, but most people say the opposite, the AI assistant will ask the user to provide a reason by selecting from the reasons submitted by others or manually entering their own. Based on the selected case and the provided reason, PolicyCraft will feed the following prompt to an LLM to generate a suggested policy revision:
\begin{quote}
{\footnotesize\fontfamily{pcr}\selectfont You are a helpful assistant focusing on supporting users' revision of the following policy: [policy description]. In a few sentences, slightly revise the policy without significant changes so that the policy <allows|disallows> the following scenario: [selected case]. Here is the reason why the policy should <allow|disallow> the scenario: [provided reason].}
\end{quote}
The LLM-generated policy will replace the red text in the right column of Figure \ref{fig:ai_workflow} and be presented to the user, who can then provide additional instructions to refine the policy further. PolicyCraft will feed the current policy suggestion and user instructions to an LLM using the following prompt to generate a refined policy:
\begin{quote}
{\footnotesize\fontfamily{pcr}\selectfont You are a helpful assistant focusing on supporting users in editing the following policy: [policy description]. In a few sentences, slightly revise the policy without significant changes based on the following instructions: [user instruction].}
\end{quote}
When the user is satisfied with the suggested policy, whether refined manually or with the help of the AI assistant, they can proceed to revise the policy.

\section{Initial Policies and Cases}\label{sec:initial_policy_and_case}
At the beginning of the field study, the instructors provided the following three initial policies to kick-start the discussion. These policies were carefully chosen to represent different types of class activities and varying levels of restriction on the use of generative AI, ranging from a complete prohibition to unrestricted use. In the full version of PolicyCraft, each policy included two initial, illustrative cases provided and labeled by the instructors. These seed policies and cases serve to minimize the cold-start problem and establish norms around aspects such as formatting and level of abstraction.
\begin{itemize}
    \item \textbf{Prohibition of AI for Reading Responses}: Absolutely no use of AI is allowed for writing reading responses.
        \begin{itemize}
            \item[$\circ$] \textit{Lukas uses AI to summarize key points from papers}: Lukas uses an AI chatbot to summarize the key points of a dense research paper from the week’s reading assignment. He then uses these AI-generated points to form the bulk of his reading response, passing off the AI's analysis as his own original thoughts and reflections. (Label: \textbf{disallowed} by the policy)
            \item[$\circ$] \textit{Ding asks AI to explain complex topics}: Ding is struggling to understand a complex topic presented in the week's readings. She turns to an AI-powered study tool, like a chatbot tutor, to explain the topic in simpler terms. She then uses the chatbot's explanation to help her formulate her reading response. Ding does not directly copy the chatbot's words but only uses its insights as a guide. (Label: \textbf{disallowed} by the policy)
        \end{itemize}
    \vfill
    \item \textbf{AI Usage Permitted for Coding Assignments}: Students may freely use AI for coding assignments with appropriate attribution.
        \begin{itemize}
            \item[$\circ$] \textit{Mark submits AI's code as his own}: Mark put his project requirements into an AI code generator. The AI spit back a flawlessly functioning, well-documented program. Mark put his name on the AI's work and hit submit for the assignment. (Label: \textbf{disallowed} by the policy)
            \item[$\circ$] \textit{Priya copies AI’s code without understanding it}: Priya enters the requirements of her coding assignment into an AI coding assistant. The AI generates a perfect code block, which Priya then directly copies and pastes into her project. She includes the comment: "Used an AI assistant for help with this part." While Priya attempted to acknowledge the AI's assistance, when the instructor asked later, she was unable to explain what purpose that code was meant to serve, or her rationale for including it. (Label: \textbf{allowed} by the policy)
        \end{itemize}
    \vfill
    \item \textbf{Guidelines for Using AI in Course Project Brainstorming}: Students may use AI to assist with brainstorming course project ideas, but the ideas have to ultimately come from students themselves.
        \begin{itemize}
            \item[$\circ$] \textit{Omar picks AI-generated ideas for course projects}: Omar inputs various prompts into an AI chatbot to help brainstorm project ideas. The chatbot generates several ideas, and Omar, without putting much thought into it, picks one he likes. Omar proceeds to develop this AI-generated idea as his own and present it as original thought during the project proposal. (Label: \textbf{disallowed} by the policy)
            \item[$\circ$]\textit{Emily uses AI to revamp her discarded ideas}: Emily had tried brainstorming course project ideas with her group, but it had yielded no good ideas. So Emily fed an AI chatbot her group’s "discarded ideas" and the assignment description. Within seconds, the chatbot generated a great idea, which the group decided to use. In this case, the “discarded” ideas that were fed to the AI chatbot technically did come from students themselves, but the final idea was AI-generated. (Label: \textbf{ambiguous} under the policy)
        \end{itemize}
\end{itemize}

\section{Post-Study Survey}\label{sec:poststudysurvey}
At the end of the field study, students completed a post-study survey where they rated their agreement or disagreement with the following five statements on a scale of 1 to 7, and briefly explained their reasoning.
\begin{itemize}
    \item Overall, I can easily \textbf{collaborate with others} in policy development using the system.
    \item I can easily \textbf{identify potential flaws} in a policy using the system.
    \item I can easily \textbf{revise or create policies} to \textbf{address potential flaws} using the system.
    \item I can easily \textbf{understand why people agree or disagree} with each other when using the system.
    \item I feel \textbf{confident and comfortable to contribute} to policy development using the system.
\end{itemize}

\section{Resulting Policies}\label{sec:finalpolicies}
Here are the full sets of policies that received majority support from each group within each class:

\begin{itemize}
    \item \small{\textbf{Class 1 -- PolicyCraft condition} (14 policies, excluding 5 without majority vote):}
        \begin{itemize}
        \footnotesize{
            \item[$\circ$] \textbf{AI Use for Conceptual Understanding}: Student is allowed to use AI for understanding different concepts, or asking for resources to understand a concept in the relevant domain.
            \item[$\circ$] \textbf{Use of Sensitive Information for AI}: Students should not enter sensitive information in their prompts/entries of GenAI models. Sensitive information includes personal identifiable information (ex. school ID, email addresses), copyrighted information from the course, and other confidential information (ex. OpenAI credentials, Azure credentials).
            \item[$\circ$] \textbf{AI for Course Understanding}: Students are permitted to utilize AI to enhance their understanding of course material, such as clarifying complex topics or visualizing key concepts. However, all submitted work must reflect the student's own analysis and understanding. While AI tools can provide guidance and support, direct copying or paraphrasing of AI-generated content is strictly prohibited (this includes drawing your comments on readings from any summary/analysis content that the AI provides.
            \item[$\circ$] \textbf{AI Guidelines for Original Work}: AI-powered tools should be used to support learning, brainstorming, and other academic activities, but they should not replace original work.
            \item[$\circ$] \textbf{AI in Collaborative Group Work}: Students may use generative AI tools to assist with tasks in collaborative group projects, such as idea generation, task delegation, and content creation. However, all group members must be actively involved in the process, and the AI’s contributions should be transparently discussed among the team. The final project must reflect the collective effort and understanding of the group, with AI being a supportive tool rather than the primary contributor.
            \item[$\circ$] \textbf{AI for Presentation Preparation}: AI should be allowed to be used for Final Project Presentation as well as other presentations needed to be prepared throughout the course. It may be employed for augmentation such as grammar/spell checking, brainstorming, and template suggestions. However, students cannot directly use AI-generated text, images, or any other content in their presentations. (They can if cited)
            \item[$\circ$] \textbf{AI Citation}: If AI contributes at all towards materials created by the student (ex: reading notes, projects, code, presentations), then the student must acknowledge that they used AI in that assignment/submission.
            \item[$\circ$] \textbf{AI Usage for Grammar Checks and Better Writing}: AI can be used to refine writing to remove grammatical errors, spelling errors etc without changing the actual content of the text.
            \item[$\circ$] \textbf{AI Usage Permitted for Coding Assignments}: Students may use AI to aid in coding assignments, but must use AI to augment their work, not create the solution for them. Students cannot use AI to create large chunks of code without verifying it themselves. AI generation of very broad high-level pseudocode is permitted, but not step-by-step pseudocode or detailed lines of code. AI can be used to add comments/documentation to already written code but students should review over them. AI usage must be appropriately attributed.
            \item[$\circ$] \textbf{Prohibition of AI for Reading Responses}: It is not permitted to use AI to summarize or generate answers for reading responses. However, the usage of AI-powered tools is permitted to edit work (syntax, spelling/grammar checks, translate), or provide explanations for readings for the purpose of understanding the text after reading on one's own. Any usage of AI tools should be cited.
            \item[$\circ$] \textbf{Prohibiting AI-Generated Grading and Feedback}: To ensure the quality and authenticity of student assessments, AI tools may not be used to generate generic grading or feedback unless the assessments are graded based on completion. This policy aims to maintain the integrity of the academic process and provide students with personalized and meaningful evaluations.
            \item[$\circ$] \textbf{AI Use in Debugging}: Students can use AI to help with debugging as long as they have considered the code themselves already and understand the small revisions made.
            \item[$\circ$] \textbf{A 3-Strike System}: Under this policy students who violate established AI usage guidelines will receive progressive consequences, starting with a 50\% reduction in assignment score and escalating to more severe penalties like 0 grades, disciplinary referrals, and loss of privileges. (We may need to adjust the specific consequences and procedures based on individual circumstances and perspectives).
            \item[$\circ$] \textbf{Guidelines for Using AI in Course Project Brainstorming}: Students may use AI to assist with brainstorming course project ideas, but must be involved in major parts of the brainstorming process by either improving AI generated ideas, doing extensive research into AI generated ideas, or using AI to improve human-generated ideas.
        }
        \end{itemize}
        \item \small{\textbf{Class 1 -- baseline condition} (7 policies, excluding 24 without majority vote):}
        \begin{itemize}
        \footnotesize{
            \item[$\circ$] \textbf{AI for Resolving Coding Errors}: Students should be allowed to use specific AI tools to fix the coding errors they come across.
            \item[$\circ$] \textbf{AI as a Tool, Not a Substitute}: Students could be taught to use AI as a resource to enhance their learning, rather than relying on it to do their work for them. AI can be used for tasks such as research, data analysis, and language translation, but it should not replace critical thinking, problem-solving, or creativity. In addition, it would be useful to know which AI tools and prompts were used that helped with the research to give credit to the tool.
            \item[$\circ$] \textbf{If You Find Cool Gen AI Application(s), Share Your Favorites with the Class?}: If a student discovers a particularly valuable or interesting application, they are encouraged to share it with the class to inspire and inform their peers. Sharing Platform: We can crearte a Miro Board Sharing Platform. Students create a card on Miro with the AI tool's name, a link, and a brief reason for sharing. Tools can be categorized by purpose (e.g., image generation, text processing). Students can explore and vote on tools at any time. Students can share directly on Miro without taking class time.
            \item[$\circ$] \textbf{Combating Hallucinations}: If using AI as a vehicle for information, we must ensure it's correct so I suggest requiring that people find sources to back info found by chatbots just to make sure information is up to date and right. This also helps make sure that people are still having to do research and not just using whatever the chatbot puts out. Students need to show the attempt of avoiding hallucination for the topics they are not familiar with by: 1) Asking the AI to provide reference for the conclusion it made in the answers 2) Adding necessary requirements in the prompt engineering, e.g. "only answer the questions if you are 100\% confident, or else return I don't know" 3) Show extra effort in doing research out of AI for validation (like providing links for reference) 4) provide an explanation of possible sources of hallucination or limitation in the answers, and propose alternative solutions
            \item[$\circ$] \textbf{Course Instructor's AI Acceptance Rules}: After finalization of policies, instructors need to provide clear guidance on what are the acceptable uses of AI and mention if any specific AI use cases are considered inappropriate or prohibited. This ensures that students and instructors are on the same page. This should be accompanied by outlining the consequences for failing to abide to the policy such as academic integrity violations.
            \item[$\circ$] \textbf{Guidelines for Using AI in Course Project Brainstorming}: Students may use AI to assist with brainstorming course project ideas, but the ideas have to ultimately come from students themselves. The AI generated ideas must be screenshotted or written out/cited if used to create your own idea.
            \item[$\circ$] \textbf{Regular Revision of AI Policies}: AI policies need to be flexible and not static. The course policies on AI should be reviewed and updated regularly to make it relevant to each class experience, expectation and intended outcomes.
        }
        \end{itemize}
        \item \small{\textbf{Class 2 -- PolicyCraft condition} (8 policies, excluding 3 without majority vote):}
        \begin{itemize}
        \footnotesize{
            \item[$\circ$] \textbf{Using AI to Combat the Language Barrier}: AI is acceptable to use for translation, improving grammar, and anything related to understanding language. However, AI translation that is directly translated from one language to another, specifically in reading responses or writing assignments, shouldn't be submitted as one's own work. Translated words, phrases, and small sentences can be used though.
            \item[$\circ$] \textbf{Image Generation}: We can use GenAI to generate images to aid our essays or responses, with only restrictions being course guidelines (no explicit nature, etc). We should cite any use of GenAI for generating images.
            \item[$\circ$] \textbf{AI to be Used as a Helping Hand or Scaffold, not a Crutch}: Having AI capabilities to help provide insight rather than completely doing our work, specifically on helping hasten the process but not making a whole new design/idea from scratch. We suggest two conditions for how to avoid using AI as a "crutch". 1) If the task a student is getting support on is related to learning goals (i.e., it shouldn't be abstracted away), students should invest their own effort independent of a generative AI before they begin to use one. For example, as one might in an internship, students might spend 30 minutes on a task, and if they are still struggling, they might seek support from an instructor, peer, or generative model. 2) Students should engage with any ideas that a model produces or inspires in them. Rather than determining who owns an idea (of course still using attribution), course policy should focus on to what extent an idea offered a concrete learning experience for a student. If a student encounters a great idea while working with AI, and thinks critically about how they want to embark on a project based on that idea, that should suffice. Their ideas will continue to evolve. In contrast, simply copying and pasting idea without engaging with it (e.g., expanding upon it, considering how to actualize it, considering how one's perspective is related to it, etc., even if one does not edit the idea itself) would be less desirable.
            \item[$\circ$] \textbf{AI Usage Permitted for Coding Assignments}: Students may freely use AI for coding assignments with appropriate attribution. However, students should show understanding of any code AI has outputted.
            \item[$\circ$] \textbf{Developing a System that Uses AI}: Students may create systems that use AI, e.g., including a foundation model package or calling a foundation model API, so long as they make users aware of any potential harms and make efforts to avoid such harms in their design of the system (e.g., sharing sensitive chat data with OpenAI's API). Students should be aware of third-party APIs' data use policies in any cases in which they anticipate that users may share sensitive data.
            \item[$\circ$] \textbf{Usage of GenAI Tools should be Referenced and Cited}: In assignments or presentations, students should declare and cite the genAI tools and prompts that they have used to create content or help that they have received openly. While some cases involve grammar checks and simple paraphrasing for fluency, the original idea comes from the student. However, when students use GenAI to generate code or ideas for responses, it is essential to add a reference.
            \item[$\circ$] \textbf{Using AI for Rapid Prototyping}: With a general Idea of where to head, sometimes creating a lo-fi or hi-fi model can be hard for development due to a lack of skills expertise, no one's fault. But to bridge the gap and allow some fruition to catapult the project forward, AI can be used to help drive the initial run and allow ideas to get out of the base stages.
            \item[$\circ$] \textbf{Using AI for Citations}: Using Generative AI for creating citations is permitted. However, the student is responsible for checking the generated citation for accuracy and ensuring that all sources are properly cited in any assignment they submit.
        }
        \end{itemize}
        \item \small{\textbf{Class 2 -- baseline condition} (7 policies, excluding 12 without majority vote):}
        \begin{itemize}
        \footnotesize{
            \item[$\circ$] \textbf{Caution Against Misinformation}: Students are strongly encouraged to verify the accuracy and conduct fact-checks on any AI-generated content before including it in their assignments, as AI can sometimes generate false information, including fabricated quotes, citations, and research papers. Students should be responsible the accuracy of AI-generated information used in assignments.
            \item[$\circ$] \textbf{AI Use for Writing a Reading Response}: AI use in Reading Responses is limited to the following: (1) refining or clarifying your own written ideas, (2) assist with grammar, syntax, and minor rephrasing on an human written draft, and (3) cannot be used in generating original content or structuring arguments.
            \item[$\circ$] \textbf{Accountability of AI Responses}: When students use AI generated responses, they should be aware that the generated responses might not be 100\% accurate. Students using information given by AI should be responsible for the accuracy of that information.
            \item[$\circ$] \textbf{Using AI for Group Organization}: AI can be used to assist in organizing group work such as drafting project outlines and summarizing meeting notes, provided that group members contribute to reviewing, editing, and finalizing these decisions
            \item[$\circ$] \textbf{Universal AI Attribution Policy}: If AI was used for a particular assignment, written notice must be given to the professor using the appropriate technology of submission for that assignment (e.g. comments in one's code if programming, the comment box of a canvas submission, etc.) outlining how AI was used in a particular work. This policy takes precedence over all other policy and is necessary to prevent any legal copyright/cheating issues.
            \item[$\circ$] \textbf{AI Usage Permitted for Coding Assignments}: Students may freely use AI for coding assignments, but must comment around (i.e. above and below) sections of code created using AI indicating AI attribution
            \item[$\circ$] \textbf{Cannot Create Segments/Personas/Archetypes}: GenAI can be used to assist in generating rough personas for general insights, pain points, or understanding the targeted audience. However, it should not be used to create detailed user segments or personas based on qualitative criteria like psychographics or behaviors, nor should it be the sole method for user profiling.
        }
        \end{itemize}
\end{itemize}

\section{Additional Detail on Data Analyses}

\subsection{Entropy Analysis}\label{sec:entropy_analysis}

\begin{figure}[h]
  \centering
  \includegraphics[width=0.95\linewidth]{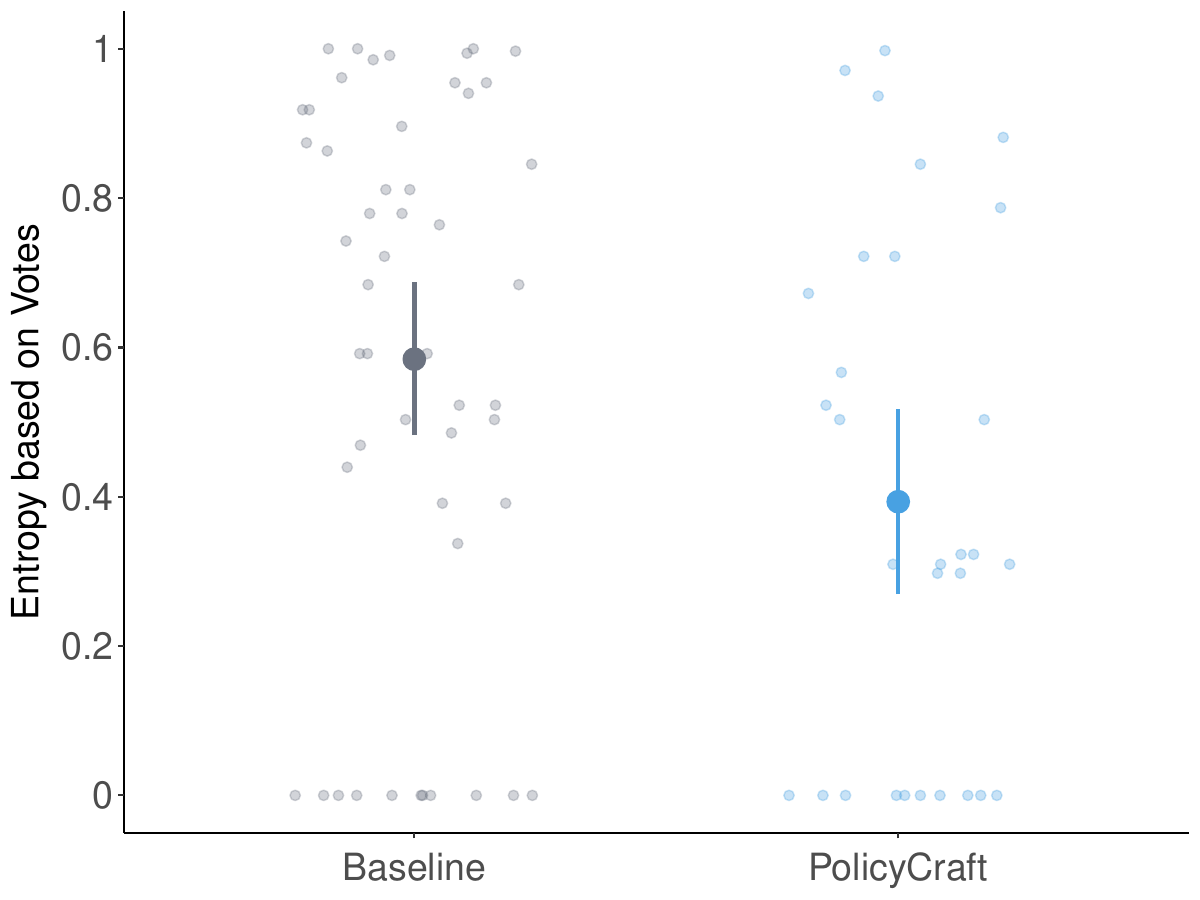}
  \caption{The distribution of entropy for individual policies calculated using the proportion of upvotes and downvotes each policy received. Lower entropy indicates a higher level of voting consensus.}
  \Description{A scatter plot showing the system participants used during the study on the x-axis (Baseline or PolicyCraft) and the entropy of individual policies on the y-axis.}
  \label{fig:policy_vote_entropy}
\end{figure}

In addition to the mean entropy reported in Table~\ref{table:policy_stats}, we fit a linear regression to analyze the impact of study condition (baseline = 0, PolicyCraft = 1) on policy entropy, controlling for between-class differences. As shown in Table~\ref{table:entropy_regression} and Figure~\ref{fig:policy_vote_entropy}, entropy was lower for policies developed using PolicyCraft compared with the baseline ($p < 0.05$).

\begin{table*}[t]
  \caption{Regression coefficients. Consensus was significantly higher (lower entropy) for policies developed in the PolicyCraft condition.}
  \label{table:entropy_regression}
  \begin{tabular}{lrrrrr}
    \toprule
    variable & estimate & std. error & t value & p value & \\
    \midrule
    class & -0.1588 & 0.0815 & -1.948 & 0.0550 & \\
    condition & -0.1930 & 0.0815 & -2.368 & \textbf{0.0204} & *\\
    \bottomrule
  \end{tabular}
\end{table*}

\subsection{Likert Data Analysis}\label{sec:likert_data_analysis}
In our analysis of participants' Likert scale ratings in Section \ref{sec:result_2}, we employed both linear and ordinal regression to ensure the robustness of our results, considering the treatment of Likert scale data as either continuous or ordinal \cite{robitzsch2020ordinal}. The appropriateness of using ordinal versus linear regression for the analysis of Likert scale data has been a subject of wide debate, with recent scholarship showing that each approach has complementary benefits and drawbacks~\cite{robitzsch2020ordinal}. In both of our models, Likert ratings are the dependent variable and study condition (baseline or PolicyCraft) is a binary independent variable. The class ID is included as an additional control variable, to account for potential between-class differences. As shown in Table \ref{table:likert_data}, our findings are robust to the choice of ordinal or linear regression.

\begin{table*}[t]
  \caption{Coefficients for both linear and ordinal regressions analyzing participants' ratings of the extent to which they could ``easily understand why people agree or disagree with each other''.}
  \label{table:likert_data}
  \begin{tabular}{llrrrrr}
    \toprule
    & variable & estimate & std. error & t value & p value & \\
    \midrule
    linear regression & class & -0.2266 & 0.3276 & -0.692 & 0.4916 & \\
    & condition & 0.9066 & 0.3190 & 2.842 & 0.0060 & **\\
    \midrule
    ordinal regression & class & -0.1908 & 0.4575 & -0.417 & 0.6768 & \\
    & condition & 1.2513 & 0.4621 & 2.708 & 0.0068 & **\\
    \bottomrule
  \end{tabular}
\end{table*}

\end{document}